\documentclass[11pt]{article}

\usepackage{epsfig,graphics}
\usepackage{amssymb,amsmath,amsthm,bm}

\newtheorem{theo}{Theorem}

\newtheorem{pro}{Proposition}

\def\R{{\mathbb R}}

\def\ga{\alpha}

\def\go{\omega}

\def\gs{\sigma}
\def\wt{\widetilde}
\def\n{\noindent}

\def\gt{\triangle}
\def\gD{\Delta}
\def\cS{\mathcal {S}}

\def\wh{\widehat}
\def\wt{\widetilde}

\begin{document}

\title{Adaptive Short-time Fourier Transform and Synchrosqueezing Transform for Non-stationary Signal Separation\thanks{This work was supported in part by the National Natural Science Foundation of China (Grant No. 61803294) and Simons Foundation (Grant No. 353185)}
}

\author{Lin Li${}^{1}$, Haiyan Cai${}^{2}$, Hongxia Han${}^{1}$,  Qingtang Jiang${}^2$
\; and Hongbing Ji${}^{1}$
}

\date{}

\maketitle

{\small 1. School of Electronic Engineering, Xidian University, Xi'an 710071, P.R. China}

{\small 2. Dept. of Math \& CS, University of Missouri-St. Louis, St. Louis,  MO 63121, USA}

\begin{abstract}
The synchrosqueezing transform, a kind of reassignment method, aims to sharpen the time-frequency representation and to separate the components of a multicomponent non-stationary signal. In this paper, we consider the short-time Fourier transform (STFT) with a time-varying parameter, called the adaptive STFT. Based on the local approximation of linear frequency modulation mode, we analyze the well-separated condition of non-stationary multicomponent signals using the adaptive STFT with the Gaussian window function.  We propose the STFT-based synchrosqueezing transform (FSST) with a time-varying parameter, named the adaptive FSST, to enhance the time-frequency concentration and resolution of a multicomponent signal, and to separate its components more accurately. In addition, we also propose the 2nd-order adaptive FSST to further improve the adaptive FSST for the non-stationary signals with fast-varying frequencies. Furthermore, we present a localized optimization algorithm based on our well-separated condition
to estimate the time-varying parameter adaptively and automatically. Simulation results on synthetic signals and the bat echolocation signal are provided to demonstrate the effectiveness and robustness of the proposed method.

\bigskip

Keywords: {\it  Instantaneous frequency, Adaptive short-time Fourier transform, Adaptive synchrosqueezing transform, Well-separated condition for  multicomponent  non-stationary signal, Component recovery of  non-stationary signal }

\end{abstract}

\section{Introduction}
To model a non-stationary signal as a superposition of locally band-limited, amplitude and frequency-modulated 
Fourier-like oscillatory modes:
\begin{equation}
\label{AHM}
x(t)=\sum_{k=1}^K x_k(t), \qquad x_k(t)=A_k(t) e^{i2\pi \phi_k(t)}, 
\end{equation}
where $A_k(t), \phi_k'(t)>0$, has been a very active research area over the past few years. 
Note that the number of component $K$ may change with time $t$, but it should be constant for long enough time intervals.
The representation of $x(t)$ in \eqref{AHM}  with $A_k(t)$ and  $\phi_k'(t)$ varying slowly or  more slowly  than $\phi_k(t)$ is called an {adaptive harmonic model (AHM) representation} of $x(t)$, where  $A_k(t)$ are called the instantaneous amplitudes  and $\phi'_k(t)$ the instantaneous frequencies (IFs).
To decompose $x(t)$ as an AHM representation \eqref{AHM} is important to extract information, such as the underlying dynamics,  hidden in $x(t)$.

Time-frequency (TF) analysis 
is widely used in engineering fields such as communication, radar and sonar as a powerful tool for analyzing time-varying non-stationary signals \cite{Leon_Cohen}.
Time-frequency analysis  is especially useful for signals containing many oscillatory components with slowly time-varying amplitudes and instantaneous frequencies.
The short-time Fourier transform (STFT), the continuous wavelet transform (CWT) and the Wigner-Ville distribution  are the most typical TF analysis, see details in \cite{Leon_Cohen}-\cite{MODFM16}.
Other TF distributions of Cohen's class include the exponential distribution \cite{Choi89}, 
a smoothed pseudo Wigner distribution \cite{Stank94} and the complex-lag distribution \cite{Stank09}. 
In addition, the TF signal analysis and synthesis using the eigenvalue decomposition method has been studied \cite{HMB04,Stank06}. In particular, an eigenvalue decomposition-based approach 
which enables the separation of non-stationary components with overlapped supports in the TF plane has been proposed in \cite{Stank18}. 

Recently a number of new TF analysis methods such as the Hilbert spectrum analysis with empirical mode decomposition (EMD) \cite{Huang98}, the reassignment method \cite{A_Flandrin_reassignment95}   and synchrosqueezed wavelet transform (SST)  \cite{Daub_Maes96} have  also been proposed to obtain  $A_k(t)$ and $\phi_k'(t)$.

EMD is a data-driven decomposition algorithm which separates the time series signal into a set of monocomponents, called intrinsic mode functions (IMFs) \cite{Huang98}. EMD has been studied by many researchers and has been used in many applications, see e.g. \cite{Flandrin04}-\cite{ShPa18}.  Because of the presence of widely disparate scales in a single IMF, or a similar scale residing in different IMF components, named as mode mixing \cite{Oberlin12a}, two close IMFs are hardly distinguished by EMD. 

The CWT-based  synchrosqueezing transform (WSST), introduced in \cite{Daub_Maes96} and future studied in \cite{Daub_Lu_Wu11},
is a special case of reassignment methods, which aims to sharpen the TF representation of the signal by allocating the CWT coefficient value to a different point in the TF  plane. A variant of WSST, the STFT-based SST (FSST) was proposed in \cite{Thakur_Wu11} and further  studied in \cite{Wu_thesis, MOM14}. 
Both WSST and FSST have been proved to be robust to noise and small perturbations \cite{Thakur_etal_Wu13}-\cite{Meignen17}.
However for frequency-varying signals, the squeezing effect of SST is not desirable. In this regard, a 2nd-order SST 
was introduced in \cite{MOM15,OM17} and further studied in \cite{FACMF17, BMO18}.
The 2nd-order SST improves the concentration of the TF representation well on perturbed linear chirps with Gaussian modulated amplitudes. The higher-order FSST was presented in \cite{Pham17}, which aims to handle signals containing more general types.

Other SST related methods include the generalized SST \cite{Li_Liang12},
a hybrid EMD-SST computational scheme \cite{Chui_Walt15},
the synchrosqueezed wave packet transform \cite{Yang15}, the S-transform-based SST \cite{S_transform_SST15},  
SST with vanishing moment wavelets \cite{Chui_Lin_Wu15},
the multitapered SST \cite{Daub_Wang_Wu15} and the demodulation-transform based SST  \cite{Wang_etal14,  Jiang_Suter17}.
In addition, the synchrosqueezed curvelet transform for two-dimensional mode decomposition was introduced in \cite{YangY14},
the signal separation operator which is related to FSST was proposed in \cite{Chui_Mhaskar15} and
the empirical signal separation algorithm was introduced in \cite{LCJJ17}.
The statistical analysis of synchrosqueezed transforms has been studied in \cite{Yang18} and a new IF estimator  within the framework of the signal's phase derivative and the linear canonical transform was introduced in \cite{ZYLD18}.
SST has been used in machine fault diagnosis \cite{Li_Liang_fault12,WCSGTZ18},
crystal image analysis \cite{Yang_Crystal_15,Yang_Crystal_18},
welding crack acoustic emission signal analysis \cite{HLY18},
and medical data analysis  \cite{Wu_breathing14}-\cite{Wu_heartbeat17}.

Most of the  FSST algorithms available in the literature are based on the short-time Fourier transform (STFT) with a  fixed window, which means high time resolution and frequency resolution cannot be obtained simultaneously. For broadband signals, a wide window is suitable for the low-frequency parts. On the contrary, a narrow window is suitable for the high-frequency parts. To enhance the TF  resolution and energy concentration, we propose in this paper the adaptive FSST 
based on the STFT  with a time-varying  window.  More precisely, let $V_x(t, \eta)$ be
the (modified) STFT of $x(t)\in L_2(\R)$ with  a window function $h(t)\in L_2(\R)$ defined by
\begin{eqnarray}
\label{def_STFT}
V_x(t, \eta) \hskip -0.6cm&&:=\int_{-\infty}^\infty x(\tau) h(\tau-t)e^{-i2\pi \eta (\tau-t)}d\tau\\
\label{STFT2} &&=\int_{-\infty}^\infty x(t+\tau)h(\tau)e^{-i2\pi \eta\tau}d\tau,
\end{eqnarray}
where $t$ and $\eta$ are the time variable and the frequency variable respectively.
In this paper we consider the STFT with a time-varying parameter $\gs(t)$ (called the adaptive STFT) defined by
\begin{eqnarray}
\label{def_STFT_para1}
\wt V_{x}(t, \eta) 
\hskip -0.6cm &&:=\int_{-\infty}^\infty x(\tau)g_{\gs(t)}(\tau-t)e^{-i2\pi \eta (\tau-t)}d\tau\\
\label{def_STFT_para2}
&&=\int_{-\infty}^\infty x(t+\tau)\frac 1{\gs(t)}g (\frac \tau{\gs(t)})e^{-i2\pi \eta\tau}d\tau,
\end{eqnarray}
where $\gs=\gs(t)$ is a positive function of $t$, and $g_{\gs(t)}(\tau)$ is defined by
\begin{equation}
\label{def_g_dilation}
g_{\gs(t)}(\tau):=\frac 1{\gs(t)}g(\frac \tau {\gs(t)}),
\end{equation}
with $g\in L_2(\R)$. The window width of $g_{\gs(t)}(\tau)$ is $\gs(t)$ (up to a constant), depending on the time variable $t$.
In this paper we consider the FSST based on $\wt V_{x}(t, \eta)$ (called the adaptive FSST) and study the choice of the time parameter $\gs(t)$ so that the adaptive FSST gives a better
instantaneous frequency estimation of the component of a multicomponent signal, and provides more accurate component recovery.

To recover/separate the components $x_k(t)$ of  a multicomponent signal as given by \eqref{AHM} with the SST approach,
\cite{Iatsenko15} indicates that  if STFTs  $V_{x_{k-1}}(t, \eta)$ and $V_{x_{k}}(t, \eta)$ of two components $x_{k-1}(t)$ and $x_k(t)$ are mixed, then FSST cannot separate these two components $x_{k-1}(t)$ and $x_k(t)$  either, and hence it cannot recover/separate components accurately. Thus it is desirable that appropriate 
window width of the window function $h(t)$
can be chosen (if possible) so that the STFTs of different components do not overlap. When $x_k(t)$ are sinusoidal signals 
$A_k e^{i2\pi c_k t}$  for some constants $A_k, c_k>0$
or they are well approximated by sinusoidal functions at any local time in the sense that
for any $t\in \R$, 
\begin{equation}
\label{sinu_model}
x_k(t+\tau)
=A_k(t+\tau)e^{i2\pi \phi_k(t+\tau)}
\approx A_k(t)e^{i2\pi \big(\phi_k(t)+\phi_k'(t) \tau\big)}
=x_k(t) e^{i2\pi \phi_k'(t) \tau}
\quad \hbox{for $\tau\approx 0$},
\end{equation}
then the STFT $V_{x_k}(t, \eta)$ of $x_k(t)$ with a window function $h(t)$ is
$$
V_{x_k}(t, \eta) \approx x_k(t)\wh h(\eta-\phi_k'(t)),
$$
where $\wh h$ denotes the  Fourier transform of $h(t)$.
Hence if  supp($\wh h)\subseteq [-\gt, \gt]$ for some $\gt>0$, then
$ V_{x_k}(t, \eta)$ lies in the TF zone given by
$$
{\cal Z}_k=\{(t, \eta): |\eta-\phi_k'(t)|< {\gt}, t\in \R\}.
$$
Therefore, if
\begin{equation}
\label{freq_resolution}
\phi'_k(t)-\phi'_{k-1}(t)\ge 2\gt, \; t\in \R, 2\le k\le K,
\end{equation}
then ${\cal Z}_k\cap {\cal Z}_{\ell}=\O, k\not=\ell$, which means the components of $x(t)$ are well separated in the TF plane. \eqref{freq_resolution} is a required condition for the study of FSST  in \cite{Wu_thesis, MOM14} and even for the study of the 2nd-order FSST in  \cite{MOM15,BMO18}. Here we call  \eqref{freq_resolution} the sinusoidal signal model-based well-separated condition for $x(t)$. 

In this paper we use the linear frequency modulation (LFM) signal to approximate a non-stationary signal at any local time to study
the TF zone of the adaptive STFT  $\wt V_{x_k}(t, \eta)$. More precisely,  we assume that each $x_k(t)$
is well approximated by an LFM at any local time: for any $t\in \R$,
\begin{eqnarray}
\nonumber &&x_k(t+\tau)
=A_k(t+\tau)e^{i2\pi \phi_k(t+\tau)}
\\
\label{LFM_model}
&&\qquad
\approx A_k(t) e^{i2\pi (\phi_k(t)+\phi_k'(t) \tau+ \frac 12 \phi_k''(t) \tau^2)}
=x_k(t) e^{i2\pi (\phi_k'(t) \tau+ \frac 12 \phi_k''(t) \tau^2)}
\quad \hbox{for $\tau\approx 0$},
\end{eqnarray}
where for a given $t$, the quantity in \eqref{LFM_model} as a function $\tau$ is called an LFM signal (or a linear chirp signal). Thus we have
\begin{eqnarray}
\wt V_{x_k}(t, \eta)
\label{linear_chirp_model}
\approx  \int_\R x_k(t)e^{i2\pi (\phi_k'(t) \tau +\frac12\phi''_k(t) \tau^2)}\frac 1{\gs(t)}g(\frac \tau{\gs(t)}) e^{-i2\pi \eta \tau}d\tau.
\end{eqnarray}
In this paper we will obtain the LFM model-based well-separated condition which guarantees that for different $k$, the quantities as functions of $(t, \eta)$ on the right-hand side of \eqref{linear_chirp_model} lie within non-overlapping zones in the TF plane when $g$ is the Gaussian window function. We will also discuss how to select the time-varying parameter $\gs(t)$ such that the corresponding adaptive FSST and 2nd-order adaptive FSST have sharp TF representation. In particular, we propose a localized optimization method based on our well-separated condition to estimate the time-varying adaptive window width $\gs(t)$.

The idea of using an optimal time-varying window parameter (window width) has been studied or considered extensively in the literature, see e.g. \cite{Jones94}-\cite{Wu17}.
In particular, the authors in \cite{Wu17} introduced  a method to select the time-varying window width for sharp SST representation by minimizing the R${\rm \acute e}$nyi entropy.
In addition, after we completed our work, we were aware of the very recent work \cite{Saito17} on the adaptive STFT-based SST with the window function containing time and frequency parameters.
Our motivation is different from others in that we do not focus on the optimal parameter such that the corresponding STFT has the sharpest representation in the TF plane. Instead, we pursue the establishment of the LFM model-based well-separated condition for multicomponent signals based on the adaptive STFT and we propose how to select window width $\gs(t)$ such that the STFTs of the components lie in non-overlapping regions of the TF plane. The R${\rm \acute e}$nyi entropy-based optimal parameter may give the overall sharp representation of STFT or FSST, but it does not guarantee all the components to be separated. The selected $\gs(t)$ proposed by us does not necessarily result in a sharp representation of the associated STFT. Instead, it is selected in such a way that the adaptive STFTs of the components are well separated, and the corresponding adaptive FSSTs have sharp representation and hence the components can be recovered more accurately. 

The remainder of this paper is organized as follows. We introduce the adaptive STFT and adaptive FSST with a time-varying parameter in Section 2, where  we also introduce the 2nd-order adaptive FSST. We derive the optimal time-varying parameter for a monocomponent signal based on the LFM  model in Section 3. In Section 4, we establish the LFM model-based well-separated condition for multicomponent signals. In Section 5 we propose a localized optimization method on the selection of window parameters based on our well-separated condition. Experimental results are provided in Section 6.  Finally we give the conclusion in Section 7.

\section{STFT and FSST with a time-varying parameter}
In this section we first provide a brief review of FSST, then we propose the adaptive FSST based on the STFT with a time-varying parameter.

\subsection{Short-time Fourier transform-based synchrosqueezed transform}

Recall that $V_x(t, \eta)$ is the STFT of $x(t)$ defined by \eqref{def_STFT}, which can be extended
to a slowly growing $x(t)$ provided that  the window function $h(t)$ is in the Schwarz class $\cS$.

The idea of FSST is to reassign the frequency variable. As in \cite{Thakur_Wu11}, for a signal $x(t)$, at $(t, \eta)$ for which $V_x(t, \eta)\not=0$, denote
\begin{equation}
\label{def_phase}
\go_x(t, \eta)=\frac{\frac {\partial}{\partial t} V_x(t, \eta)}{2\pi i V_x(t, \eta)}. 
\end{equation}
When $x(t)=A e^{i 2\pi c t}$, where $A, c$ are constants with $c>0$, then  $\go_x(t, \eta)$ is exactly $c$, the IF of $x(t)$.
The quantity $\go_x(t, \eta)$ is called the ``phase transformation"  \cite{Daub_Lu_Wu11}. 
FSST is to reassign the frequency variable $\eta$ by transforming STFT $V_x(t, \eta)$ of $x(t)$ to a quantity, denoted by $R_x(t, \eta)$, on the TF plane:
\begin{equation}
\label{def_FSST_simple}
R_x(t, \xi):=\int_{\{\zeta: V_x(t, \zeta)\not=0\}} V_x(t, \zeta) \delta\big(\go_x(t, \zeta)-\xi\big) d \zeta,
\end{equation}
where $\xi$ is the frequency variable.

For a multicomponent signal $x(t)$ given by \eqref{AHM}, when $A_k(t), \phi_k(t)$ satisfy certain conditions, each component $x_k(t)$ can be recovered from its FSST: 
\begin{equation}
\label{reconst_SST_component}
x_k(t)\approx \frac 1{h(0)}\int_{|\xi-\phi'_k(t)|< \Gamma} R_x(t, \xi)d\xi,
\end{equation}
for certain $\Gamma>0$. For more mathematically precise definition of FSST and the conditions on $A_k(t), \phi_k(t)$ for \eqref{reconst_SST_component}, see  \cite{Thakur_Wu11}-\cite{MOM14}.

\subsection{Adaptive STFT with a time-varying parameter}

We consider the window function given by
\begin{equation*}
g_\gs(t)=\frac 1\gs g(\frac t\gs),
\end{equation*}
where $\gs>0$ is a parameter, $g(t)$ is a positive function in $L_2(\R)$ with $g(0)\not=0$ and having certain decaying order as $t\rightarrow \infty$. If
\begin{equation}
\label{def_g}
g(t)=\frac 1{\sqrt {2\pi}} e^{-\frac {t^2}2},
\end{equation}
then $g_\gs(t)$ is the Gaussian window function. The parameter $\gs$ is also called the window width in the time-domain of the window function $g_\gs(t)$ since the time duration  $\Delta_{g_\gs}$ of $g_\gs$ is $\gs$ (up to a constant): $\Delta_{g_\gs}=\gs\Delta_{g}$, where $\Delta_{g}$ is the time duration of $g$ .
The parameter $\gs$ affects the shape of $g_\gs$ and hence, the representation of the STFT of a signal with $g_\gs$.
As mentioned in Section 1, \cite{Iatsenko15} states that
for a multicomponent signal as given by \eqref{AHM}, if STFTs  $V_{x_{k-1}}(t, \eta)$ and $V_{x_{k}}(t, \eta)$ of two components $x_{k-1}$ and $x_k$ are mixed, then FSST cannot separate these two components $x_{k-1}(t)$ and $x_k(t)$  either. Thus it is desirable that an appropriate $\gs$ can be chosen so that the STFTs of different components do not overlap.
In this paper we introduce STFT with a time-varying parameter and then establish the separability condition of a multicomponent signal  based on this type of STFT.

The STFT of $x(t)$ with a time-varying parameter  $\wt V_{x}(t, \eta)$ (called the adaptive STFT) we consider
 is defined by \eqref{def_STFT_para1}. One can verify that $\wt V_{x}(t, \eta)$  can be written as
\begin{eqnarray}
\label{STFT_para_freqdomain}
\wt V_{x}(t, \eta)=\int_{-\infty}^\infty \wh x(\xi)\wh g_{\gs(t)}(\eta-\xi)e^{i2\pi t \xi}d\xi =
\int_{-\infty}^\infty \wh x(\xi)\wh g\big(\gs(t)(\eta-\xi)\big)e^{i2\pi t \xi}d\xi.
\end{eqnarray}
where for a signal $x(t)$, its Fourier transform $\wh x(\xi)$ is defined by
$$
\wh x(\xi)=\int_{-\infty}^\infty x(t)e^{-i2\pi \xi t} dt.
$$
We can obtain that $x(t)$ can be recovered from $\wt V_{x}(t, \eta)$ 
as shown in the following theorem.
\begin{theo}
\label{theo:recover_STFT_para}
Let $\wt V_{x}(t, \eta)$  be the time-varying STFT of $x(t)\in L_2(\R)$ defined by \eqref{def_STFT_para1}. Suppose $\wh x, \wh g \in L_1(\R)$.  Then
\begin{equation}
\label{STFT_recover_para}
x(t)=\frac {\gs(t)}{g(0)} \int_{-\infty}^\infty \wt V_{x}(t, \eta) d\eta.
\end{equation}
If in addition $g(t)$ is real-valued, then for a real-valued $x(t)$, we have
\begin{equation}
\label{STFT_recover_real_para}
x(t)= \frac{2\gs(t)}{g(0)}  {\rm Re} \Big( \int_0^\infty \wt V_{x}(t, \eta)  d\eta\Big).
\end{equation}
\end{theo}

The proof of Theorem \ref{theo:recover_STFT_para} is presented in Appendix.

\subsection{Adaptive FSST with a time-varying parameter}

Next we introduce the  synchrosqueezing transform (SST) associated with the adaptive STFT. First we need to define the phase transformation $\go^{adp}_x$ associated with the adaptive STFT.
In the following we use 
$g(\tau)$ replaced by $\tau g'(\tau)$, namely,
\begin{eqnarray}
\label{def_STFT2_para}
&&\wt V^{\tau g'(\tau)}_x(t, \eta):=
\int_{-\infty}^\infty x(t+\tau)\frac \tau{\gs^2(t)}{g'(\frac \tau{\gs(t)})}e^{-i2\pi \eta \tau} d\tau.
\end{eqnarray}

To define the phase transformation $\go^{adp}_x$, we first consider  $s(t)=A  e^{i2\pi c t}$.
From
$$
\wt V_s(t, \eta)=\int_{-\infty}^\infty s(t+\tau) {g_{\gs(t)}(\tau)}e^{-i2\pi \eta \tau} d\tau=A\int_{-\infty}^\infty e^{i2\pi c (t+\tau)} \frac 1{\gs(t)}g(\frac \tau{\gs(t)})e^{-i2\pi \eta \tau} d\tau,
$$
we have
\begin{eqnarray*}
&&\frac {\partial} {\partial t} \wt V_s(t, \eta)=A\int_{-\infty}^\infty (i2\pi c)e^{i2\pi c (t+\tau)}\;\frac 1{\gs(t)}g(\frac \tau{\gs(t)})e^{-i2\pi \eta \tau} d\tau
\\
&&\qquad + A \int_{-\infty}^\infty e^{i2\pi c (t+\tau)}(-\frac {\gs'(t)}{\gs(t)^2})g(\frac \tau{\gs(t)})e^{-i2\pi \eta \tau} d\tau
 + A\int_{-\infty}^\infty e^{i2\pi c (t+\tau)} (-\frac {\gs'(t) \tau}{\gs(t)^3}) {g'(\frac \tau{\gs(t)})}e^{-i2\pi \eta \tau} d\tau\\
&&=i2\pi c\; \wt V_s(t, \eta)- \frac {\gs'(t)}{\gs(t)}\; \wt V_s(t, \eta) -\frac {\gs'(t)}{\gs(t)}\; \wt V^{\tau g'(\tau)}_s(t, \eta).
\end{eqnarray*}
Thus, if $\wt V_s(t, \eta)\not=0$, we have
\begin{equation*}
\frac {\frac{\partial}{\partial t} \wt V_s(t, \eta)}{i 2\pi \wt V_s(t, \eta)}=
c- \frac {\gs'(t)}{i2\pi \gs(t)}- \frac {\gs'(t)}{\gs(t)} \frac {\wt V^{\tau g'(\tau)}_s(t, \eta)}{i2\pi \wt V_s(t, \eta)}.
\end{equation*}
Therefore, the IF of $s(t)$, which is $c$, can be obtained by
\begin{equation}
\label{para_derivation2}
c={\rm Re}\Big\{\frac {\frac{\partial}{\partial t} \wt V_s(t, \eta)}{i2\pi \wt V_s(t, \eta)}\Big\}
+ \frac {\gs'(t)}{\gs(t)} {\rm Re}\Big\{ \frac {\wt V^{\tau g'(\tau)}_s(t, \eta)}{i2\pi \wt V_s(t, \eta)} \Big\}, \quad \hbox{for $\wt V_s(t, \eta)\not=0$}.
\end{equation}
Hence, for a general $x(t)$, at $(t, \eta)$ for which $\wt V_x(t, \eta)\not=0$,  the quantity in the right-hand side of the above equation is a good candidate for the IF of $x$. This quantity is also called the {phase transformation},
and we denote it by $\go^{adp}_x(t, \eta)$:
\begin{equation}
\label{def_phase_para}
\go^{adp}_x(t, \eta)={\rm Re}\Big\{\frac{\partial_t\big(\wt V_x(t, \eta)\big)}{i2\pi \wt V_x(t, \eta)}\Big\}+ \frac {\gs'(t)}{\gs(t)} {\rm Re}\Big\{ \frac {\wt V^{\tau g'(\tau)}_x(t, \eta)}{i2\pi \wt V_x(t, \eta)} \Big\},  \quad \hbox{for $\wt V_x(t, \eta)\not=0$}.
\end{equation}

The FSST with a time-varying parameter (called the adaptive FSST of $x(t)$) is defined by
\begin{equation}
\label{def_FSST_para_simple}
R^{adp}_x(t, \xi) :=\int_{\{\eta \in \R: \; \wt V_x(t, \eta)\not=0\}}\wt V_x(t, \eta)
 \delta\big(\go^{adp}_x(t, \eta)-\xi\big) d\eta,
\end{equation}
where $\xi$ is the frequency variable. The reconstruction formulas in \eqref{STFT_recover_para} and
\eqref{STFT_recover_real_para} lead to that $x(t)$ can be reconstructed from its adaptive FSST:
\begin{equation}
\label{FSST_recover_para}
x(t)=\frac {\gs(t)}{g(0)} \int_{-\infty}^\infty R^{adp}_x(t, \xi) d\xi;
\end{equation}
and if in addition $g(t)$ is real-valued, then for real-valued $x(t)$, we have
\begin{equation}
\label{FSST_recover_real_para}
x(t)= \frac{2\gs(t)}{g(0)}  {\rm Re} \Big( \int_0^\infty R^{adp}_x(t, \xi) d\xi\Big). 
\end{equation}
One can use the following formula to recover the $k$th component $x_k(t)$ of a multicomponent signal from the adaptive FSST:
\begin{equation}
\label{FSST_recover_para_component}
x_k(t)= \frac{2\gs(t)}{g(0)}   {\rm Re} \Big( \int_{|\xi-\phi_k'(t)|<\Gamma_1} R^{adp}_x(t, \xi) d\eta\Big)
\end{equation}
for some $\Gamma_1>0$.

\subsection{Second-order adaptive FSST}
The 2nd-order FSST was introduced in \cite{MOM15}.
The main idea is to define a new phase transformation $\go_x^{2nd}$ such that when $x(t)$ is a linear frequency modulation (LFM) signal, 
then $\go_x^{2nd}$ is exactly the IF of $x(t)$. We say $s(t)$ is an LFM signal or a linear chirp if
\begin{equation}
\label{def_chirp}
s(t)=A e^{i2\pi \phi(t)}=A e^{i2\pi (ct +\frac 12 r t^2)}
\end{equation}
with phase function $\phi(t)=ct +\frac 12 r t^2$, IF  $\phi'(t)=c +r t$ and chirp rate  $\phi''(t)=r$.

Recall that in Section 2.3, we use $\wt V^{\tau g'(\tau)}_x(t, \eta)$ to denote the adaptive STFT defined by \eqref{def_STFT_para1} with  $g(\tau)$ replaced by $\tau g'(\tau)$. In the following, we use
 $\wt V^{\tau g(\tau)}_x(t, \eta)$ to denote the adaptive STFT defined by \eqref{def_STFT_para1} with  $g(\tau)$ replaced by $\tau g(\tau)$. That is,
\begin{eqnarray}
\label{def_STFT1_para}
&&\wt V^{\tau g(\tau)}_x(t, \eta):=
\int_{-\infty}^\infty x(t+\tau) \frac \tau{\gs^2(t)}g(\frac \tau{\gs(t)})e^{-i2\pi \eta \tau}d\tau.
\end{eqnarray}

For a signal $x(t)$, we define the phase transformation for the 2nd-order adaptive FSST as
\begin{equation}
\label{2nd_phase_para}
\go^{adp, 2nd}_x(t, \eta)=\left\{
\begin{array}{l}
{\rm Re}\Big\{\frac {\frac{\partial}{\partial t} \wt V_x(t, \eta)}{i2\pi \wt V_x(t, \eta)}\Big\}
+ \frac {\gs'(t)}{\gs(t)} {\rm Re}\Big\{ \frac {\wt V^{\tau g'(\tau)}_x(t, \eta)}{i2\pi \wt V_x(t, \eta)} \Big\}
- {\rm Re}\Big\{ \frac{\wt V^{\tau g(\tau)}_x(t, \eta)}{i2\pi \wt V_x(t, \eta)} P_0(t, \eta)\Big\},\\
\hskip 5cm \hbox{if $\frac{\partial}{\partial \eta}\Big(\frac {\wt V^{\tau g(\tau)}_x(t, \eta)}{\wt V_x(t, \eta)}\Big)\not=0$ and $\wt V_x(t, \eta)\not=0;$}\\
{\rm Re}\Big\{\frac {\frac{\partial}{\partial t} \wt V_x(t, \eta)}{i2\pi \wt V_x(t, \eta)}\Big\}+ \frac {\gs'(t)}{\gs(t)} {\rm Re}\Big\{ \frac {\wt V^{\tau g'(\tau)}_x(t, \eta)}{i2\pi \wt V_x(t, \eta)} \Big\},
\hbox{if $\frac{\partial}{\partial \eta}\Big(\frac {\wt V^{\tau g(\tau)}_x(t, \eta)}{\wt V_x(t, \eta)}\Big)=0$,  $\wt V_x(t, \eta)\not=0$,}
\end{array}
\right.
\end{equation}
where
\begin{equation}
\label{def_R0}
P_0(t, \eta)=\frac 1{\frac{\partial}{\partial \eta}\Big( \frac {\wt V^{\tau g(\tau)}_x(t, \eta)}{\wt V_x(t, \eta)}\Big) }\Big\{\frac{\partial}{\partial \eta}\Big(\frac {\frac{\partial}{\partial t} \wt V_x(t, \eta)}{\wt V_x(t, \eta)}\Big)+ \frac {\gs'(t)}{\gs(t)}
\frac{\partial}{\partial \eta}\Big(\frac {\wt V^{\tau g'(\tau)}_x(t, \eta)}{\wt V_x(t, \eta)}\Big)\Big\}.
\end{equation}
Then we have the following theorem with its proof given in Appendix.
\begin{theo}
\label{theo:2nd_phase_para}
If $x(t)$ is an LFM signal
given by \eqref{def_chirp}, then at $(t, \eta)$ where $\frac{\partial}{\partial \eta}\Big(\frac {\wt V^{\tau g(\tau)}_x(t, \eta)}{\wt V_x(t, \eta)}\Big)\not=0$ and $\wt V_x(t, \eta)\not=0$,
$\go^{adp, 2nd}_x(t, \eta)$ defined by \eqref{2nd_phase_para} is the IF of $x(t)$, namely $\go^{adp, 2nd}_x(t, \eta)=c+r t$.
\end{theo}

Observe that when $\gs(t)\equiv \gs$ is a constant function,  $\go^{adp, 2nd}_x(t, \eta)$ is reduced to $\go^{2nd}_x(t, \eta)$ given by
\begin{equation}
\label{2nd_phase}
\go^{2nd}_x(t, \eta)=\left\{
\begin{array}{ll}
{\rm Re}\Big\{\frac {\frac{\partial}{\partial t} V_x(t, \eta)}{i2\pi V_x(t, \eta)}\Big\}
- {\rm Re}\Big\{ \frac{V^{\tau g(\tau)}_x(t, \eta)}{i2\pi V_x(t, \eta)}p_0(t, \eta)\Big\},&\hbox{if $\frac{\partial}{\partial \eta}\Big(\frac{V^{\tau g(\tau)}_x(t, \eta)}{V_x(t, \eta)}\Big)\not=0, V_x(t, \eta)\not=0;$}\\
{\rm Re}\Big\{\frac {\frac{\partial}{\partial t} V_x(t, \eta)}{i2\pi V_x(t, \eta)}\Big\},&\hbox{if $\frac{\partial}{\partial \eta}\Big(\frac{V^{\tau g(\tau)}_x(t, \eta)}{V_x(t, \eta)}\Big)=0, V_x(t, \eta)\not=0,$}
\end{array}
\right.
\end{equation}
where
$$
p_0(t, \eta) =\frac 1{\frac{\partial}{\partial \eta}\Big(\frac{V^{\tau g(\tau)}_x(t, \eta)}{V_x(t, \eta)}\Big)} \frac{\partial}{\partial \eta}
\Big(\frac {\frac{\partial}{\partial t} V_x(t, \eta)}{V_x(t, \eta)}\Big).
$$
$\go_x^{2nd}$ in \eqref{2nd_phase} is one of the phase transformations considered in \cite{Pham17} for the conventional 2nd-order FSST.

With the phase transformation $\go^{adp, 2nd}_x(t, \eta)$ in \eqref{2nd_phase_para}, we define the 2nd-order FSST with a time-varying parameter, called the 2nd-order adaptive FSST, of a signal $x(t)$ as in  \eqref{def_FSST_para_simple}:
\begin{equation}
\label{def_2ndFSST_para_simple}
R^{adp, 2nd}_x(\xi, t) 
:=\int_{\{\eta \in \R: \; \wt V_x(t, \eta)\not=0\}} \wt V_x(t, \eta) \delta\big(\go^{adp, 2nd}_x(t, \eta)-\xi\big) d\eta,
\end{equation}
where $\xi$ is the frequency variable. We also have the reconstruction formulas for $x(t)$ and $x_k(t)$ similar to \eqref{FSST_recover_para},  \eqref{FSST_recover_real_para} and \eqref{FSST_recover_para_component} with $R^{adp}_x(\xi, t)$ replaced by $R^{adp, 2nd}_x(\xi, t)$. 
Note that the conventional 2nd-order FSST is defined by
 \begin{equation}
\label{def_2ndFSST_simple}
R^{2nd}_x(\xi, t)
:=\int_{\{\eta \in \R: \; V_x(t, \eta)\not=0\}} V_x(t, \eta) \delta\big(\go^{2nd}_x(t, \eta)-\xi\big) d\eta,
\end{equation}
where one can use $\go^{2nd}_x(t, \eta)$ defined by \eqref{2nd_phase} or choose one of
several different $\go_x^{2nd}$ in \cite{MOM15}.

\section{Support zones of STFTs of LFM signals}

The parameter $\gs$ for the window function $g_\gs$ affects the sharpness of the STFT of a signal. In this section, we study how the time-varying parameter $\gs(t)$ controls the representation of STFT $\wt V_x(t, \eta)$ of a monocomponent signal $x(t)$
and provide the parameter $\gs(t)$ with which STFT has the sharpest representation in the TF plane. In the next section, we will consider the following problem: under which condition (if any) for a multicomponent signal as given by \eqref{AHM}, with a suitable choice of $\gs(t)$, the STFTs of $\wt V_{x_k}(t, \eta), 1\le k\le K$ are
well separated. 

To study the sharpness of the STFT of a monocomponent signal or the separability of STFTs (including STFTs with a time-varying parameter) 
of different components $x_k$ of $x(t)$, we need to consider the support zone of STFT $V_{x_k}(t, \eta)$ in the TF plane, 
the region outside which $V_{x_k}(t, \eta)\approx 0$. 
Since the support zone of $V_{x_k}(t, \eta)$ is determined by the support of $\wh g$ outside which $\wh g(\xi)\approx 0$, 
first of all, we need to define the ``support'' of  $\wh g$ if $g$ is not band-limited.
More precisely, for a given threshold $0<\epsilon<1$, if
$|\wh g(\xi)|/\max_\xi|\wh g(\xi)|<\epsilon$ for $|\xi|\ge \xi_0$, then we say $\wh g(\xi)$ is ``supported" in $[-\xi_0, \xi_0]$. We use $L_{\wh g}=2\xi_0$ to denote the length of the ``support" interval of $\wh g$ and we call it the duration of $\wh g$. Note that $\xi_0=\xi_{0, \epsilon}$ depends on $\epsilon$. For simplicity, here and below we drop the subscript $\epsilon$.  Also
in applications, $\epsilon$ is quite small.

In the remainder of this paper, we consider $g$ given by \eqref{def_g} and thus $g_\gs(t)$ is the Gaussian window function defined by
\begin{equation}
\label{def_Gaussian_time}
g_\gs(\tau)=\frac 1{\gs \sqrt {2\pi}}e^{-\frac {\tau ^2}{ 2 \gs^2}},
\end{equation}
with its Fourier transform given by
\begin{equation}
\label{def_Gaussian}
\wh g_\gs(\xi)=e^{-2\pi^2 \gs^2 \xi^2}.
\end{equation}

For $g$ given by  \eqref{def_g},
$|\wh g(\xi)|=e^{-2\pi^2 \xi ^2}<\epsilon$ if and only if  $|\xi|> \ga$, where
\begin{equation}
\label{def_ga}
\ga=\frac 1{2\pi}\sqrt{2\ln (1/\epsilon)}.
\end{equation}
Thus we regard that $\wh g$ is ``supported" in $[-\ga, \ga ]$. Hence, $\wh g_\gs$, given by \eqref{def_Gaussian}, 
is ``supported'' in $[-\frac \ga{\gs}, \frac \ga{\gs}]$, and $L_{\wh g_\gs}=\frac {2\ga}\gs$.

For $s(t)=Ae^{i 2\pi ct}$, since its STFT with $g_\gs$ is
$$
V_{s}(t, \eta)=A e^{i2\pi t c }\wh g_\gs(\eta-c),
$$
and $\wh g_\gs(\eta-c)$ is ``supported" in $c-\frac \ga\gs\le \eta \le c+\frac \ga\gs$,
$V_{s}(t, \eta)$ concentrates around $\eta=c$ and lies within the zone (a strip) of the TF plane $(t, \eta)$:
\begin{equation}
\label{zone_constant}
\big\{(t, \eta): \quad  c-\frac \ga\gs\le \eta \le c+\frac \ga\gs, \; t\in \R\big\}.
\end{equation}

Next we consider LFM signals with IF $\phi'(t)=c+rt>0$. First we find the  STFT of  an LFM signal.
\begin{pro}
\label{pro:STFT_LinearChip}
Let $s(t)$ be an LFM given by \eqref{def_chirp}.
The STFT of $s(t)$ with the Gaussian window function $g_\gs(\tau)$ is given by
\begin{equation}
\label{STFT_LinearChip}
V_{s}(t, \eta)=\frac {A}{\sqrt{1-i2\pi \gs^2 r}}\;
e^{i2\pi(ct +rt^2/2)} \; h\big(\eta-(c+rt)\big),
\end{equation}
where
\begin{equation*}
h(\xi)= e^{-\frac{2\pi^2 \gs^2}{1+(2\pi r \gs^2)^2}(1+i2\pi \gs^2 r) \xi^2}.
\end{equation*}
\end{pro}

One can obtain \eqref{STFT_LinearChip} by applying the following formula
(see \cite{Leon_Cohen}):  for real $\ga$ and $\beta$ with $\ga>0$
\begin{equation}
\label{STFT_LinearChip1}
\int_{-\infty}^\infty e^{-(\ga+i\beta)t^2+i \go t}dt =\frac{\sqrt \pi}{\sqrt{\ga+i\beta}} e^{-\frac{\go^2}{4(\ga+i\beta)}}.
\end{equation}

Observe that $|h(\xi)|$ is a Gaussian function with duration
$$
L_{|h|}=2\ga \sqrt{\frac{1+(2\pi r \gs^2)^2}{\gs^2}}=2\ga \sqrt{\frac1{\gs^2}+(2\pi r \gs)^2}.
$$
Thus the ridge of $V_{s}(t, \eta)$ concentrates around  $\eta=c+rt$ in the TF plane, and  $V_{s}(t, \eta)$ lies within the zone of TF plane of $(t, \eta)$:
$$
-\frac 12 L_{|h|}\le c+rt-\eta \le \frac 12 L_{|h|},
$$
or equivalently
\begin{equation}
\label{def_zone}
 c+rt-\ga  \sqrt{\frac1{\gs^2}+(2\pi r \gs)^2} \le \eta \le  c+rt+ \ga \sqrt{\frac1{\gs^2}+(2\pi r \gs)^2}.
\end{equation}

$L_{|h|}$ gains its minimum when $ \frac1{\gs^2}=(2\pi r \gs)^2$, namely,
\begin{equation}
\label{opt_gs}
\gs=\frac1{\sqrt{2\pi |r|}}=\frac1{ \sqrt{2\pi |\phi''(t)|}}.
\end{equation}
The choice of $\gs$ given in \eqref{opt_gs} results in the sharpest representation of $V_{s}(t, \eta)$.

For a monocomponent signal $x(t)=A(t)e^{i2\pi \phi(t)}$, if its STFT with $g_\gs$, which is  also given by (refer to \eqref{STFT2}),
$$
V_x(t, \eta)=\int_{-\infty}^\infty A(t+\tau) e^{i2\pi \phi(t+\tau)}g_\gs(\tau)e^{-i2\pi \eta \tau}d\tau
$$
can be well approximated by
\begin{equation*}
V_x(t, \eta)\approx
\int_{-\infty}^\infty A(t) e^{i2\pi \big(\phi(t)+\phi'(t) \tau  +\frac 12 \phi''(t) \tau ^2\big)}
g_\gs(\tau)e^{-i2\pi \eta \tau}d\tau,
\end{equation*}
then  the choice of $\gs$ given
\begin{equation}
\label{opt_gs1}
\gs=\frac1{\sqrt{2\pi |\phi''(t)|}}
\end{equation}
results in the sharpest representation of $V_x(t, \eta)$. The choice of  $\gs=\gs(t)$ in \eqref{opt_gs1} coincides with the result derived in \cite{Leon_Cohen}.

Observe that $\gs$ in \eqref{opt_gs1} is the optimal parameter for the representation of a monocomponent signal. In the next section, we will use the obtained TF zone in \eqref{def_zone} for the STFT of an LFM  signal to study the well-separated condition for a multicomponent signal.

\section{Separability of multicomponent signals and selection of time-varying parameter}

 In this section, we will consider the problem that under which condition (if any), for a multicomponent signal as given by \eqref{AHM}, with a suitable choice of $\gs(t)$, STFTs $\wt V_{x_k}(t, \eta), 1\le k\le K$ of different components $x_k$ defined in \eqref{def_STFT_para1} are well separated, and  the associated adaptive FSST of $x(t)$ has a sharp representation. 

\subsection{Sinusoidal signal model}

First we consider 
the sinusoidal signal model.
Recall that the STFT of  $s(t)=A e^{i 2\pi c t}$ with $g_\gs(t)$ is supported in the zone of the  TF plane given by \eqref{zone_constant}. Suppose $x(t)$ is a finite summation of sinusoidal signals:
\begin{equation}
\label{const_AHM}
x(t)=\sum_{k=1}^K A_k e^{i2\pi c_k t},
\end{equation}
where $A_k, c_k$  are positive constants with $0<c_k<c_{k+1}$. 
Since the STFT of the $k$-component of $x(t)$ lies within the zone of the TF plane $(t, \eta)$:
$c_k-\ga/\gs\le \eta \le c_k+\ga/\gs$ for any $t$, the components of $x(t)$ will be well-separated in the  TF plane if
  $$
c_{k-1}+\frac \ga\gs\le c_k-\frac \ga\gs,
  $$
or equivalently
$$
  \gs\ge \frac {2\ga}{c_k-c_{k-1}}, \quad  \hbox{for $k=2, 3, \cdots, K$.}
 $$

More generally, for $x(t)$ given by \eqref{AHM}, suppose each $x_k(t)$ is well approximated by sinusoidal functions at any local time, that is \eqref{sinu_model} holds. Then the time-varying STFT
$\wt V_{x_k}(t, \eta)$  of $x_k(t)$ 
  can be well-approximated by
  \begin{eqnarray*}
\wt V_{x_k}(t, \eta) \hskip -0.6cm &&\approx \int_{-\infty}^\infty x_k(t) e^{i2\pi \phi'_k(t) \tau}
g_{\gs(t)}(\tau)e^{-i2\pi \eta \tau}d\tau\\
&&=x_k(t)\wh g\big(\gs(t)(\eta -\phi'_k(t))\big).
\end{eqnarray*}
Hence $\wt V_{x_k}(t, \eta)$ lies within the zone of the TF plane $(t, \eta)$:
$$
O_k=\{(t, \eta): \; -\ga\le \gs(t)\big(\eta-\phi'_k(t)\big)\le \ga\} = \{(t, \eta): \;
\phi'_k(t)-\frac \ga{\gs(t)}\le \eta\le \phi'_k(t)+\frac \ga{\gs(t)}\}.
$$
Thus,  the components of $x(t)$ will be well-separated in the  TF plane (namely,  $O_k, 1\le k\le K$ do not overlap) if
 $$
 \phi'_{k-1}(t)+\frac \ga{\gs(t)}\le  \phi'_k(t)-\frac \ga{\gs(t)}, \; t\in \R, k=2, \cdots, K,
 $$
or equivalently
\begin{equation}
 \label{sinu_model_condition}
  \gs(t)\ge \frac {2\ga}{\phi'_k(t)-\phi'_{k-1}(t)}, \; t\in \R, k=2, 3, \cdots, K.
\end{equation}
\eqref{sinu_model_condition} is the sinusoidal signal model-based well-separable condition for $x(t)$ with the adaptive STFT.
When $\gs(t)\equiv\gs$ is a positive constant function, \eqref{sinu_model_condition} is reduced to
\eqref{freq_resolution} with $\gt=\ga/\gs$.

We observe in our experiments that in general a big $\gs$ will result in low time-resolution and unreliable representation of the FSST of a signal $x(t)$. Actually, the error bounds derived in
\cite{Daub_Lu_Wu11, Thakur_Wu11, Wu_thesis, MOM14, BMO18}
imply that for a signal, its synchrosqueezed representation is sharper when
the window width in the time domain of the window function $g_\gs$, which is $\gs$ (up to a constant), is smaller.
Thus we should choose $\gs(t)$ as small as possible.
Hence, we propose the sinusoidal signal model-based choice for $\gs$, denoted by $\gs_1(t)$, to be
  \begin{equation}
\label{def_gs1}
  \gs_1(t)=\max_{2\le k\le K}\Big\{\frac {2\ga}{\phi'_k(t)-\phi'_{k-1}(t)}\Big\}.
\end{equation}

\subsection{ Linear frequency modulation (LFM) model}
In this subsection we will derive the well-separated condition
based on the LFM model.
More precisely, we consider $x(t)=\sum_{k=1}^K x_k(t)$, where each $x_k(t)$ is an LFM signal, namely,
$$
x_k(t)=A_k e^{i2\pi (c_kt +\frac 12 r_k t^2)}
$$
with the phase function $\phi_k(t)=c_kt +\frac 12 r_k t^2$ and $\phi'_{k-1}(t)<\phi'_k(t)$.

From \eqref{def_zone},  STFT $V_{x_k}(t, \eta)$ of $x_k$
with Gaussian window function $g_\gs$ lies within the zone of  TF plane $(t, \eta)$:
 \begin{equation}
\label{def_zone_k}
 c_k+r_k t-\ga  \sqrt{\frac1{\gs^2}+(2\pi r_k \gs)^2} \le \eta \le  c_k+r_k t+ \ga  \sqrt{\frac1{\gs^2}+(2\pi r_k \gs)^2},
\end{equation}
for all $t$.
Thus $x_{k-1}(t)$ and $x_k(t)$ are separable in the TF plane if
\begin{equation}\label{separable_k}
 c_{k-1}+r_{k-1} t+\ga  \sqrt{\frac1{\gs^2}+(2\pi r_{k-1} \gs)^2}
 \le  c_k+r_k t-\ga  \sqrt{\frac1{\gs^2}+(2\pi r_k \gs)^2}.
\end{equation}

The condition that \eqref{separable_k} holds for $k=2, \cdots, K$ is the well-separated condition for a multicomponent signal consisting of LFM signals. 
One of the main goals of this paper to obtain an explicit $\gs(t)$ such that $V_{x_k}(t, \eta), 1\le k\le K$ lie within non-overlapping TF zones.
To this end, we replace the TF zone of $V_{x_k}(t, \eta)$ in \eqref{def_zone_k}
by a larger zone for $V_{x_k}(t, \eta)$ by
using $\frac1{\gs}+2\pi |r_k| \gs$ to replace $\sqrt{\frac1{\gs^2}+(2\pi r_k \gs)^2}$ in \eqref{def_zone_k}:
 \begin{equation}
\label{def_zone_k_larger}
 c_k+r_k t-\ga  (\frac1{\gs}+2\pi |r_k| \gs)\le \eta \le  c_k+r_k t+ \ga  (\frac1{\gs}+2\pi |r_k| \gs).
\end{equation}
Since
 $$
\frac {\sqrt 2}2(\frac1{\gs}+2\pi |r_k| \gs) \le \sqrt{\frac1{\gs^2}+(2\pi r_k \gs)^2}\le
\frac1{\gs}+2\pi |r_k| \gs,
$$
the zone given by \eqref{def_zone_k_larger} is slightly larger than that given by \eqref{def_zone_k}.
Clearly, $x_{k-1}(t)$ and $x_k(t)$ are separable in the TF plane if
\begin{equation*}
 c_{k-1}+r_{k-1} t+\ga  (\frac1{\gs}+2\pi |r_{k-1}| \gs)
  \le  c_k+r_k t-\ga(\frac1{\gs}+2\pi |r_k| \gs).
\end{equation*}

More generally, for $x(t)$ given by \eqref{AHM},  suppose each $x_k$ is well approximated by an LFM at any local time, namely \eqref{LFM_model} holds. Then
the time-varying STFT $\wt V_{x_k}(t, \eta)$ of $x_k$ with $g_{\gs(t)}$, which is (refer to \eqref{def_STFT_para2})
 \begin{equation*}
  \int_{-\infty}^\infty A_k(t+\tau) e^{i2\pi \phi_k(t+\tau)}g_{\gs(t)}(\tau)e^{-i2\pi \eta \tau}d\tau,
\end{equation*}
can be well-approximated by the quantity on the right-hand side of \eqref{linear_chirp_model}, which is (by applying Proposition \ref{pro:STFT_LinearChip} or  \eqref{STFT_LinearChip1})
$$
\frac{x_k(t)}{\sqrt{1-i2\pi \gs(t)^2 \phi_k''(t)}} e^{-\frac{2\pi^2}{\frac 1{\gs(t)^2}+(2\pi \phi_k''(t) \gs(t))^2}(1+i2\pi \gs^2(t)  \phi_k''(t)) (\eta- \phi_k'(t))^2}.
$$
Thus $\wt V_{x_k}(t, \eta)$ lies within the zone of  TF plane: 
\begin{equation}
\label{def_zone_k_general}
\phi_k'(t)-\ga  \sqrt{\frac1{\gs(t)^2}+(2\pi  \phi_k''(t) \gs(t))^2} \le \eta \le
\phi_k'(t)+\ga  \sqrt{\frac1{\gs(t)^2}+(2\pi  \phi_k''(t) \gs(t))^2},
\end{equation}
for $t\in \R$, and  the well-separable condition for $x(t)$ is
\begin{equation}
\label{LFM_separate}
\phi_{k-1}'(t)+\ga  \sqrt{\frac1{\gs(t)^2}+(2\pi  \phi_{k-1}''(t) \gs(t))^2} \le
\phi_k'(t)-\ga  \sqrt{\frac1{\gs(t)^2}+(2\pi  \phi_k''(t) \gs(t))^2}, \;  2\le k\le K,
\end{equation}
for $t\in \R$.

As above, we replace the TF zone \eqref{def_zone_k_general} of $\wt V_{x_k}(t, \eta)$
 by a larger zone given by
 \begin{equation*}
\phi_k'(t)-\ga\big (\frac1{\gs(t) }+2\pi |\phi''_k(t)| \gs(t)\big) \le \eta \le  \phi_k'(t)+\ga \big(\frac1{\gs(t)}+2\pi |\phi''_k(t)| \gs(t)\big).
\end{equation*}
Then the corresponding well-separable condition for $x(t)$ is
\begin{equation}
\label{separable_k_larger_general}
\phi_{k-1}'(t)+\ga \big(\frac1{\gs(t) }+2\pi |\phi''_{k-1}(t)| \gs(t)\big) \le  \phi_k'(t)-\ga \big(\frac1{\gs(t) }+2\pi |\phi''_k(t)| \gs(t)\big), \; 2\le k\le K,
\end{equation}
which is equivalent to
\begin{equation}
\label{separable_k_larger_general_short}
a_k(t) \gs(t)^2-b_k(t)\gs(t)+2\ga \le 0, \;  2\le k\le K, 
\end{equation}
where
\begin{equation}
\label{def_akbk}
a_k(t)=2\pi\ga (|\phi''_{k-1}(t)| +|\phi''_k(t)|), \; b_k(t)=\phi_{k}'(t)-\phi_{k-1}'(t).
\end{equation}

If
$$
b_k(t)^2-8\ga a_k(t)=\big(\phi'_k(t)-\phi'_{k-1}(t)\big)^2-16\pi \ga^2 \big(|\phi''_k(t)|+|\phi''_{k-1}(t)|\big)\ge 0,
$$
then \eqref{separable_k_larger_general} (or  \eqref{separable_k_larger_general_short}) is equivalent to
\begin{equation}
\label{gs_ineq}
\frac{4\ga}{b_k(t)+\sqrt{b_k(t)^2-8\ga a_k(t)}}\le \gs(t)\le \frac{4\ga}{b_k(t)-\sqrt{b_k(t)^2-8\ga a_k(t)}}, \; 2\le k\le K,
\end{equation}
for $t\in \R$. 
Otherwise, if $b_k(t)^2-8\ga a_k(t)<0$, then there is no suitable solution of the parameter $\gs$ for \eqref{separable_k_larger_general} or equivalently
\eqref{separable_k_larger_general_short}. In this case we say that components  $x_{k-1}(t)$ and $x_k(t)$ of multicomponent signal $x(t)$ cannot be separated in the  TF plane.
Note that when $a_k(t)=0$, i.e. $\phi''_k(t)=\phi''_{k-1}(t)=0$, \eqref{gs_ineq} is reduced to \eqref{sinu_model_condition}.
In the next theorem, we summarize the LFM model-based well-separated condition 
we have derived above.

\begin{theo} Let
$x(t)=\sum_{k=1}^K x_k(t)$, where each $x_k(t)=A_k(t)e^{i2\pi\phi_k(t)}$ is an LFM signal 
or its adaptive STFT
$\wt V_{x_k}(t, \eta)$
with $g_{\gs(t)}$ can be well approximated by \eqref{linear_chirp_model}, and $\phi'_{k-1}(t)<\phi'_k(t)$.
If 
\begin{eqnarray}
\label{LinearChirp_sep_cond1}
&& 4\ga \sqrt{\pi} \sqrt {|\phi''_k(t)|+|\phi''_{k-1}(t)|}\le \phi'_k(t)-\phi'_{k-1}(t),
\quad k=2, \cdots, K, \; \hbox{and}\\
\label{LinearChirp_sep_cond2}
 &&
 \max_{2\le k\le K}\Big\{\frac{4\ga}{b_k(t)+\sqrt{b_k(t)^2-8\ga a_k(t)}}\Big\}\le \min_{2\le k\le K}\Big\{\frac{4\ga}{b_k(t)-\sqrt{b_k(t)^2-8\ga a_k(t)}}\Big\},
 \end{eqnarray}
 for $t\in \R$, then the components of $x(t)$ are well-separable in TF plane in the sense that $\wt V_{x_k}(t, \eta), 1\le k\le K$ with $\gs(t)$ chosen to satisfy \eqref{gs_ineq} lie in non-overlapping regions in the TF plane.
\end{theo}

We call \eqref{LinearChirp_sep_cond1}-\eqref{LinearChirp_sep_cond2} the LFM model-based well-separated condition for a multicomponent signal $x(t)$.
Observe that our LFM model-based well-separated condition \eqref{LinearChirp_sep_cond1}
requires the boundedness of the 2nd-order derivatives $\phi''_k(t)$, while it seems the sinusoidal signal-based well-separated condition \eqref{freq_resolution} 
or \eqref{sinu_model_condition} does  not have such a constraint.  Actually the sinusoidal signal model assumption \eqref{sinu_model} 
requires $\phi''_k(t)$ be small. In addition, to make the recovery error in \eqref{reconst_SST_component}  small, 
$\phi''_k(t)$ must be very small (see \cite{Wu_thesis, MOM14, BMO18} for the details about the recovery error estimates).

Any $\gs(t)$ between the two quantities in the two sides of the inequality \eqref{LinearChirp_sep_cond2} can separate the components of $x(t)$ in the TF plane.
As discussed above, since a smaller $\gs(t)$ gives a sharper synchrosqueezing representation,
we should choose $\gs(t)$ as small as possible.
Hence,  we propose the LFM model-based 
choice for $\gs$, denoted by $\gs_2(t)$, to be
\begin{equation}
\label{def_gs2}
\gs_2(t)=
\max\Big\{
\frac{4\ga}{b_k(t)+\sqrt{b_k(t)^2-8\ga a_k(t)}}:
\; 2\le k\le K \Big\}, 
 \end{equation}
where $a_k(t)$ and $b_k(t)$ are defined by \eqref{def_akbk}, and $\ga$ is defined by \eqref{def_ga}.

\begin{figure}[th]
\centering
\begin{tabular}{ccc}
\resizebox{2.0in}{1.5in}{\includegraphics{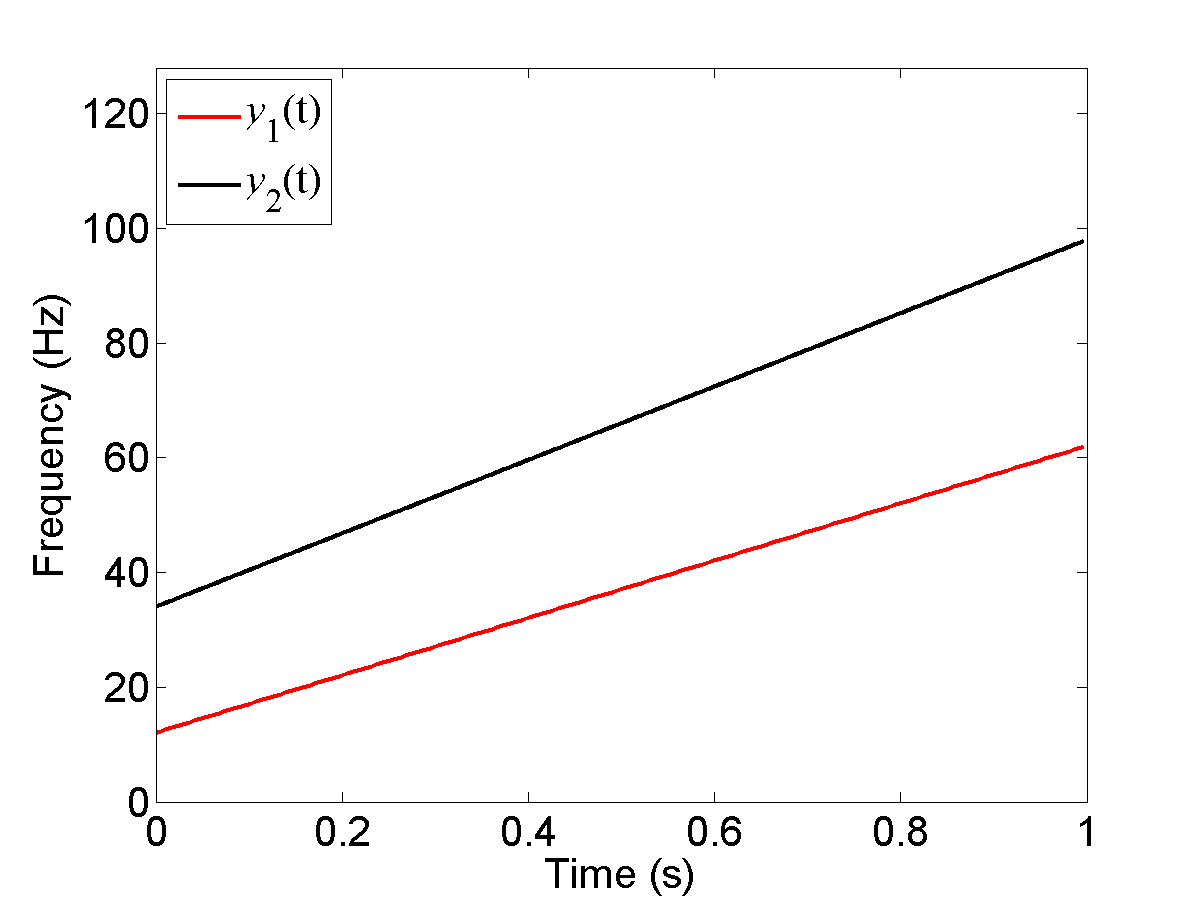}}
& \resizebox{2.0in}{1.5in}{\includegraphics{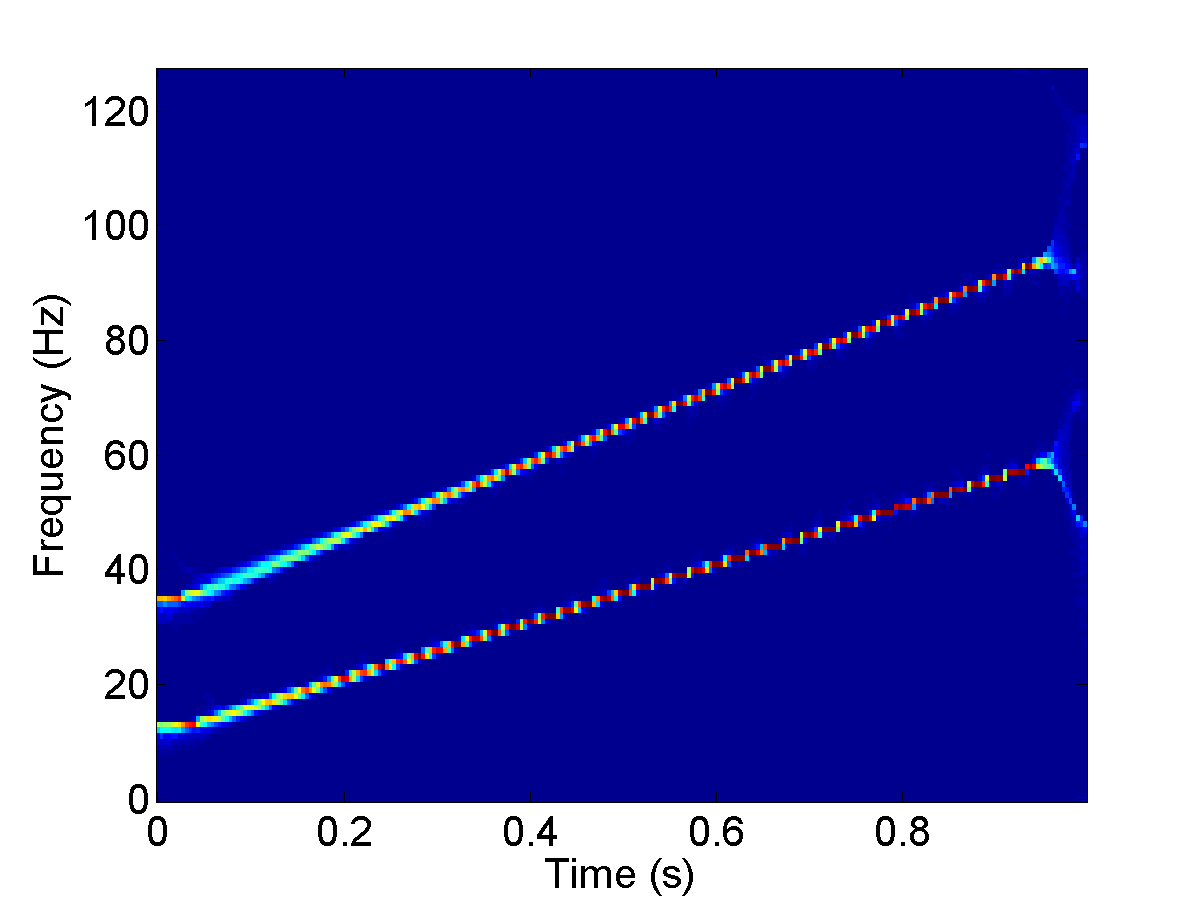}}
&\resizebox{2.0in}{1.5in}{\includegraphics{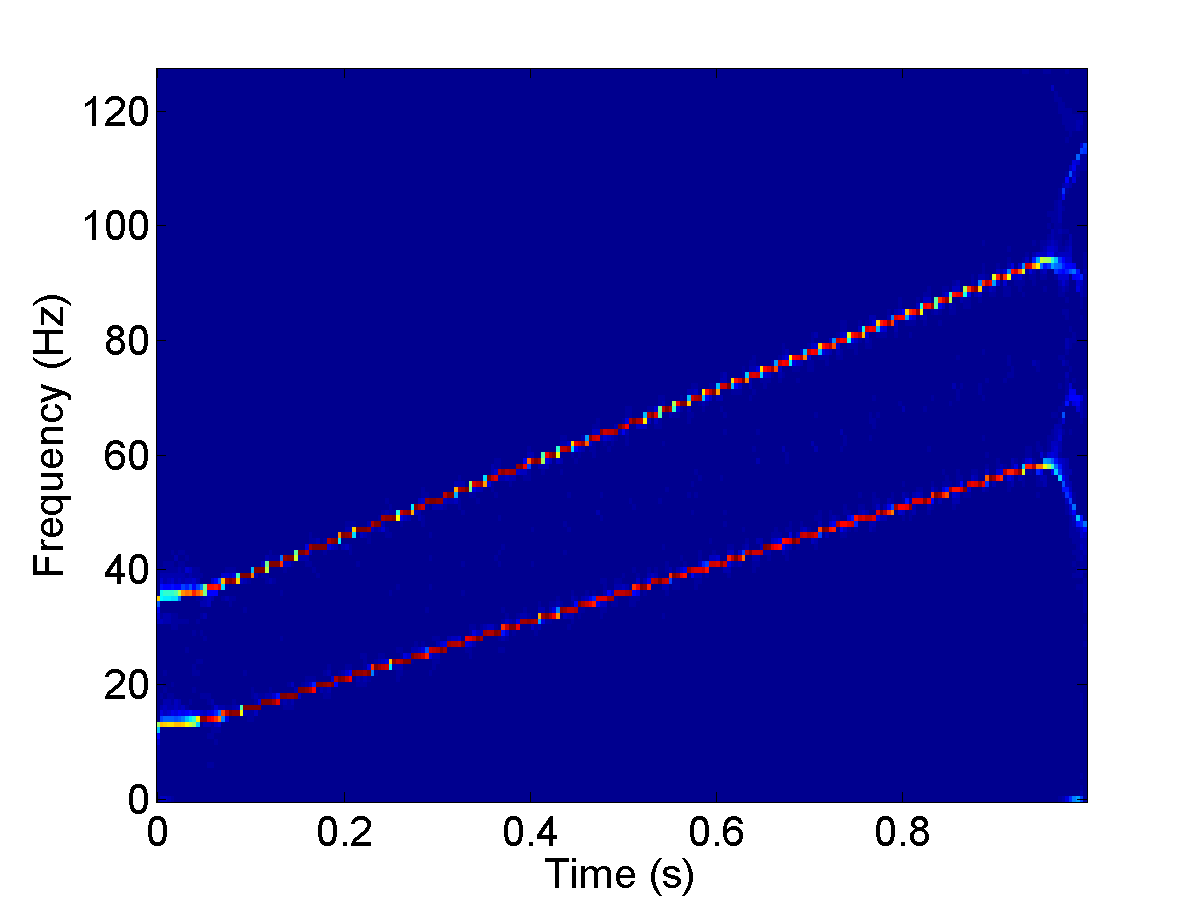}} \\
&\resizebox{2.0in}{1.5in}{\includegraphics{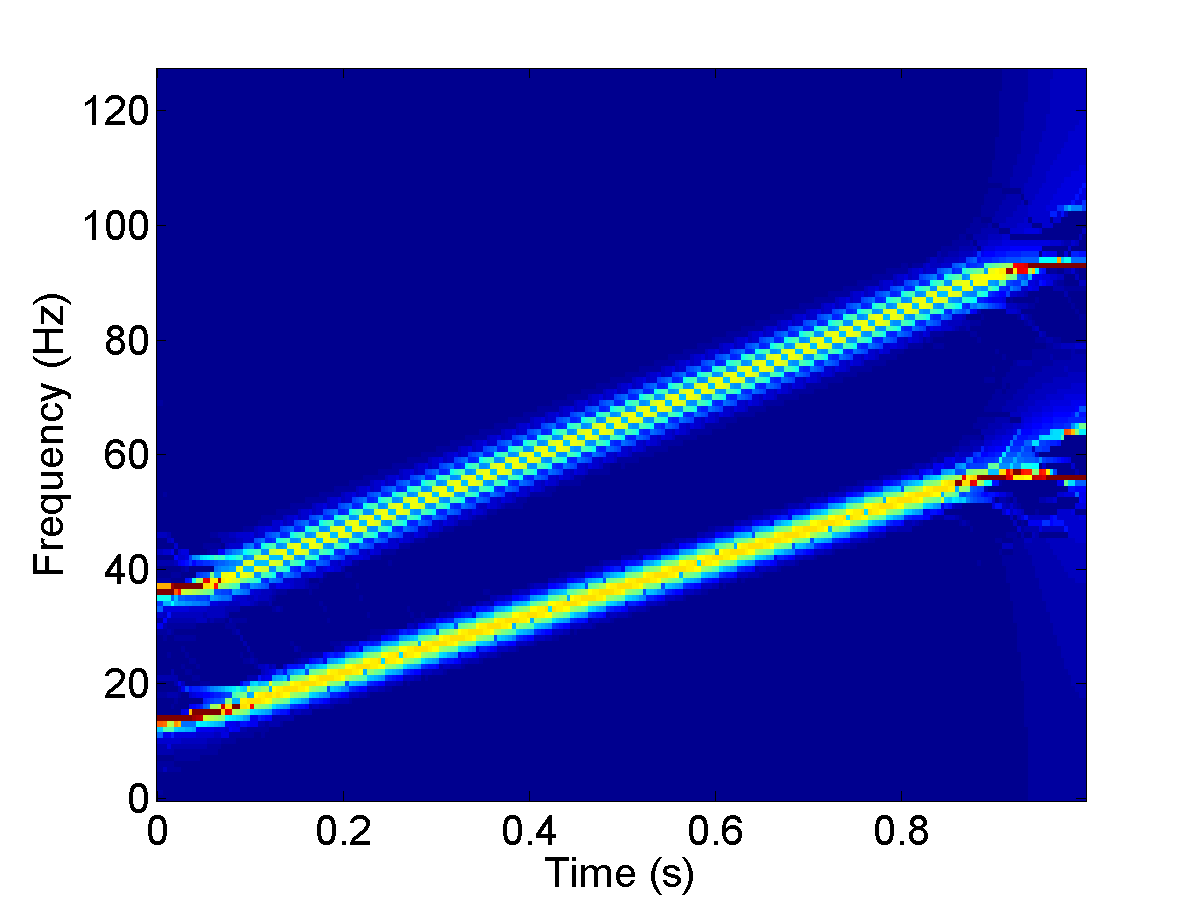}} 
&\resizebox {2.0in}{1.5in} {\includegraphics{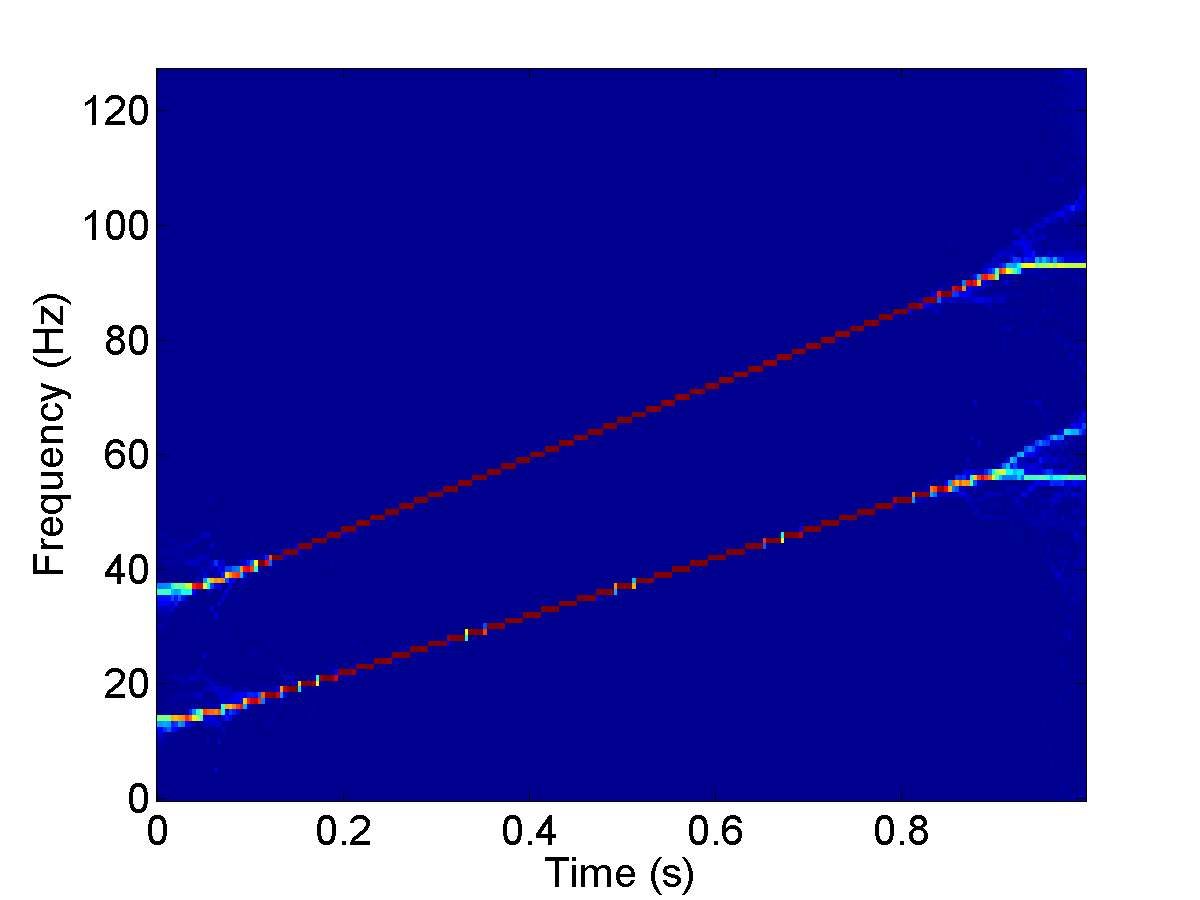}}
\end{tabular}
\caption{\small  Experimental results on the two-component LFM signal in \eqref {two_chirps_12_34}:
IFs (Top-left); adaptive FSST with time-varying parameter $\gs_2(t)$ (Top-middle); 2nd-order adaptive FSST  with time-varying parameter $\gs_2(t)$ (Top-right);
  conventional  FSST (Bottom-left) and conventional 2nd-order FSST (Bottom-right) with  $\gs=0.057$
 }
\label{figure: SST_two_chirps_1234}
\end{figure}

Next we show some experimental results.
We consider a two-component LFM signal,
\begin{equation}
\label{two_chirps_12_34}
y(t)=y_1(t)+y_2(t)=\cos(2\pi(12t+25t^2))+\cos(2\pi(34t+32t^2)), \quad t\in [0, 1],
\end{equation}
where the number of sampling points is 256, namely the sampling rate is 256Hz.
The IFs of  $y_1(t)$ and $y_2(t)$ are  $\phi'(t)=12+50t$ and $\phi'_2(t)=34+64t$, respectively.
The top-left panel of Fig.\ref{figure: SST_two_chirps_1234} shows the instantaneous frequencies of $y_1(t)$ and $y_2(t)$.
With $\gs_2(t)$, both the proposed adaptive FSST  defined by \eqref{def_FSST_para_simple}
and 2nd-order adaptive FSST  defined by \eqref{def_2ndFSST_para_simple} can represent this signal sharply.
Here and below, we choose $\epsilon=\frac 15$, and hence  $\ga$ which is defined by \eqref{def_ga} and used in
\eqref{def_gs2} is $\ga\approx 0.2855$.
Observe that the 2nd-order adaptive FSST further improves the TF energy concentration of the adaptive FSST.  Here we also give the results of conventional FSST studied in
\cite{Thakur_Wu11}-\cite{MOM14}, and conventional 2nd-order FSST defined in \cite{MOM15} with 
$\gs=0.057$. This $\gs$ is obtained by minimizing the R${\rm \acute e}$nyi entropy of the STFT (refer to the next section about the definition of R${\rm \acute e}$nyi entropy). 
Observe that the 2nd-order FSST is better than the FSST with the same $\gs$. When $\gs=0.057$
the TF representation of the conventional 2nd-order FSST is not as sharp or clear as that of the 2nd-order adaptive FSST.

\section{Selecting the time-varying parameter automatically}

Suppose $x(t)$ given by \eqref{AHM} is separable, meaning \eqref{LinearChirp_sep_cond1} and \eqref{LinearChirp_sep_cond2} hold.
If we know $\phi'_k(t)$ and $\phi''_k(t)$, then we can choose a $\gs(t)$ such as $\gs_2(t)$ in \eqref{def_gs2}
 to satisfy  \eqref{gs_ineq} to define  the adaptive STFT and adaptive FSST for sharp representations of $x_k(t)$ in the TF plane and for 
accurate recovery of $x_k(t)$. However in practice, we in general have no prior knowledge of $\phi'_k(t)$ and $\phi''_k(t)$. 
Hence, we need to have a method which provides suitable $\gs(t)$. 
In this section, we propose an algorithm to estimate $\gs(t)$ which is based on the well-separated condition of \eqref{separable_k_larger_general}.

First for temporarily fixed $t$ and $\gs$,  denote $V_{x, (t, \gs)}(\eta)=V_{x}(t, \eta, \gs)$, 
the STFT of $x(t)$ with a time-varying parameter defined by \eqref{def_STFT_para1}. We extract the peaks (local maxima) of $|V_{x, (t, \gs)}(\eta)|$ with certain height. 
More precisely, assuming $\gamma_1>0$ is a given threshold, we find local maximum points $\eta_1, \eta_2, \cdots, \eta_m$ of $|V_{x, (t, \gs)}(\eta)|$ 
at which $|V_{x, (t, \gs)}(\eta)|$ attains local maxima with
\begin{equation}
\label{threhold_gamma1}
\frac{|V_{x, (t, \gs)}(\eta_k)|}{\max _{\eta}|V_{x, (t, \gs)}(\eta)|} > \gamma_1, \; k=1, \cdots, m.
\end{equation}
Note that $m$ may depend on $t$ and $\gs$. We assume $\eta_1<\eta_2< \cdots <\eta_m$. 
The threshold $\gamma_1$ is used to remove the local maxima with smaller amplitudes, which are regarded as noises and interferences.

For each local maximum point $\eta_k$, we regard $\eta_k$ is the local maximum of the adaptive STFT $V_{x_k, (t,\gs)}(\eta)$ of a potential component, 
denoted by $x_k(t)$ of $x(t)$.  To check whether $x_k$ is indeed a component of $x(t)$ or not, we consider the support interval 
$[l_k, h_k]$ for $V_{x_k, (t,\gs)}(\eta)$ with $|V_{x_k,(t, \gs)}(\eta)|>0$ for $\eta\in [l_k, h_k]$. 
If there is no overlap among  $[l_k, h_k]$, $[l_{k-1}, h_{k-1}]$, $[l_{k+1}, h_{k+1}]$, then we decide that  $x_k(t)$ is 
indeed a component of $x(t)$, where $[l_{k-1}, h_{k-1}]$, $[l_{k+1}, h_{k+1}]$ are the support intervals for $x_{k-1}$ and $x_{k+1}$ defined similarly.  
With our LFM model, if the estimated IF $\phi_k'(t)$ of $x_k(t)$ is
 $\wh c_k+ \wh r_kt$, then by \eqref{separable_k_larger_general},
 \begin{eqnarray}
\label{estimate_hk}
&&h_k=\wh c_k+\ga \big(\frac1{\gs }+2\pi |\wh r_k| \gs\big), \\
&&l_k=\wh c_{k-1}-\ga \big(\frac1{\gs }+2\pi |\wh r_{k-1}| \gs\big).
\label{estimate_lk}
\end{eqnarray}

 Notice that $\wh c_k=\eta_k$. Thus we need to estimate the chirp rate $\wh r_k$ of $x_k(t)$.
 To this end, we extract a small piece of curve in the TF plane passing through $(t, \eta_k)$ which corresponds to the local ridge on $|V_{x_k,(t, \gs)}(\eta)|$. More precisely, letting
$$
t_{k1}=t-\frac{1}{2}L_{g_\gs}=t-2\pi \ga \gs, \quad
t_{k2}=t-\frac{1}{2}L_{g_\gs}=t+2\pi \ga \gs,
$$
define
$$
\wh d_k(\tau)=\underset{ \hbox{$\eta$: $\eta$ {\small is near $\eta_k$}}}{\rm argmax}   |V_{(\tau, \gs)}(\eta)|, \quad \tau\in [t_{k1},t_{k2}].
$$
In the above we have used the fact  that the duration of  $g_\gs(t)$ is (refer to \eqref{def_ga})
\begin{equation*}
L_{g_\gs}=2\gs \sqrt{2\ln(1/\epsilon)}= 4\pi\gs\ga.
\end{equation*}

Note that $\wh d_k(t)=\eta_k$ and $(t, \eta_k)$ is a point lying on the curve  in the TF of $(\tau, \eta)$ given by
$$
L=\{(\tau, \wh d_k(\tau)): \tau\in [t_{k1},t_{k2}]\}=\{(\tau, \eta): \eta=\wh d_k(\tau),  \tau \in [t_{k1},t_{k2}]\}.
$$
Most importantly, $\{|V_{x_k, (\tau, \gs)}(\eta)|: (\tau, \eta)\in L\}$ is the local ridge on $|V_{x_k, (\tau, \gs)}(\eta)|$  
near $(t, \eta_k)$, and thus, it is also the local ridge on $|V_{x}(t, \eta, \gs)|$. Observe that from the STFT of an LFM given by  Proposition
\ref{pro:STFT_LinearChip}, the local ridge on $|V_{x_k}(t, \eta, \gs)|$ occurs when $\phi'_k(t)=c_k+r_k t$. Thus we use the linear function
\begin{equation*}
\label{local_L_functions}
d_k(\tau) =\wh r_k(\tau-t)+  \wh c_k , \tau \in [t_{k1},t_{k2}]
\end{equation*}
 to fit $\wh d_k(\tau)$. The obtained $\wh r_k$ is the estimated chirp rate $r_k$ of $x_k(t)$. With this $\wh r_k$ and $\wh c_k={\eta_k}$ as given above, 
we have $h_k, l_k$ given in \eqref{estimate_hk} and \eqref{estimate_lk}.  Especially when $\wh r_k=0$, 
recalling the support zone of a sinusoidal signal mode in \eqref{zone_constant}, we have
$$
h_k=\wh c_k+\frac \ga\gs,  \; l_k=\wh c_k-\frac \ga\gs.
$$

This way we obtain the collection of support intervals for $V_{x}(t, \eta, \gs)$ for fixed $t$ and $\gs$:
\begin{equation}
\label{def_s_intervals}
{\bf s}=\{[l_1, h_1], \cdots, [l_m, h_m] \}.
\end{equation}
If adjacent intervals of ${\bf s}$ do not overlap, namely,
\begin{equation}
\label{nonoverlap_hkgk}
h_k \le l_{k+1}, \; \hbox{ for all $k=1, 2, \cdots, m-1$}
\end{equation}
holds, then this $\gs$  is a right parameter to separate the components and such a $\gs$  is a good candidate which we consider to select. 
Otherwise, if a pair of adjacent intervals of ${\bf s}$ overlap, namely, \eqref{nonoverlap_hkgk} does not hold, then this $\gs$ is not the parameter 
we shall choose and we need to consider a different  $\gs$.

In the above description of our idea for the algorithm, we start with a $\gs$ and (temporarily fixed) $t$, then we decide whether 
this $\gs$ is a good candidate to select or not based our proposed criterion: \eqref{nonoverlap_hkgk} holds or does not. 
The choice of the initial $\gs$ plays a critical role for the success of our algorithm due to that on one hand, as we have mentioned above, in general a smaller $\gs$ will result in
a sharper representation of SST, and hence, we should find $\gs$  as small as possible such that \eqref{nonoverlap_hkgk} holds; and
on the other hand,  different $\gs$ with which \eqref{nonoverlap_hkgk} holds may result in different number of intervals $m$ in
\eqref {def_s_intervals} even for the same time instance $t$.  To keep the number $m$ (an estimation of the number of modes $K$ 
for a given time $t$) unchanged when we search for different $\gs$ with a fixed $t$,
the initial $\gs$ is required to provide a good estimate of the number of the components of a multicomponent signal $x(t)$. 
To this end, in this paper we propose to use the R${\rm \acute e}$nyi entropy to determine the initial $\gs(t)$.

The R${\rm \acute e}$nyi entropy is a commonly used measurement to evaluate the concentration of a TF representation such as STFT, SST, etc. of a signal of $x(t)$, see \cite{ Stankovic01, Baraniuk01, SP18}.
Taking STFT $V_x(t,\eta)$ of a signal $x(t)$ as an example, the R${\rm \acute e}$nyi entropy with $V_x(t,\eta)$ is
  \begin{equation}
 \label{def_renyi_entropy_spec}
E_{\zeta} (t): =
 \frac{1}{{1 - \ell }}\log _2 \frac{{\int_{t - \zeta }^{t + \zeta } {\int_0^\infty  {\left| {V_x(b,\eta )} \right|^{2\ell } d\eta db} } }}{{\left( {\int_{t - \zeta }^{t + \zeta }
{\int_0^\infty  {\left| {V_x(b,\eta )} \right|^2 d\eta db} } } \right)^\ell  }},
\end{equation}
where $\ell$ is usually greater than 2.  In this paper we choose $\ell=2.5$, a common value used in other papers, see for examples \cite{Stankovic01, Wu17}. 
Parameter $\zeta>0$ determines the local duration to  calculate the local R${\rm \acute e}$nyi entropy. 
We choose $\zeta=4$. One may choose some large $\zeta$ for non-stationary signals with slow-varying IFs.
Note that the smaller the R${\rm \acute e}$nyi entropy, the better the TF resolution. So for a fixed time $t$, 
we can use \eqref{def_renyi_entropy_spec} to find a $\gs$ (denoted as $\gs_u(t)$) with the best TF concentration of  $V_x(t, \eta, \gs)$, where $V_x(t, \eta, \gs)$ is 
the regular STFT of $x(t)$ defined by \eqref{def_STFT} with the window function $h(\tau)=g_\gs(\tau)=\frac 1{\gs}g(\frac \tau\gs)$ having a parameter $\gs$.
More precisely, replacing $V_x(b,\eta)$ in \eqref{def_renyi_entropy_spec} by
$V_x(b, \eta, \gs)$, we define the R${\rm \acute e}$nyi entropy $E_{\zeta} (t, \sigma)$ of $V_x(t, \eta, \gs)$, and then, obtain
  \begin{equation}
 \label{def_renyi_entropy_best}
\sigma _u (t) = \mathop {\rm argmin }\limits_{\sigma > 0}  \left\{ {E_{\zeta  } (t, \sigma)} \right\}.
\end{equation}
We set $\gs_u(t)$ as the upper bound of $\gs(t)$ for a fixed $t$.

With these discussions, we propose an algorithm to estimate $\gs(t)$  as follows.


{\bf Algorithm 1.} (Separability parameter estimation)
Let  $\{\gs_j, j=1, 2, \cdots, n\}$ be an uniform discretization of $\gs$  with $\gs_1>\gs_2>\cdots>\gs_n>0$ and sampling step $\gD\gs = \gs_{j-1}-\gs_j$. The discrete sequence $s(t),$ $t=t_1, t_2, \cdots , t_N $  (or $t=0, 1, \cdots, N-1$) is the signal to be analyzed.
 \begin{itemize}
\item[] {\bf Step 1.} Let $t$ be one of $t_1, t_2, \cdots , t_N $. Find $\gs_u$ in \eqref{def_renyi_entropy_best} with $\gs$ ranging over 
$\{\gs_j, j=1, 2, \cdots, n\}$. 
\item[] {\bf Step 2.} Let ${\bf s}$ be the set of the intervals given by \eqref{def_s_intervals} with  $\gs=\gs_u$. Let $z=\gs_u$.
If \eqref{nonoverlap_hkgk} holds, then go to Step 3. Otherwise, go to Step 5.
\item[]{\bf Step 3.} Let $\gs=z-\gD\gs$. If the number of intervals $m$ in \eqref{def_s_intervals} with this new $\gs$ remains unchanged, 
$\gs\ge\gs_n$ and \eqref{nonoverlap_hkgk} holds, then go to Step 4. Otherwise, go to Step 5.
\item[]{\bf Step 4.} Repeat Step 3 with $z=\gs$.
\item[]{\bf Step 5.} Let $C(t)=z$, and do Step 1 to Step 4 for different time $t$ of $t_1, t_2, \cdots , t_N $.
\item[]{\bf Step 6.} Smooth  $C(t)$ with a low-pass filter $B(t)$:
\begin{equation}
\label{smooth_C}
\gs_{est}(t)=(C*B)(t).
\end{equation}
\end{itemize}

We call $\gs_{est}(t)$ the estimation of the separability time-varying parameter $\gs_2(t)$ in \eqref{def_gs2}.
In Step 6, we use a low-pass filter $B(t)$ to smooth $C(t)$. This is because of the assumption of the continuity condition for $A_k(t)$ and $\phi_k(t)$.
With the estimated $\gs_{est}(t)$, we can define the adaptive STFT, the adaptive FSST and the 2nd-order adaptive FSST with a time-varying parameter $\gs(t)=\gs_{est}(t)$.

\begin{figure}[th]
\centering
\begin{tabular}{ccc}
\resizebox{2.0in}{1.5in}{\includegraphics{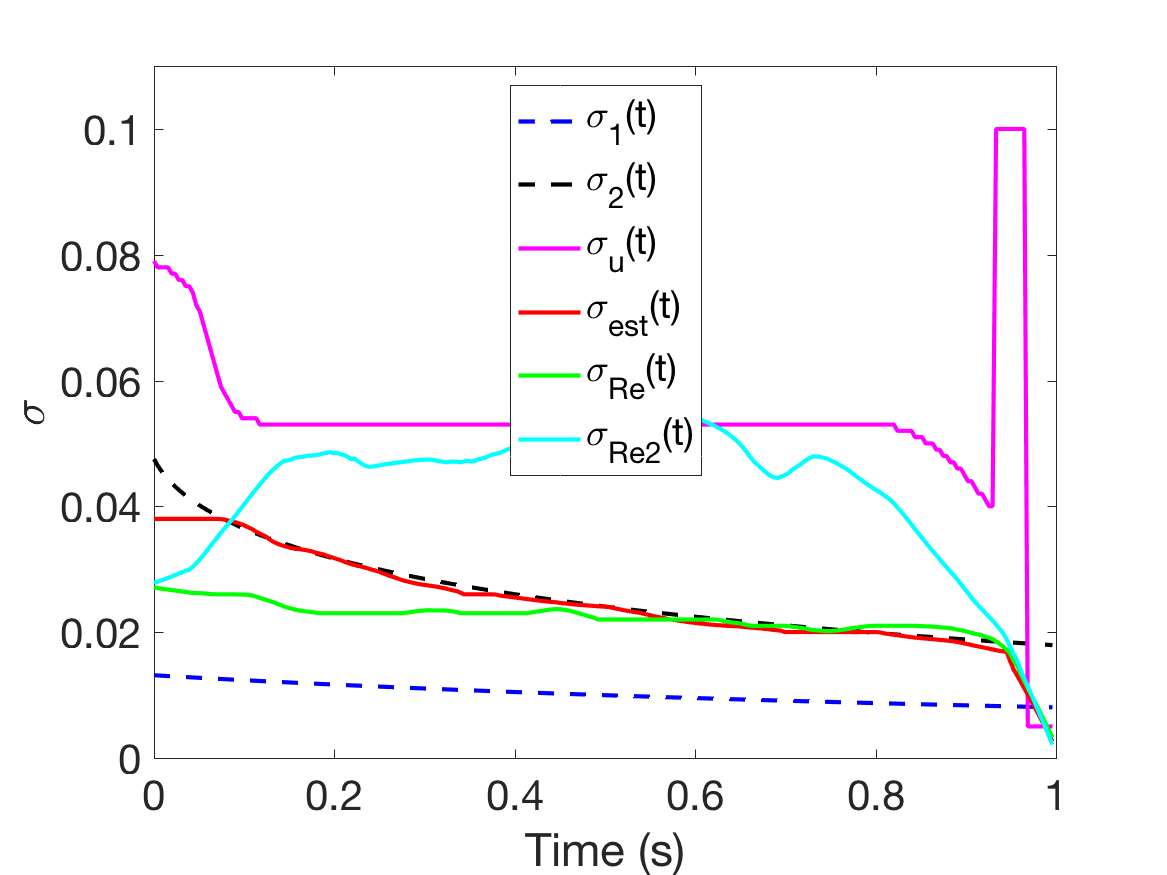}}  
 &
\resizebox{2.0in}{1.5in}{\includegraphics{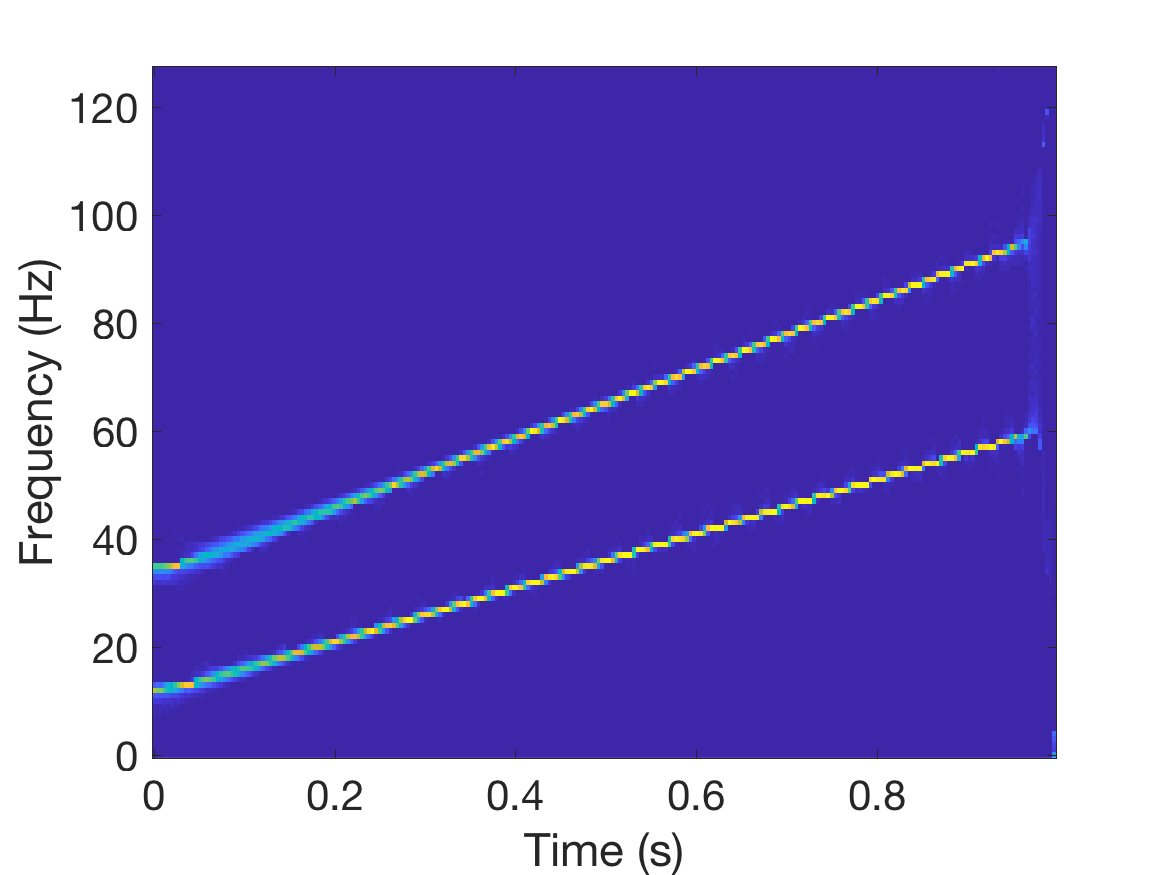}}
& \resizebox{2.0in}{1.5in}{\includegraphics{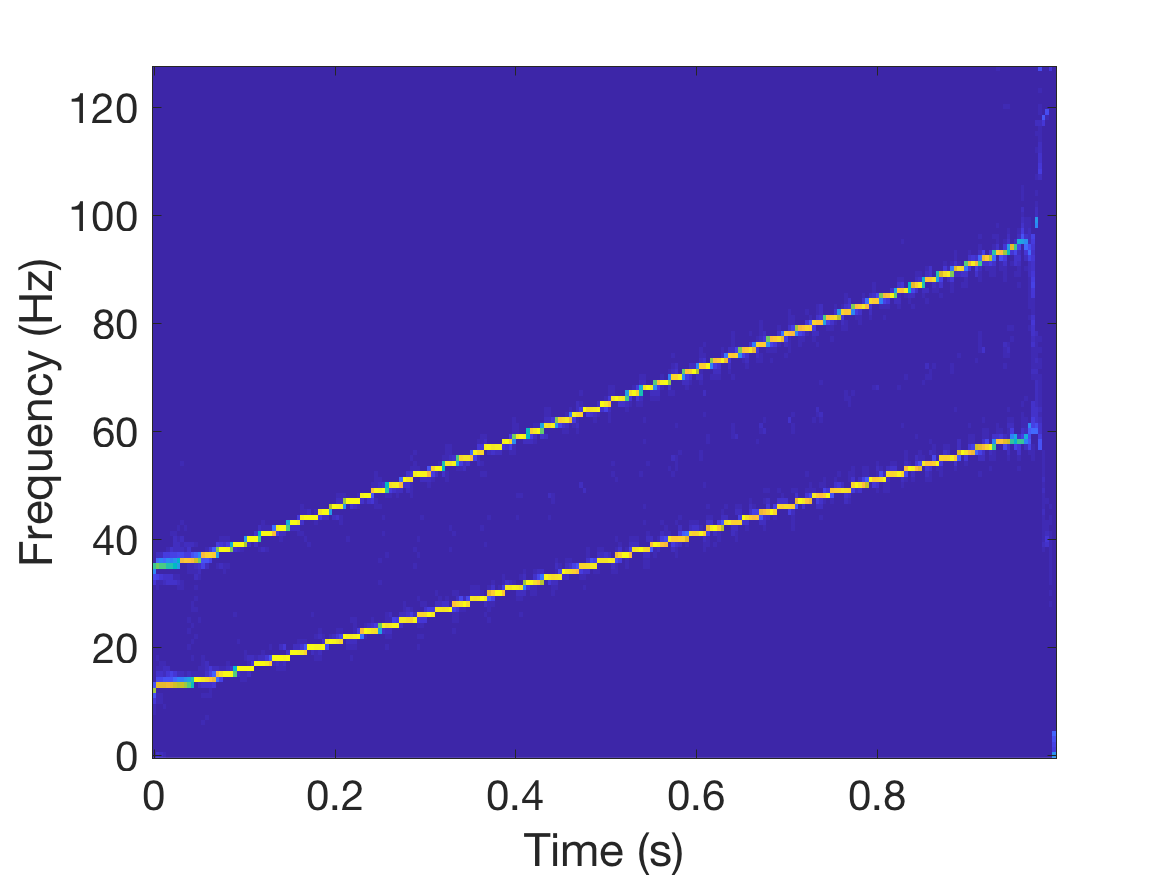}} \\
&
\resizebox{2.0in}{1.5in}{\includegraphics{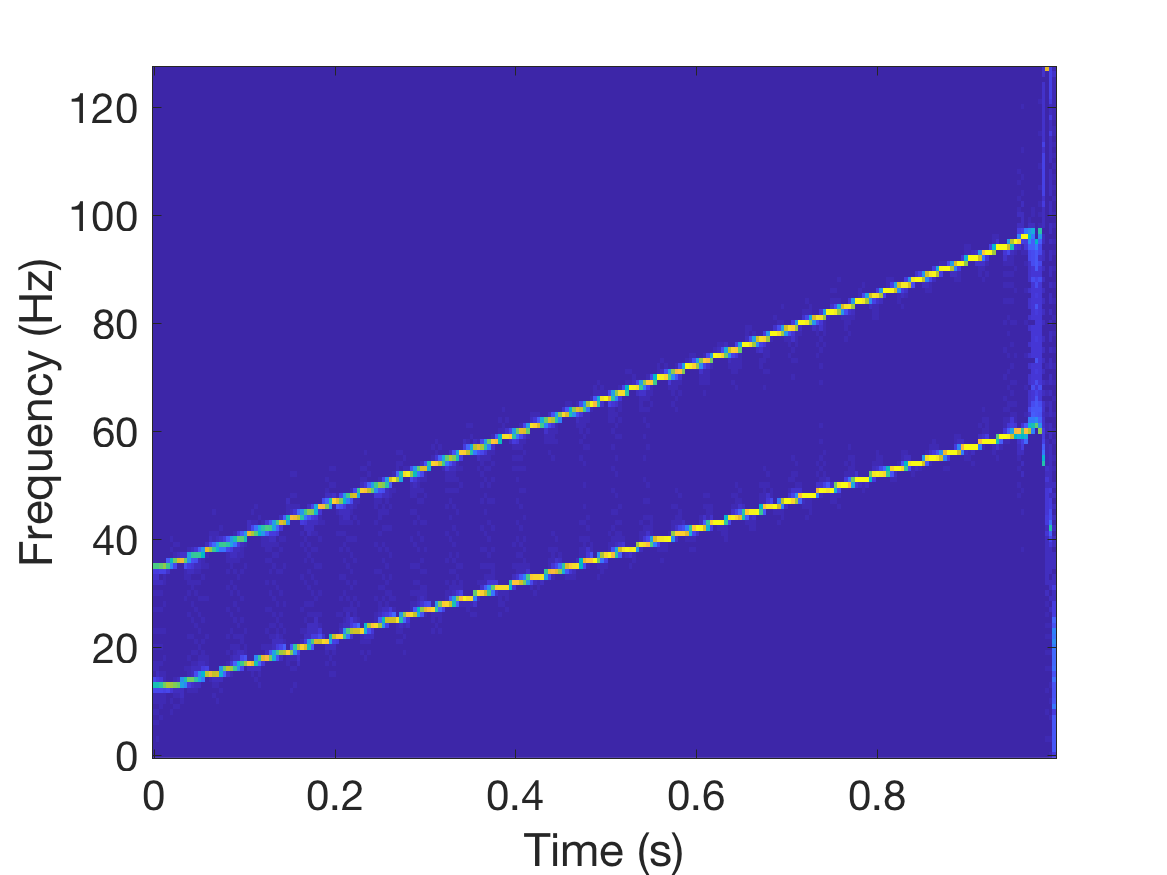}}
&
\resizebox{2.0in}{1.5in}{\includegraphics{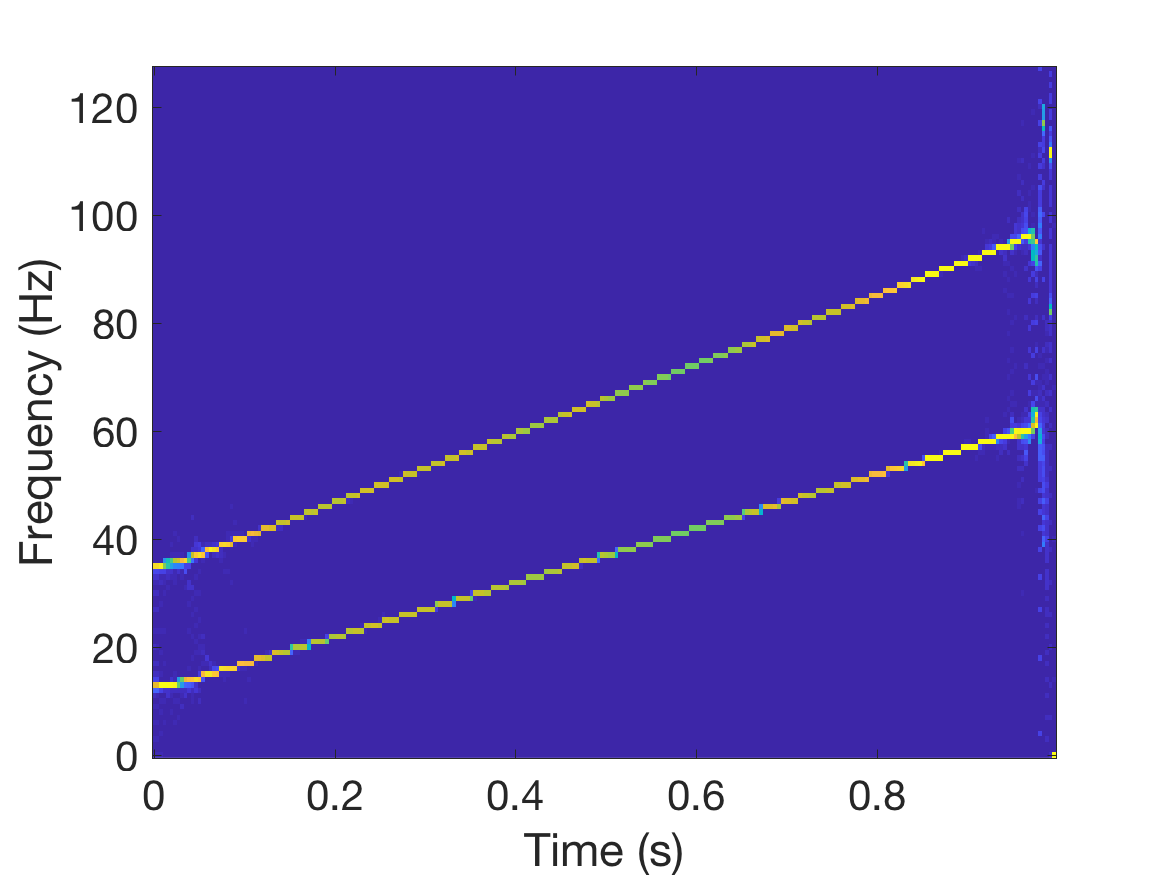}}
\end{tabular}
\caption{\small Experimental results of different time-varying parameters for the two-component LFM signal in \eqref {two_chirps_12_34}: various time-varying parameters (Top-left);
adaptive FSST with $\gs_{est}(t)$ (Top-middle) and 2nd-order adaptive FSST with $\gs_{est}(t)$ (Top-right);
regular-PT adaptive FSST with $\gs_{Re}(t)$ (Bottom-left) and 2nd-order regular-PT adaptive FSST with $\gs_{Re2}(t)$ (Bottom-right).}
\label{figure:STFT_two_chirp_est}
\end{figure}

In \cite{Wu17}, the  R${\rm \acute e}$nyi entropy-based optimal time-varying window was proposed for the sharp representation of SST.
More precisely, let $R_x(b,\xi, \gs)$  and $R_x^{2nd}(b,\xi, \gs)$ be  the regular
FSST and the regular 2nd-order FSST of $x(t)$ (with the phase transformation $\go_x^{2nd}(a, b)$
given in \cite{MOM15}) defined by \eqref{def_FSST_simple} and  \eqref{def_2ndFSST_simple} respectively
with the window function $h(t)=g_\gs(t)$ given by 
\eqref{def_Gaussian_time} containing $\gs>0$.
Denote the R${\rm \acute e}$nyi entropies of $R_x(b,\xi, \gs)$  and $R_x^{2nd}(b,\xi, \gs)$
by $E^{SST}_{\ell,\zeta, \gs} (t)$  and $E^{SST2}_{\ell ,\zeta, \gs} (t)$ respectively, which are defined by  \eqref{def_renyi_entropy_spec} with $V_x(b,\xi)$ to be replaced by $R_x(b,\xi, \gs)$  and $R_x^{2nd}(b,\xi, \gs)$ for certain fixed $\ell, \zeta$. The optimal time-varying parameter is obtained by minimizing $E^{SST}_{\ell ,\zeta, \gs} (t)$  and $E^{SST2}_{\ell ,\zeta, \gs} (t)$:
\begin{equation}
\label{def_Wu_optimal_SST_para}
\gs_{Re}(t)=\mathop {\rm argmin }\limits_{\gs>0}E^{SST}_{\ell ,\; \zeta, \; \gs} (t),  \;
\gs_{Re2}(t)=\mathop {\rm argmin }\limits_{\gs>0}E^{SST2}_{\ell ,\; \zeta, \; \gs} (t).
\end{equation}
With $\gs_{Re}(t)$ and $\gs_{Re2}(t)$ obtained by \eqref{def_Wu_optimal_SST_para}, \cite{Wu17} defines the time-varying-window FSST with $\gs=\gs_{Re}(t)$ by
\eqref{def_FSST_para_simple} but with the phase transformation $\go^{adp}_x(t,\xi)$ in \eqref{def_phase_para} replaced by the regular 
phase transformation $\go_x(t,\xi)$ defined by the formula  \eqref{def_phase} for the conventional FSST. Similarly, the 2nd-order time-varying-window FSST  with $\gs=\gs_{Re2}(t)$ in \cite{Wu17} is defined by
\eqref{def_2ndFSST_para_simple} but with the phase transformation $\go^{adp, 2nd}_x(t,\xi)$ in \eqref{2nd_phase_para} replaced by a regular phase transformation $\go^{2nd}_x(t,\xi)$ defined by a formula in \cite{MOM15} for the conventional 2nd-order FSST. With PT representing phase transformation,
we call them the regular-PT adaptive FSST and the 2nd-order regular-PT adaptive FSST, respectively.

We use the  R${\rm \acute e}$nyi entropy-based  adaptive FSST and our proposed adaptive FSST with $\gs=\gs_{est}(t)$ to 
process the two-component linear chirp signal in \eqref{two_chirps_12_34}.
The different time-varying parameters are shown in the top-left panel of Fig.\ref{figure:STFT_two_chirp_est}, where $\gs_1(t)$, 
$\gs_2(t)$, $\gs_u(t)$, $\gs_{est}(t)$, $\gs_{Re}(t)$ and $\gs_{Re2}(t)$ are defined by \eqref{def_gs1}, \eqref{def_gs2}, \eqref{def_renyi_entropy_best}, 
\eqref{smooth_C} and \eqref{def_Wu_optimal_SST_para}, respectively. Here we let $\gs \in [0.001, 0.2]$ with $\gD\gs=0.001$, namely $\gs_1 = 0.2$ in {Algorithm 1}. 
We set $\ell=2.5$, $ \zeta=4$ (sampling points, for discrete signal) and $\gamma_1$ in \eqref{threhold_gamma1} to be 0.3. 
Note that we set the same values of $\ell$, $ \zeta$, and $\gamma_1$ for all the following experiments.
We use a simple rectangular window $B=\{1/5, 1/5, 1/5, 1/5, 1/5\}$ as the low-pass filter. One can use some other filters, such as an FIR filter or a window of Gaussian or Hamming.
Note that the length of the filter we use is 5, which is related to the parameter $\zeta=4$.
And we use a constant $\epsilon=1/5$ in \eqref{def_ga}, namely constant $\alpha$ in \eqref{estimate_lk}.
The estimation $\gs_{est}(t)$ by  {Algorithm 1} is very close to $\gs_2(t)$ except for the start near $t=0$. 
So the estimation algorithm is an efficient method to estimate the well-separation time-varying parameter $\gs_2(t)$.
Fig.\ref{figure:STFT_two_chirp_est} shows the proposed 
adaptive FSST and 2nd-order adaptive FSST with $\gs_{est}(t)$. The proposed 2nd-order adaptive FSST  gives energy concentration.
In Fig.\ref{figure:STFT_two_chirp_est}, we also provide the regular-PT adaptive FSST with $\gs=\gs_{Re}(t)$ and the 2nd-order regular-PT adaptive FSST 
with $\gs_{Re2}(t)$ as described above.
The regular-PT adaptive FSST performs well in  the TF energy concentration of this
two-component signal.
The Matlab routines for {Algorithm 1}, FSST,  the adaptive FSST and regular-PT adaptive FSST can be downloaded 
at the website of one of the authors:  www.math.umsl.edu/$\sim$jiang .


For most well-separated signals, {\bf Algorithm 1} results in a suitable $\gs_{est}(t)$ with which the 2nd-order adaptive FSST is clear, sharp and concentrated. 
However, when IFs of different components are too close, 
then two adjacent components at Step 2 of  {\bf Algorithm 1} may merge into one,
which results in component mixing. In addition,  the regular-PT adaptive FSST method is unable
to separate such components either, see an experimental example in the next section.
To tackle this problem, we propose to use a varying $\epsilon$ or $\alpha$ in \eqref{def_ga}, 
which defines the bandwidth of $\wh g(\xi)$ and hence determines the support zones of STFTs.
Although a greater $\epsilon$ may result in a  larger recovery error, some components with extremely close IFs can be separated with a large $\epsilon$.
Suppose $\epsilon \in [\epsilon_s,\epsilon_o]$ for some $0<\epsilon_s<\epsilon_o<1$. Our method is first 
we choose the maximum $\epsilon=\epsilon_o$ for fixed $t$ and $\gs$ first, and obtain the support intervals ${\bf s}$ 
in \eqref {def_s_intervals} satisfying \eqref{nonoverlap_hkgk}. Then we decrease $\epsilon$ step by step. 
This way the support intervals in \eqref {def_s_intervals} will increase gradually. 
We stop our procedure when $\epsilon$ reaches the minimum value $\epsilon_s$ or the condition in \eqref{nonoverlap_hkgk} does not hold. 
The following is the revised algorithm to estimate $\gs(t)$.

\bigskip

{\bf Algorithm 2.} 
Let  $\{\gs_j, j=1, 2, \cdots, n\}$ be an uniform discretization of $\gs$  with $\gs_1>\gs_2>\cdots>\gs_n>0$ and sampling step $\gD\gs = \gs_{j-1}-\gs_j$. Let  $\{\epsilon_j, j=1, 2, \cdots, m\}$ be an uniform discretization of $\epsilon$  with $\epsilon_1>\epsilon_2>\cdots>\epsilon_m>0$ and sampling step $\gD\epsilon = \epsilon_{j-1}-\epsilon_j$.
The discrete sequence $s(t),$ $t=t_1, t_2, \cdots , t_N $  (or $t=0, 1, \cdots, N-1$) is the signal to be analyzed.
 \begin{itemize}
\item[] {\bf Step 1.} Let $t$ be one of $t_1, t_2, \cdots , t_N $. Find $\gs_u$ in \eqref{def_renyi_entropy_best} with $\gs\in \{\gs_j, j=1, 2, \cdots, n\}$. 
\item[] {\bf Step 2.} Let ${\bf s}$ be the set of the intervals given by \eqref{def_s_intervals} with  $\gs=\gs_u$ and $\epsilon = \epsilon_1$. Let $z=\gs_u$.
\item[] {\bf Step 3.} If \eqref{nonoverlap_hkgk} holds and $\epsilon > \epsilon_m$, update $\epsilon$ with $\epsilon-\gD\epsilon$, and repeat Step 3.
\item[] {\bf Step 4.} If $\epsilon > \epsilon_m$, go to Step 7. Otherwise, if $\epsilon = \epsilon_m$, go to Step 5.
\item[]{\bf Step 5.} Let $\gs=z-\gD\gs$. If the number of intervals $m$ in \eqref{def_s_intervals} with this new $\gs$ remains unchanged, $\gs\ge\gs_n$ and \eqref{nonoverlap_hkgk} holds, then go to Step 6. Otherwise, go to Step 7.
\item[]{\bf Step 6.} Let $\epsilon = \epsilon_1$, go to Step 3 with $z=\gs$.
\item[]{\bf Step 7.} Let $C(t)=z$, and do Step 1 to Step 6 for different time $t$ of $t_1, t_2, \cdots , t_N $.
\item[]{\bf Step 8.} Smooth  $C(t)$ with a low-pass filter $B(t)$:
\begin{equation*}
\gs_{est2}(t)=(C*B)(t).
\end{equation*}
\end{itemize}

\section{Further experiments and results}

In this section, we provide more numerical examples to further illustrate the effectiveness and robustness of our method in the IF estimation and component recovery.

\subsection{Experiments with a three-component synthetic signal}

\begin{figure}[th]
\centering
\begin{tabular}{cc}
\resizebox{2.0in}{1.5in}{\includegraphics{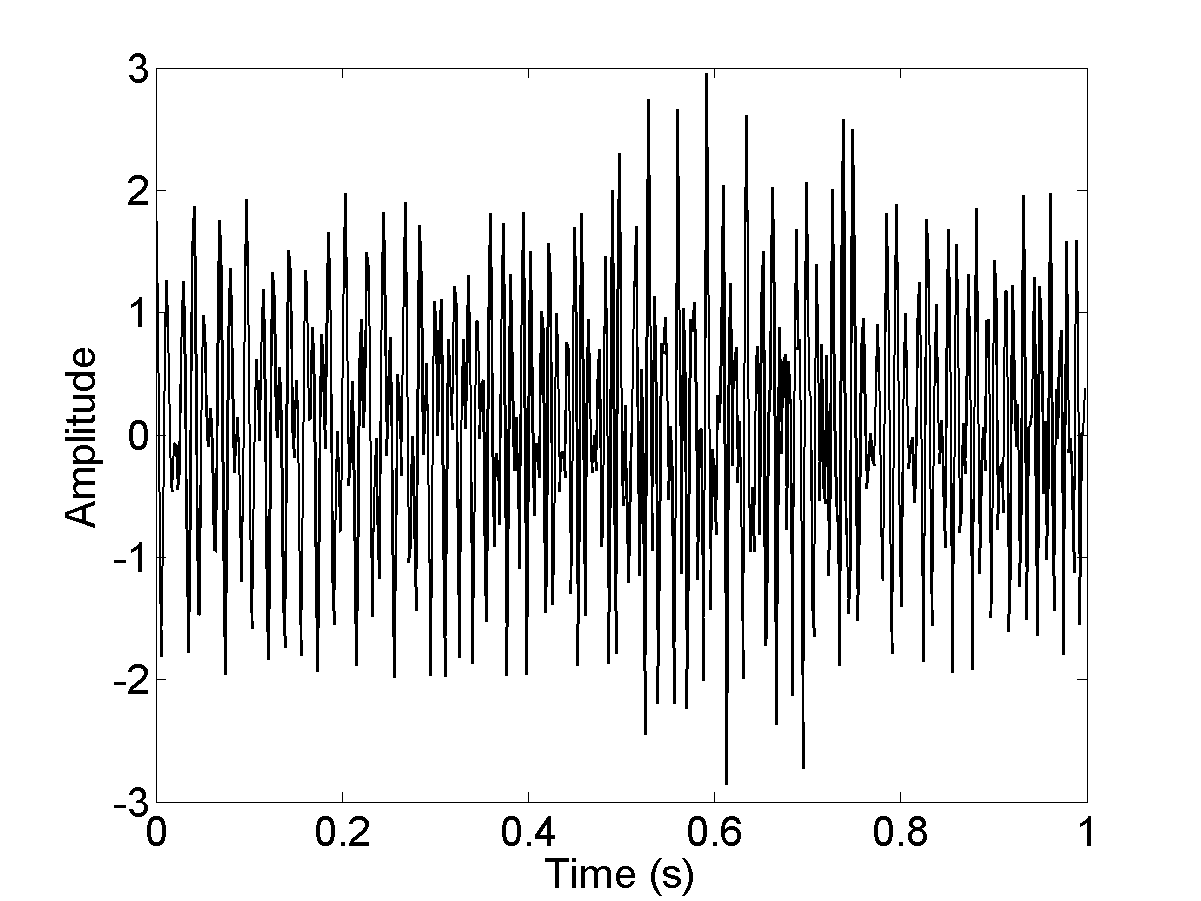}} \quad &
\resizebox{2.0in}{1.5in}{\includegraphics{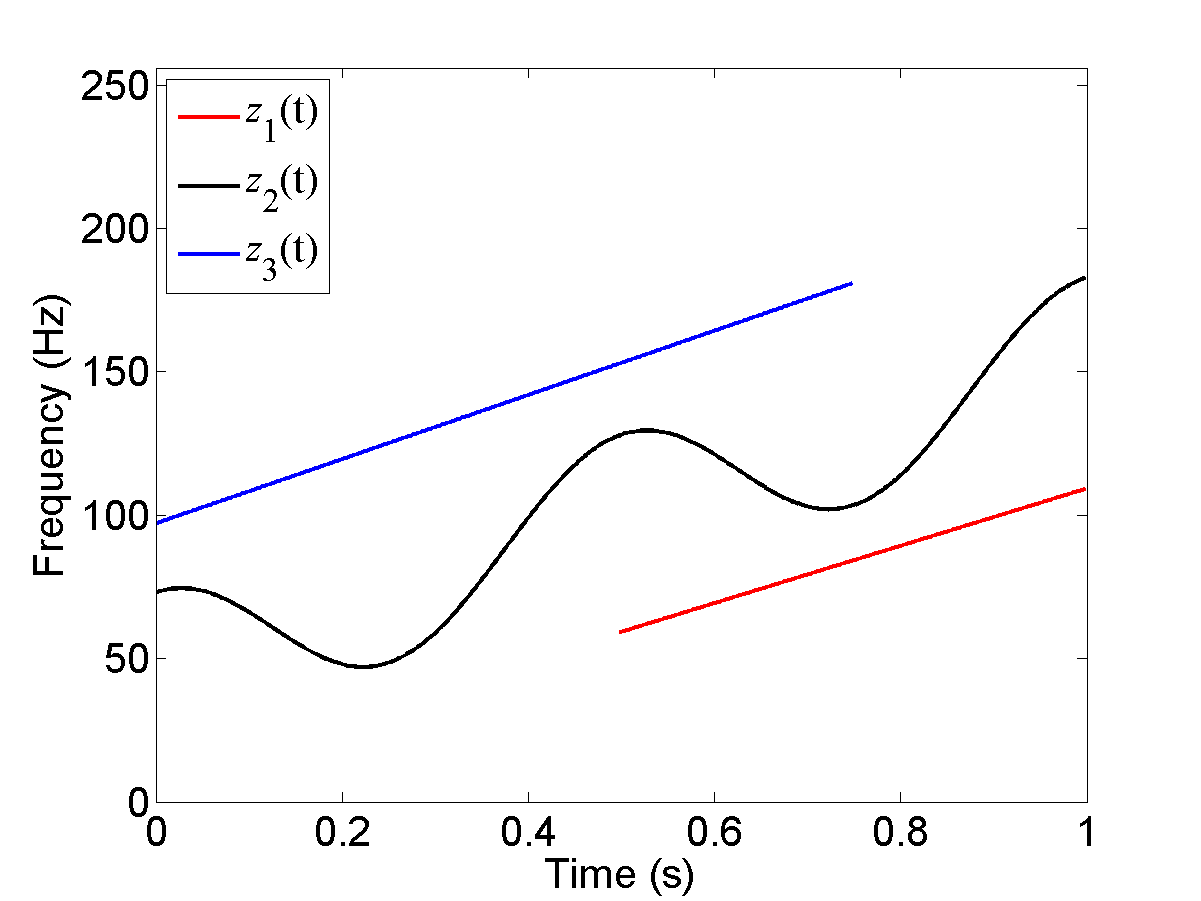}}\\
\end{tabular}
\caption{\small Three-component signal in \eqref{three_component_signal}: its waveform (Left panel) and  IFs of its components (Right panel).}
\label{fig:waveform_IFs_three_component}
\end{figure}

The three-component signal we consider is given by
\begin{equation}
\label{three_component_signal}
z(t)=z_1(t)+z_2(t)+z_3(t),
\end{equation}
where
\[
\begin{split}
z_1(t) & = \cos \left(118\pi(t-1/2)+100\pi(t-1/2)^2\right), \quad t\in [1/2, 1],\\
z_2(t) & = \cos \left(94\pi t+13\cos(4\pi t-\pi/2)+110\pi t^2\right), \quad t\in [0, 1],\\
z_3(t) & = \cos \left(194\pi t+112\pi t^2\right), \quad t\in [0, 3/4].
\end{split}
\]
Note that the durations of the three components in \eqref {three_component_signal} are different, that is $K$ in \eqref{AHM} can be time-varying. 
The sampling rate for this experiment is 512Hz, namely we have 512 discrete samples for $z(t)$.
The IFs of the three components are $\phi_1'(t)=59+100(t-1/2)$, $\phi_2'(t)=47-26\sin(4\pi t-\pi/2)+110t$ and $\phi_3'(t)=97+112t$,  respectively.
Fig.\ref{fig:waveform_IFs_three_component} shows the waveform and IFs of $z(t)$.

\begin{figure}[th]
\centering
\begin{tabular}{ccc}
\resizebox{2.0in}{1.5in}{\includegraphics{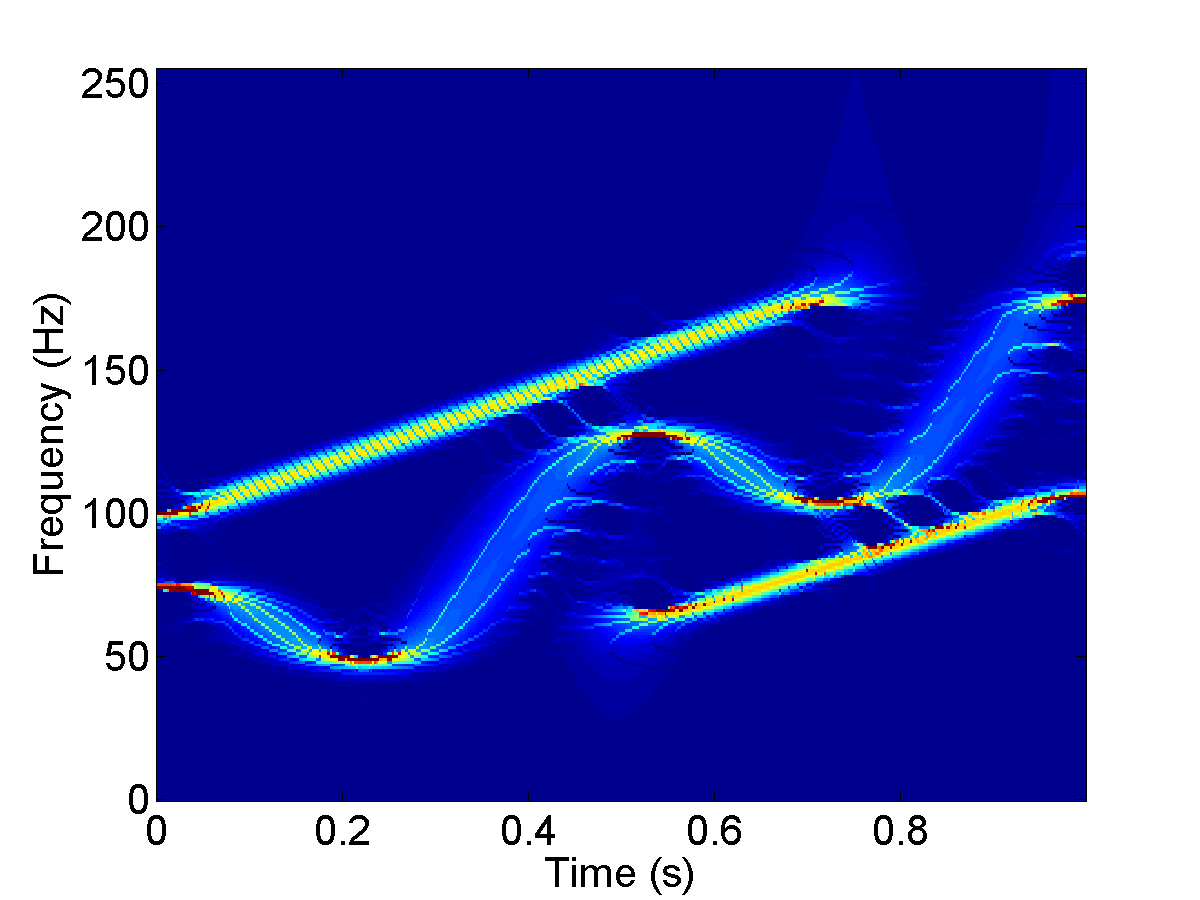}}&
\resizebox{2.0in}{1.5in}{\includegraphics{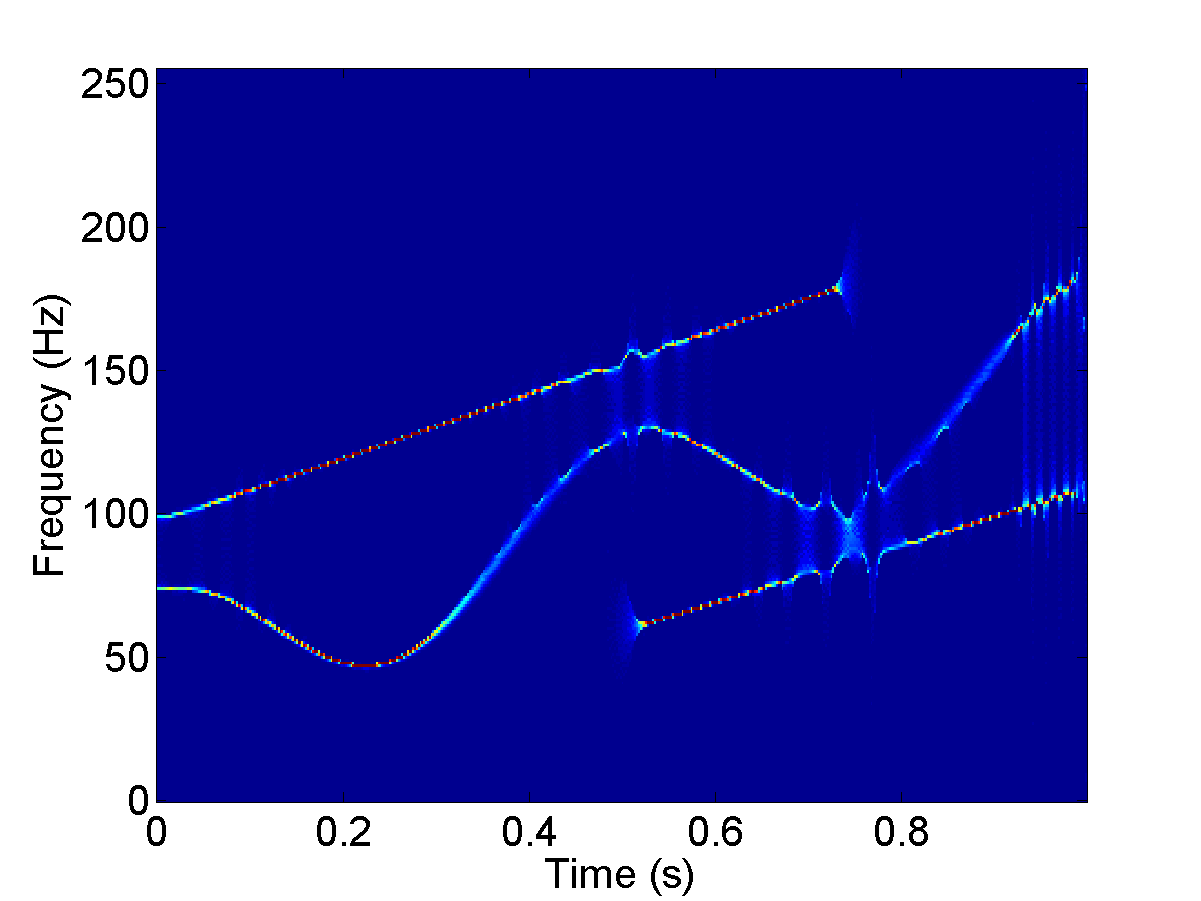}} 
&\resizebox{2.0in}{1.5in}{\includegraphics{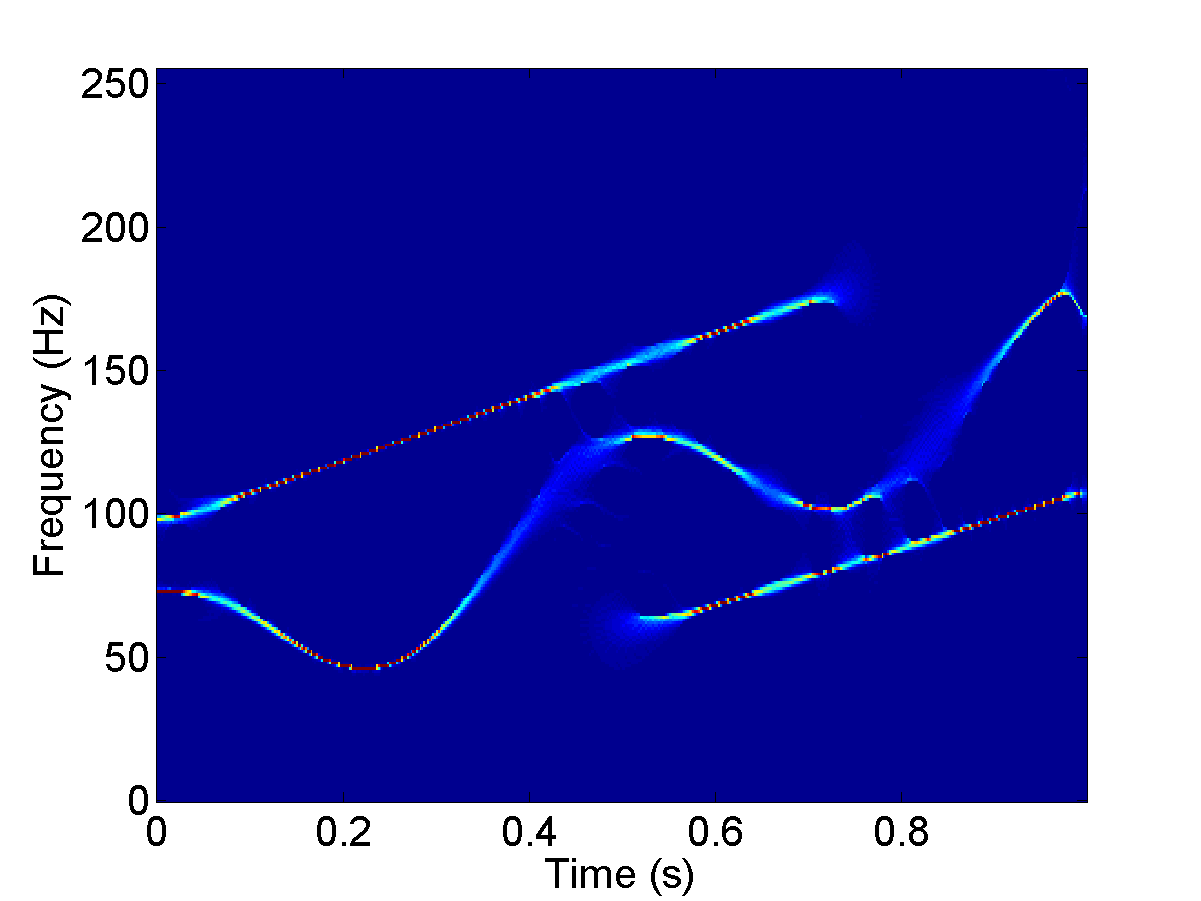}} \\
\resizebox{2.0in}{1.5in}{\includegraphics{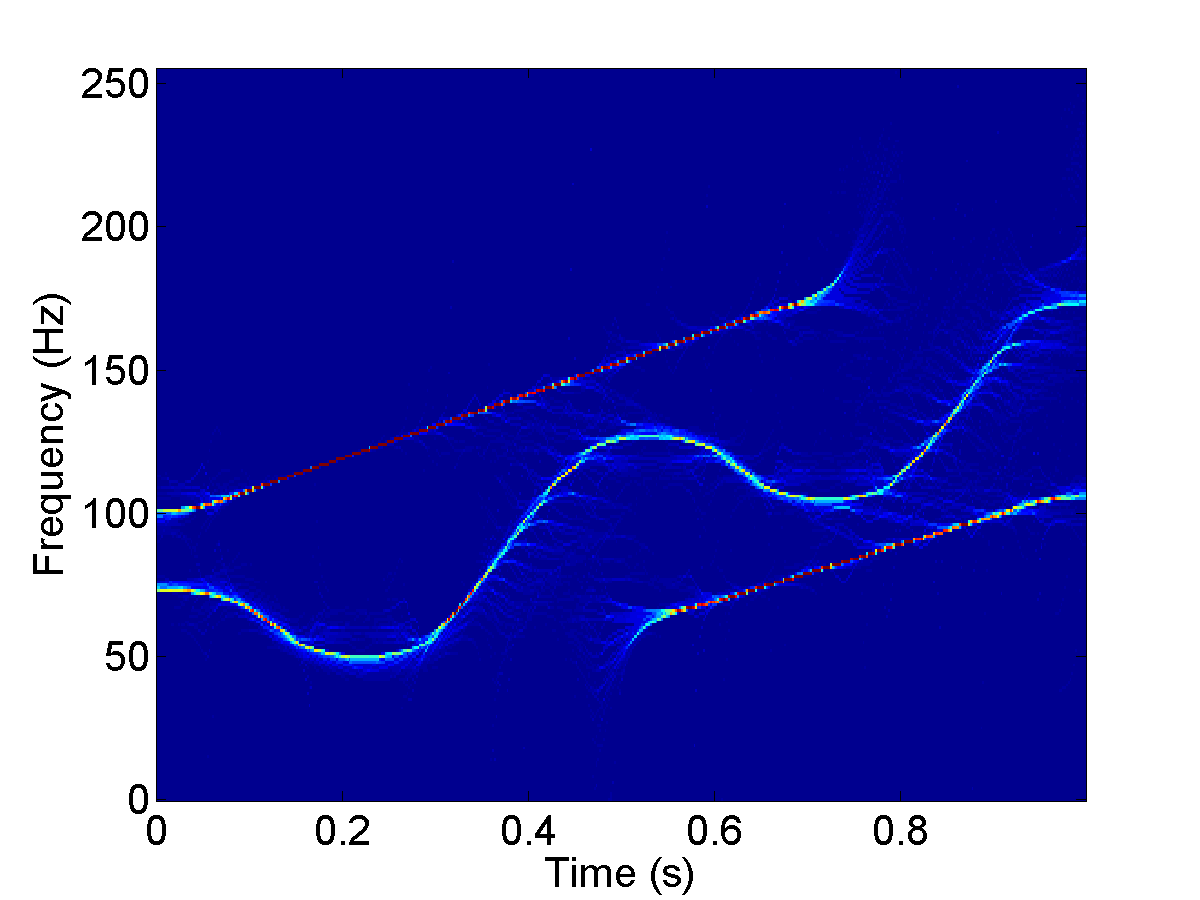}}
&\resizebox{2.0in}{1.5in}{\includegraphics{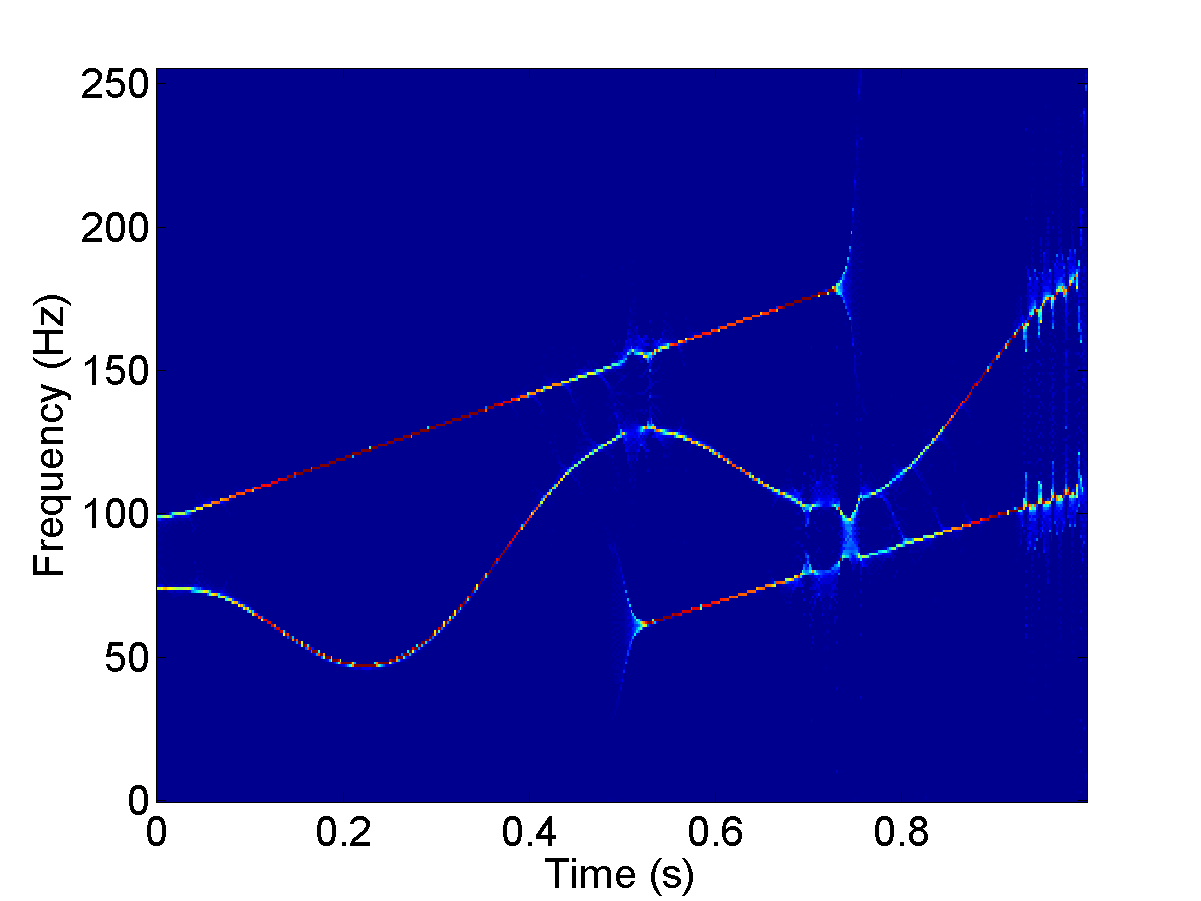}} 
&\resizebox{2.0in}{1.5in}{\includegraphics{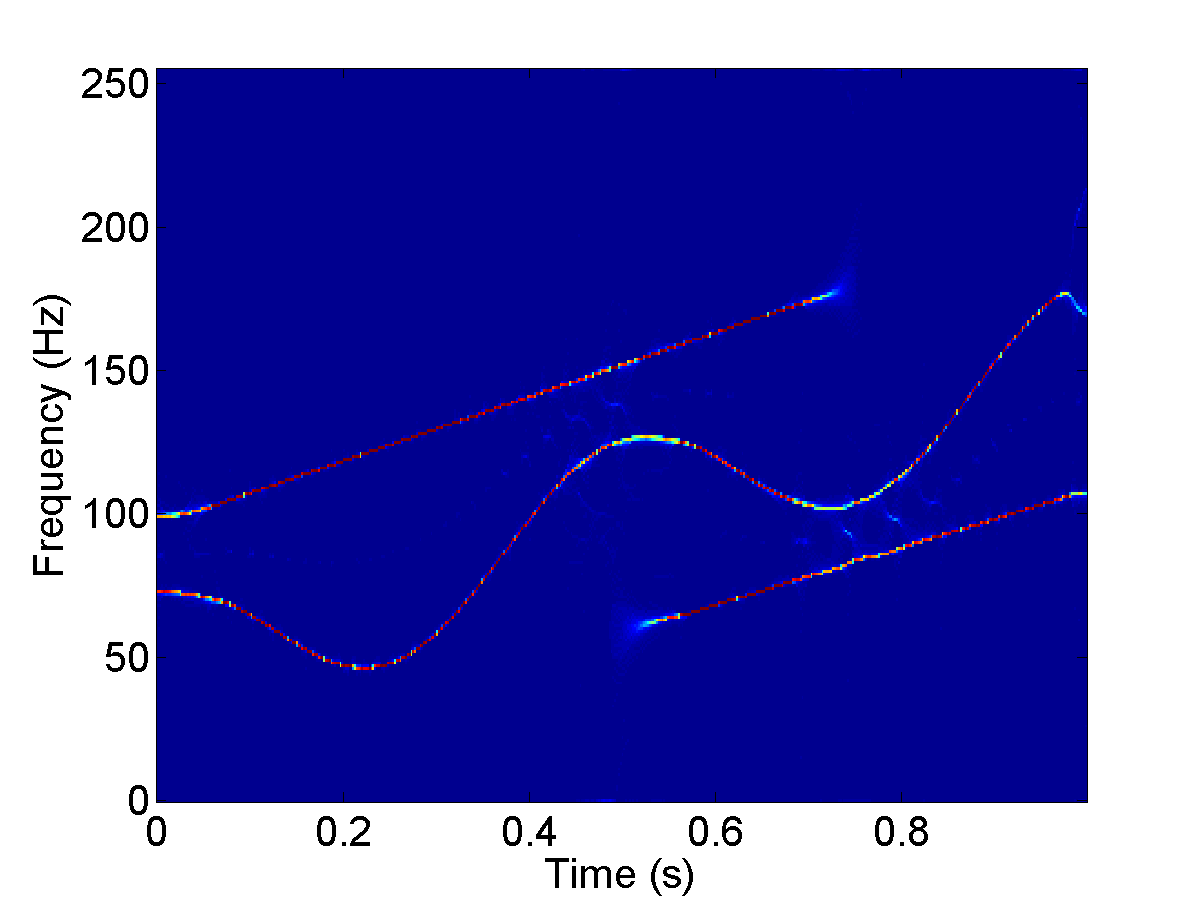}} 
\end{tabular}
\caption{\small Experimental results on the three-component signal in \eqref{three_component_signal}. 
Top-row (from left to right): conventional FSST with constant $\gs=0.04$, regular-PT adaptive FSST with $\gs_{Re}(t)$ and adaptive FSST with $\gs_{est2}(t)$; 
Bottom-row (from left to right): conventional 2nd-order FSST with constant $\gs=0.04$, 2nd-order regular-PT adaptive FSST 
with $\gs_{Re2}(t)$ and 2nd-order adaptive FSST with $\gs_{est2}(t)$.}
\label{fig:FSST_three_component}
\end{figure}

We calculate various time-varying parameters as those shown in Fig.\ref {figure:STFT_two_chirp_est} and $\gs_{est2}(t)$ as well.
In this experiment, to obtain $\gs_{est2}(t)$ with Algorithm 2, we consider a time-varying $\epsilon(t)$ with $\epsilon \in [0.2, 0.8]$ and $\gD\epsilon = 0.01$. 
For this three-component signal, we observe that 
the conventional FSST, regular-PT adaptive FSST and adaptive FSST with $\gs_{est}(t)$
cannot separate the three components well due to that the frequencies of two components are close to each other, see Fig.\ref{fig:FSST_three_component}, 
while the 2nd-order adaptive
FSST with $\gs_{est2}(t)$ provides quite sharp and clear representations of the three components.
In Fig.\ref{fig:FSST_three_component}, for the conventional FSST and  conventional 2nd-order FSST, we use 
$\sigma=0.04$ which is obtained by minimizing the R${\rm \acute e}$nyi entropy of the STFT. 

\begin{figure}[th]
	\centering
	\begin{tabular}{ccc}
		\resizebox{2.1in}{1.5in}{\includegraphics{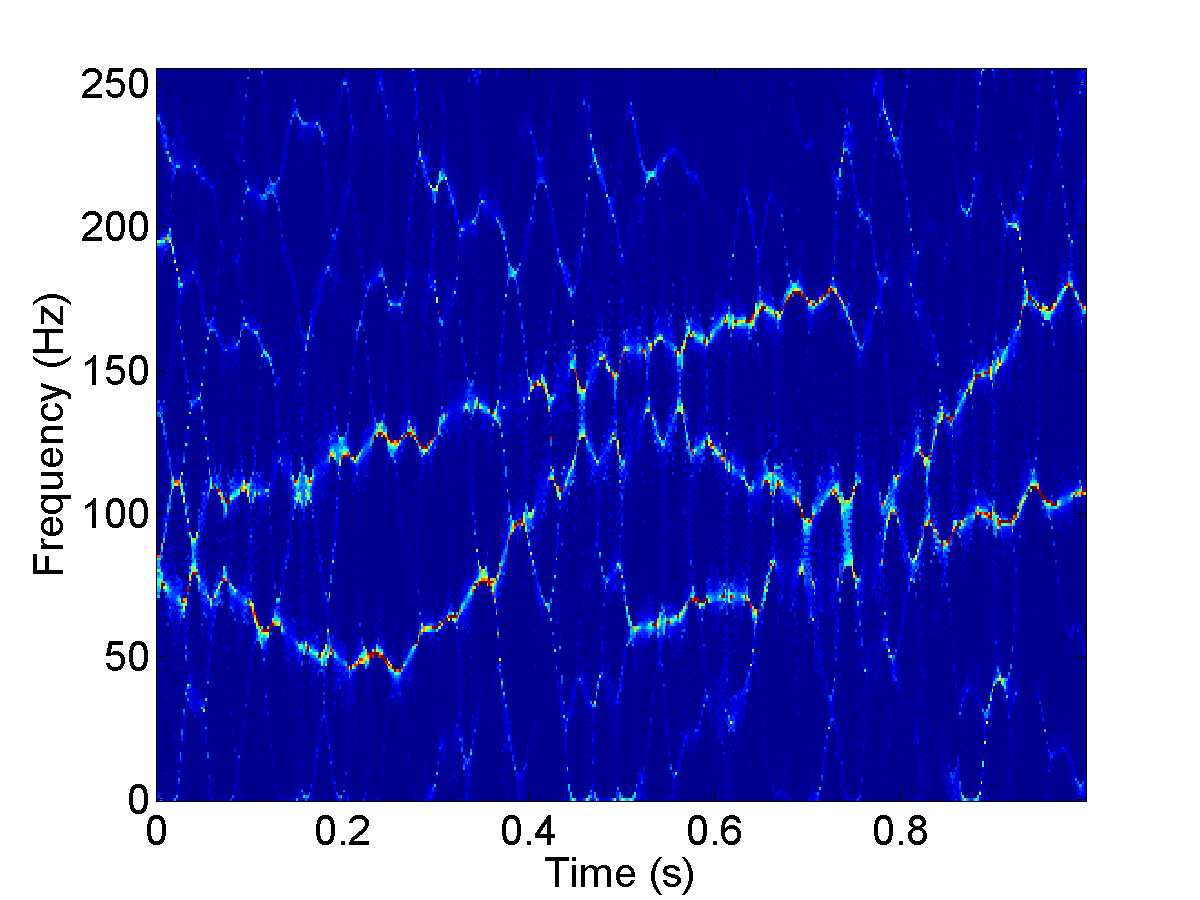}} \quad &
		\resizebox{2.1in}{1.5in}{\includegraphics{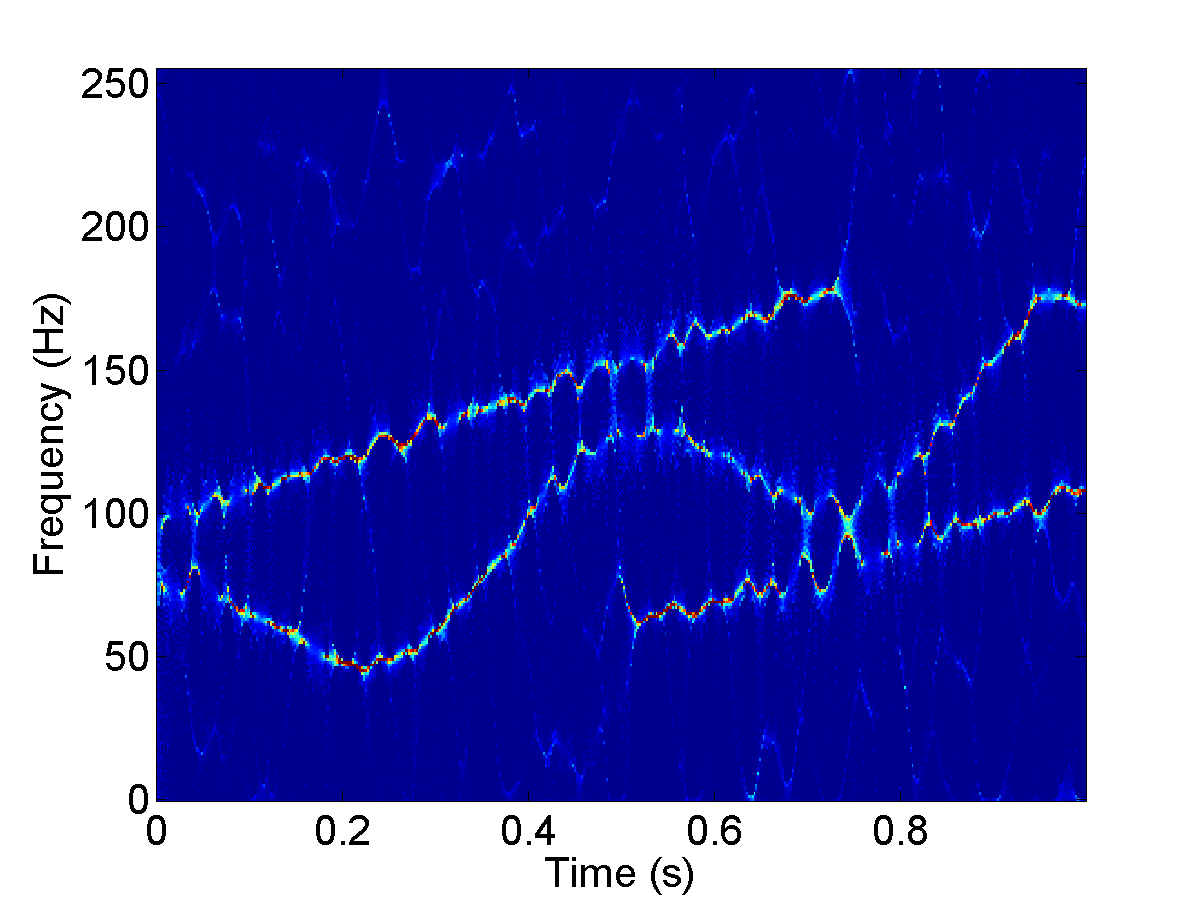}} \quad &
		\resizebox{2.1in}{1.5in}{\includegraphics{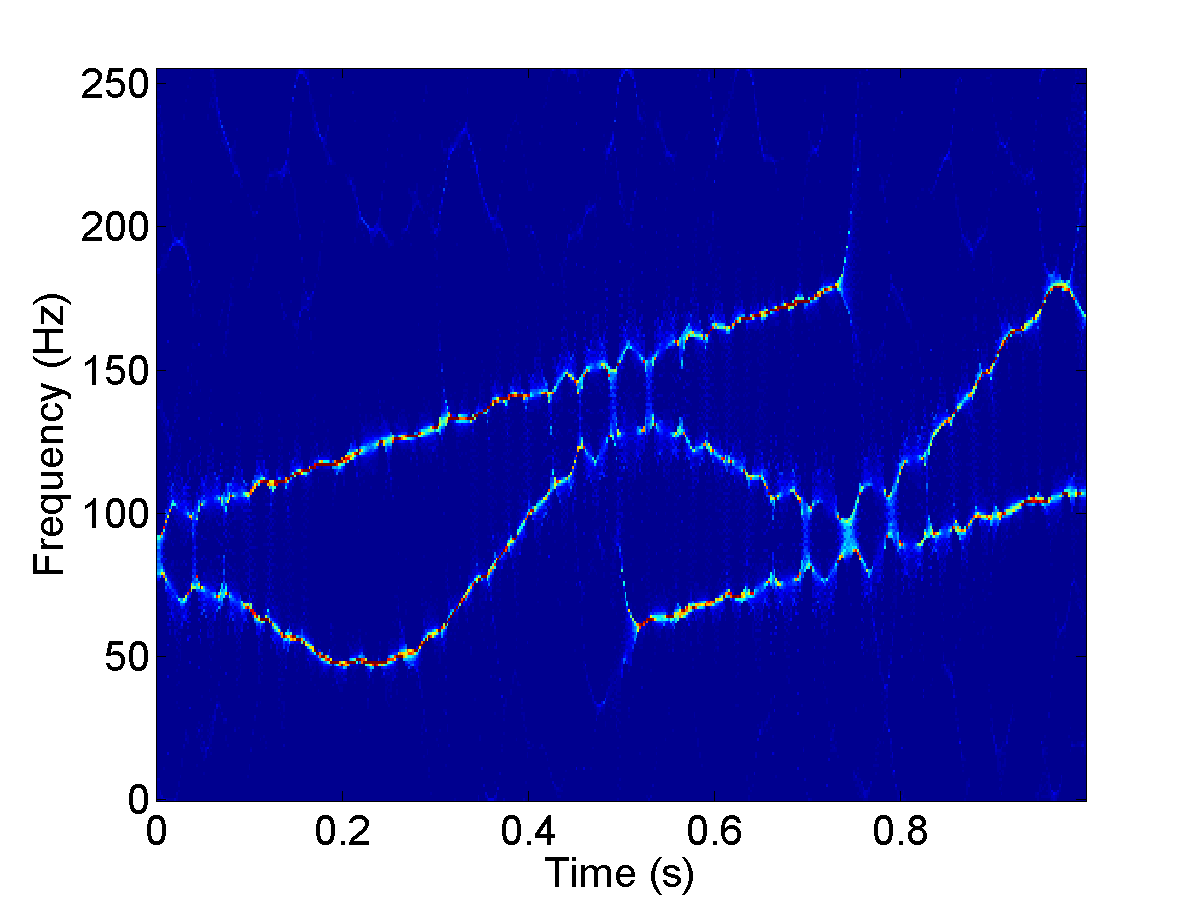}}\\
		\resizebox{2.1in}{1.5in}{\includegraphics{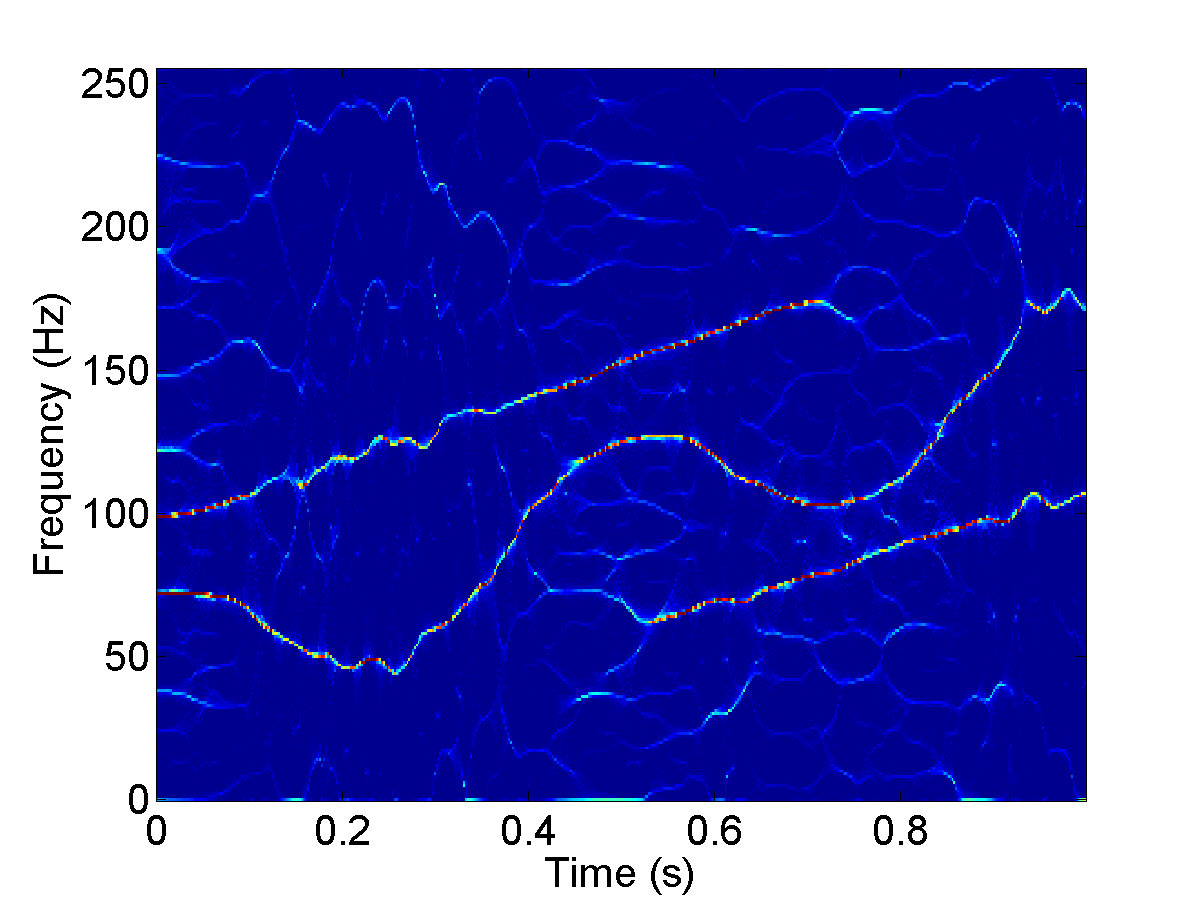}} \quad &
		\resizebox{2.1in}{1.5in}{\includegraphics{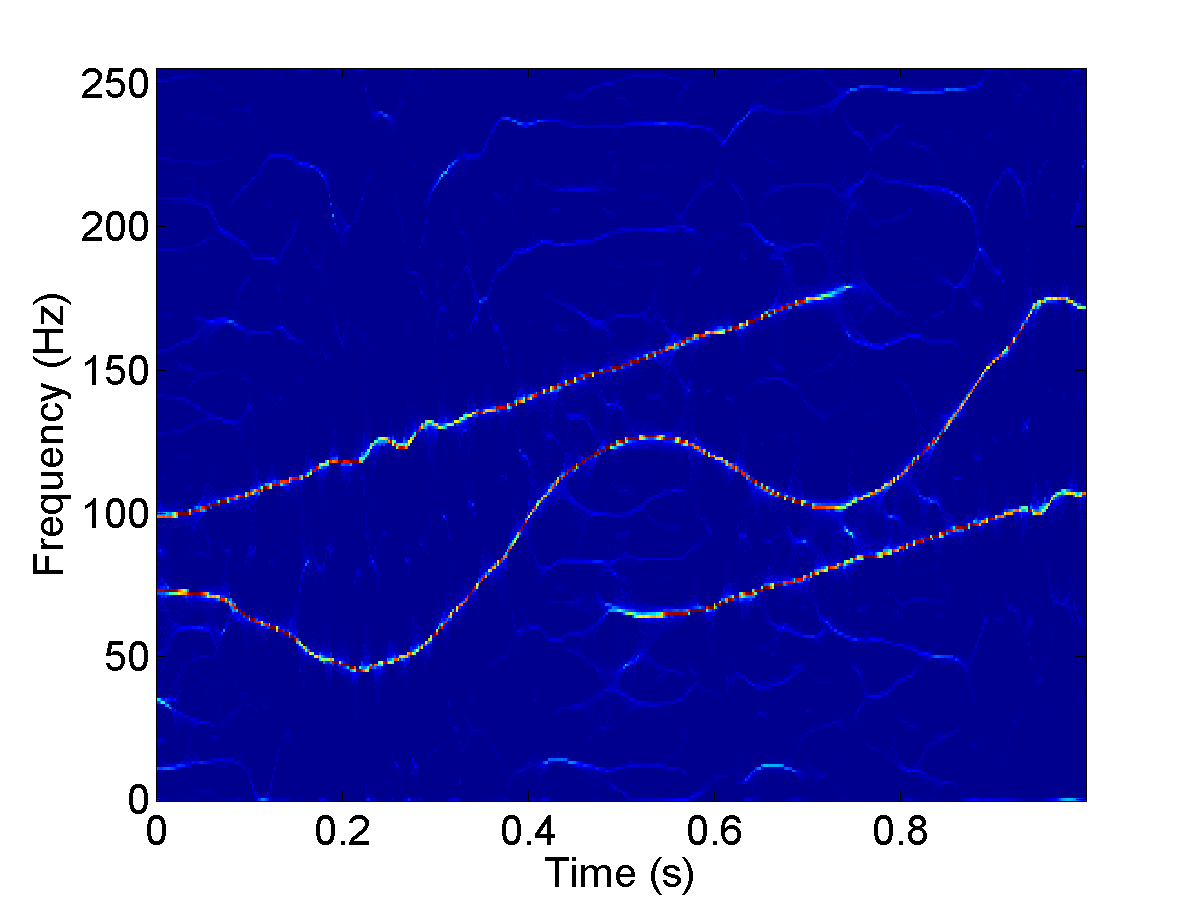}} \quad &
		\resizebox{2.1in}{1.5in}{\includegraphics{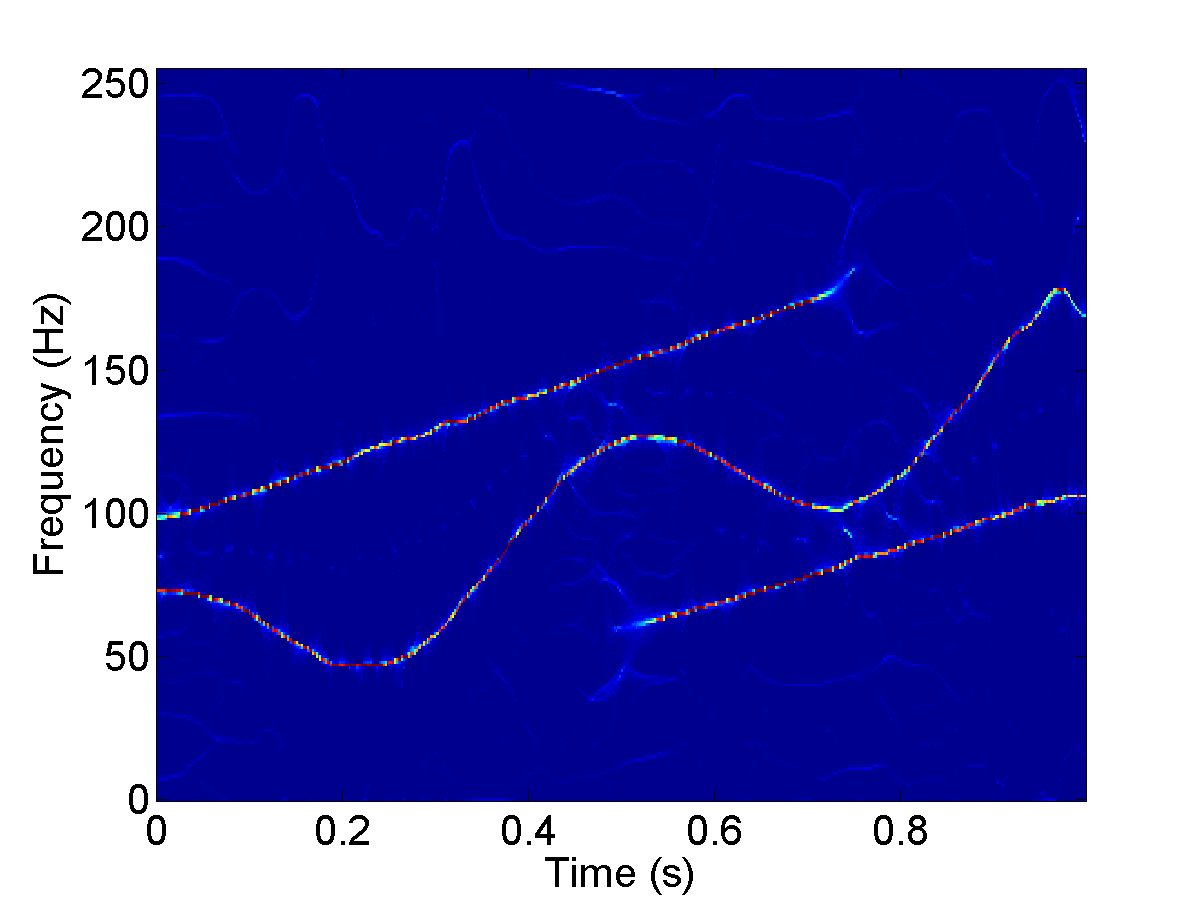}}
	\end{tabular}
	\caption{\small FSSTs of three-component signal in (66) under different noise levels. 
		Top row (from left to right): Conventional 2nd-order FSSTs with constant $\sigma=0.01$ under SNRs of 5dB, 10dB and 15dB. 
		Bottom row (from left to right): 2nd-order adaptive FSSTs under SNRs of 5dB, 10dB and 15dB.}
	\label{fig:TFs_three_component_noises}
\end{figure}

We also consider FSSTs in noise environment.  We add Gaussian noises to the original signal given in  \eqref{three_component_signal} with different signal-to-noise ratios (SNRs).
Fig.\ref{fig:TFs_three_component_noises} shows the conventional 2nd-order FSST and our proposed 2nd-order adaptive FSST 
under different noise levels. Observe that our method works well under noisy environment.

\begin{figure}[th]
\centering
\begin{tabular}{cc}
\resizebox{2.4in}{1.6in}{\includegraphics{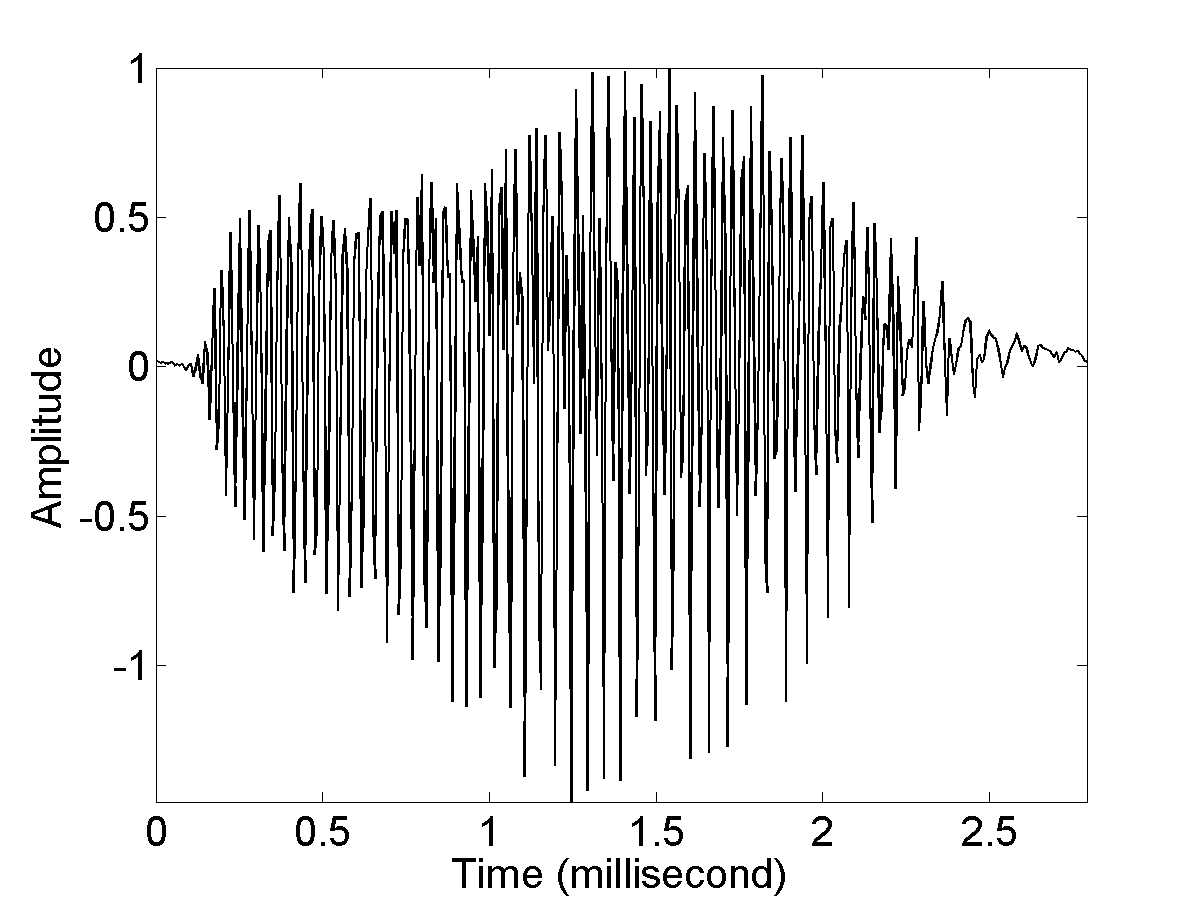}}
\quad &
\resizebox{2.4in}{1.6in}{\includegraphics{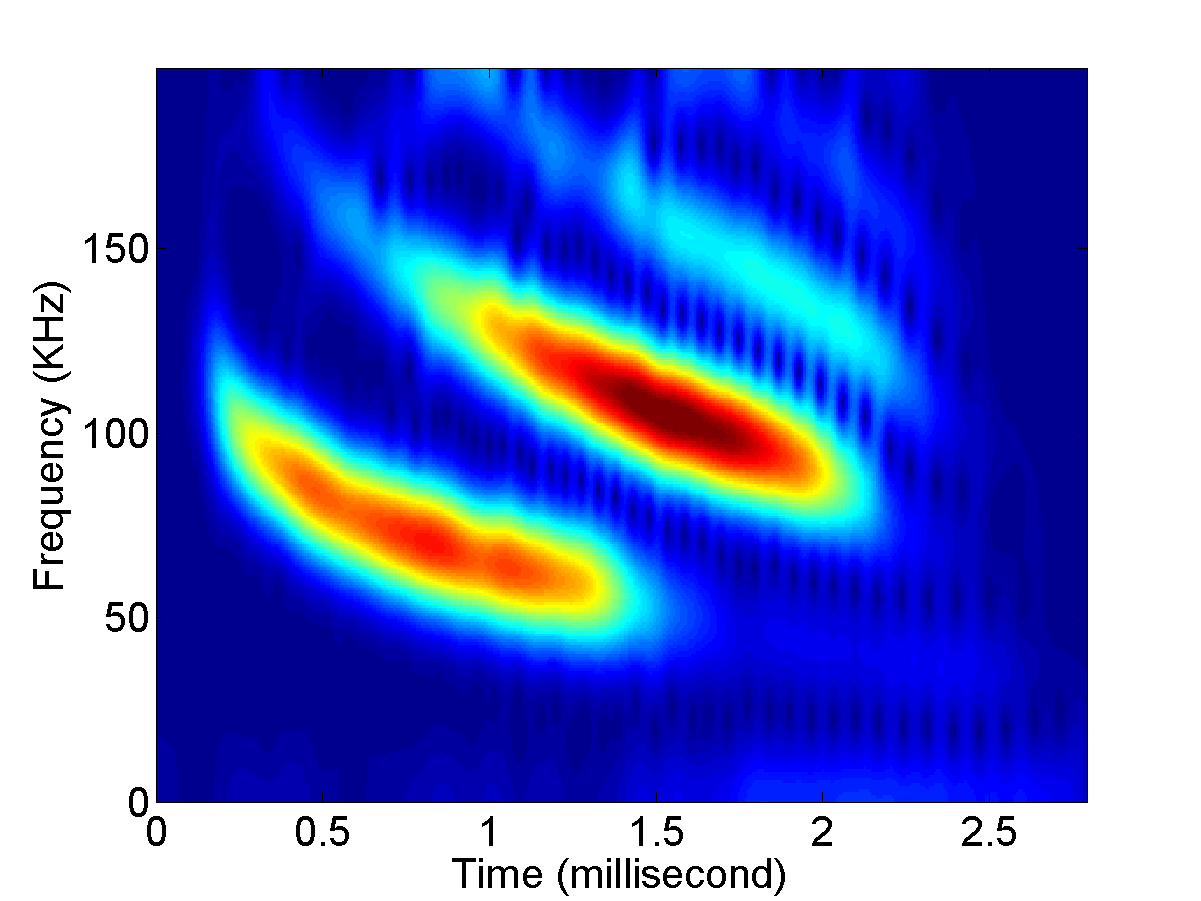}}\\ 
\resizebox{2.4in}{1.6in}{\includegraphics{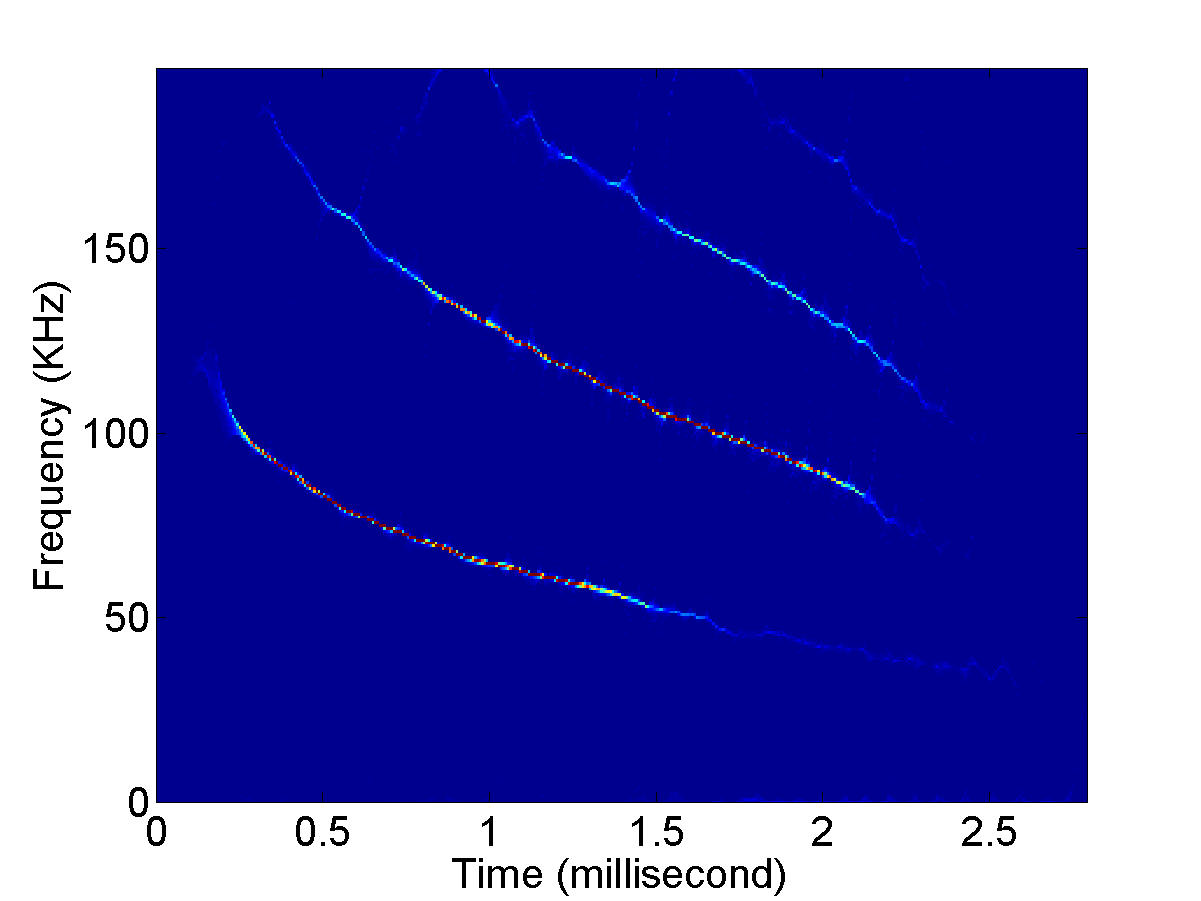}} \quad &
\resizebox{2.4in}{1.6in}{\includegraphics{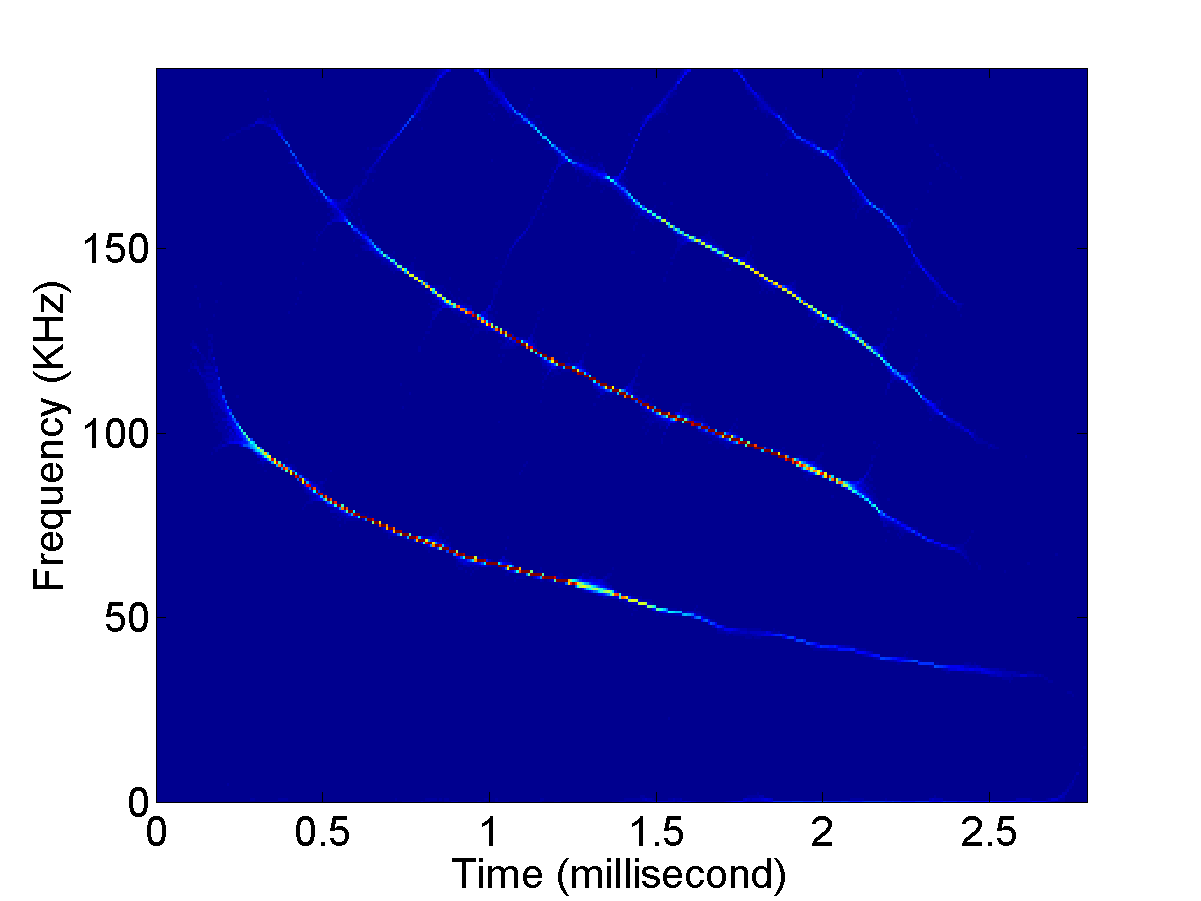}}
\end{tabular}
\caption{\small Example of the bat echolocation signal. Top-left:  waveform; Top-right: conventional STFT with $\gs=8\times10^{-5}$;
Bottom-left: conventional 2nd-order FSST with $\gs=8\times10^{-5}$; Bottom-right: 2nd-order adaptive FSST with time-varying parameter $\gs_{est}(t)$ obtained by our proposed Algorithm 1.}
\label{fig:adaptiveSST_bat}
\end{figure}

\subsection{Application to bat echolocation signal}

In order to further verify the reliability of the proposed algorithm, we test our method on a bat echolocation signal emitted by a large brown bat. There are 400 samples with the sampling period 7 microseconds (sampling rate $F_s\approx142.86$ KHz). 
For a given real-world signal, how to select an appropriate constant $\gs$ such that the resulting conventional SST or 2nd-order  
SST has a sharp representation is probably not very simple.
Here we choose $\gs=8 \times 10^{-5}$, which is close to the mean of $\gs_{est}(t)$ obtained by Algorithm 1.  Fig.\ref{fig:adaptiveSST_bat} 
shows the TF representations of the echolocation signal: STFT,
conventional 2nd-order FSST with $\gs=8 \times10^{-5}$ 
and the 2nd-order adaptive FSST with time-varying parameter $\gs_{est}(t)$.
Unlike the three-component signal in Fig.\ref{fig:FSST_three_component}, the four components in the bat signal are much well separated.
Thus, both the conventional 2nd-order FSST and the 2nd-order adaptive FSST can separate well the components of the signal. In addition, they give sharp representations in the TF plane.
Comparing with the conventional 2nd-order FSST, the  2nd-order adaptive FSST with $\gs_{est}(t)$ 
gives a better representation for the fourth component (the highest frequency component) and the two ends of the signal.
Furthermore, $\gs_{est}(t)$ provides a hint how to select $\gs$ for the conventional 2nd-order FSST.

\subsection{Signal separation}
Finally we consider the separation of a multicomponent signal: to recover/reconstruct its components.
We use \eqref{reconst_SST_component}, \eqref{FSST_recover_para_component} and similar formulas to recover the signal components for conventional FSST and adaptive FSST. Here we use the maximum values on the FSST plane to search for the IF ridges $\phi'_k(t)$ one by one, see details in \cite{Meignen17}. Then integrate around the ridges with $\Gamma=\Gamma_1=15$ (discrete value, unitless). We use the relative ``root mean square error" (RMSE) to evaluate the separation performance, which is defined by
\begin{equation}
\label{relative_root_MSE}
RMSE=\frac{1}{K}\sum\limits_{k = 1}^K {\frac{{\left\| {z_k  - \hat z_k } \right\|_2 }}{{\left\| {z_k } \right\|_2 }}} ,
\end{equation}
where $\hat z_k$ is the reconstructed $z_k$, $K$ is the number of components.
We also consider signal separation in noise environment.  As before we add Gaussian noises to the original signal given in \eqref{three_component_signal} with different signal-to-noise ratios (SNRs).

\begin{figure}[th]
\centering
\begin{tabular}{cc}
\resizebox{2.40in}{1.6in}{\includegraphics{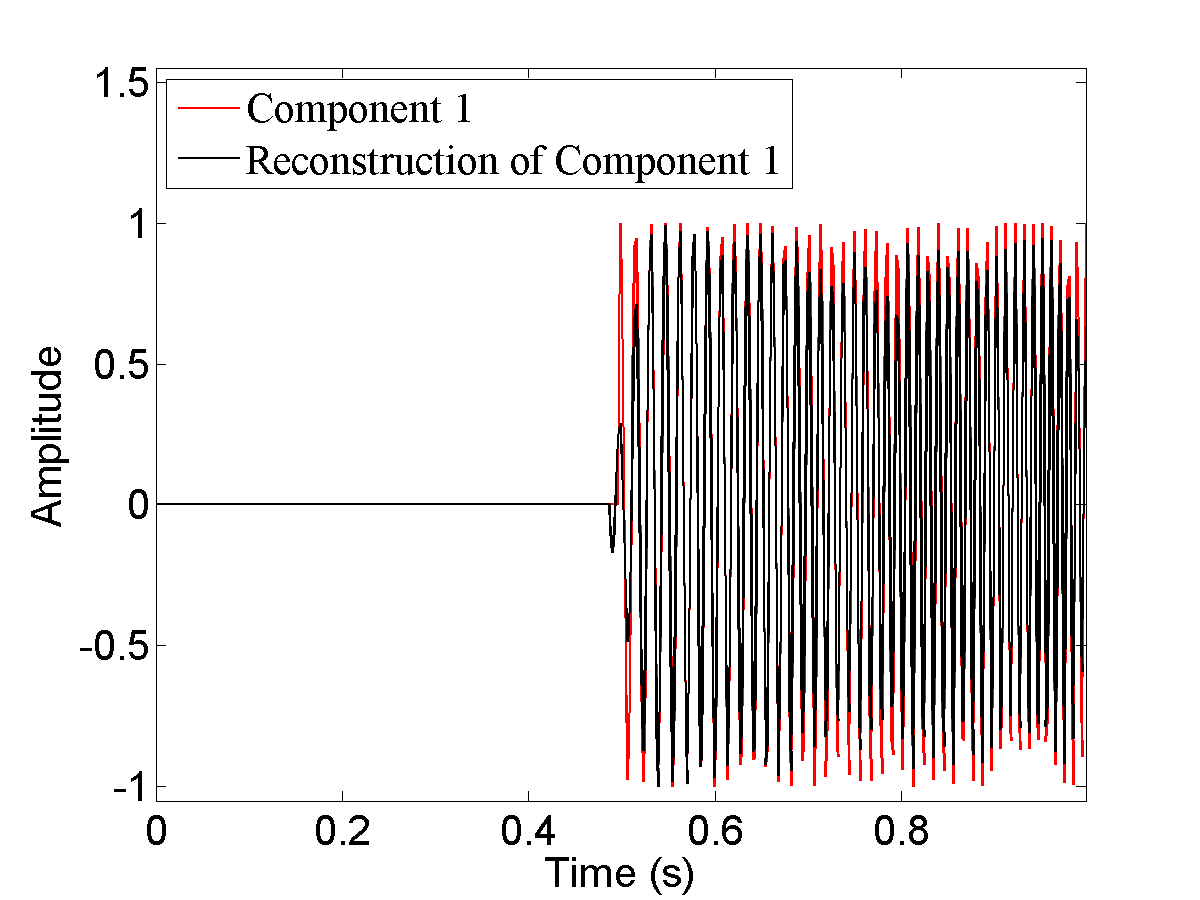}} &
\resizebox{2.40in}{1.6in}{\includegraphics{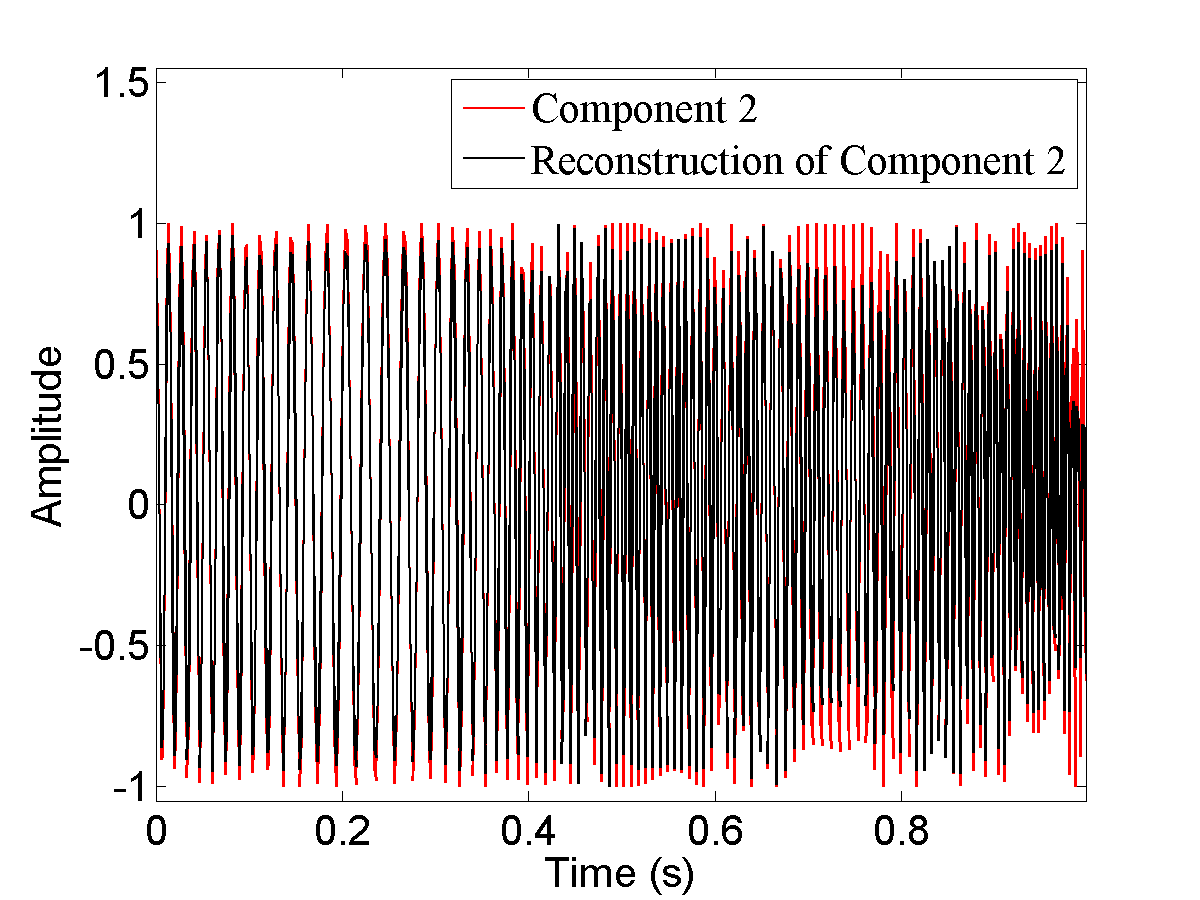}} \\
\resizebox{2.40in}{1.6in}{\includegraphics{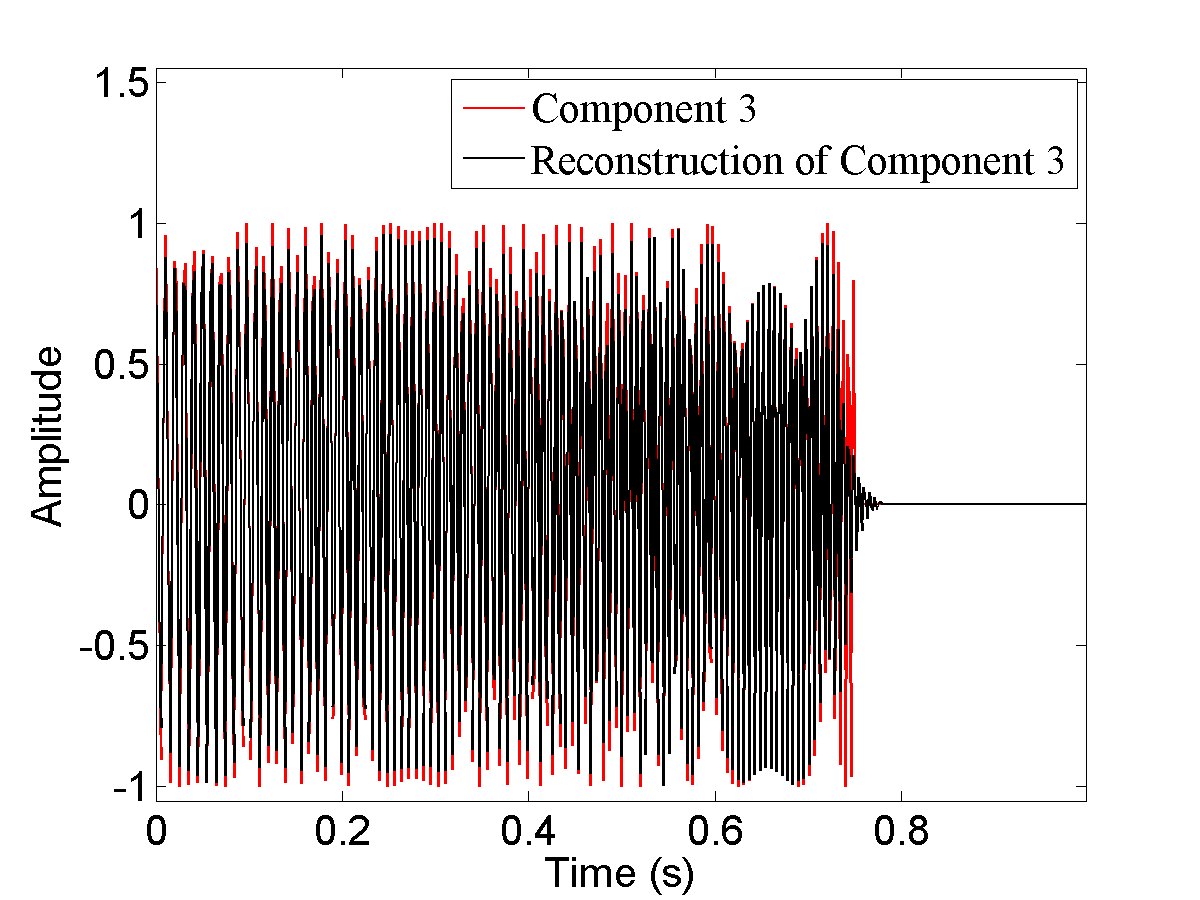}} &
\resizebox{2.40in}{1.6in}{\includegraphics{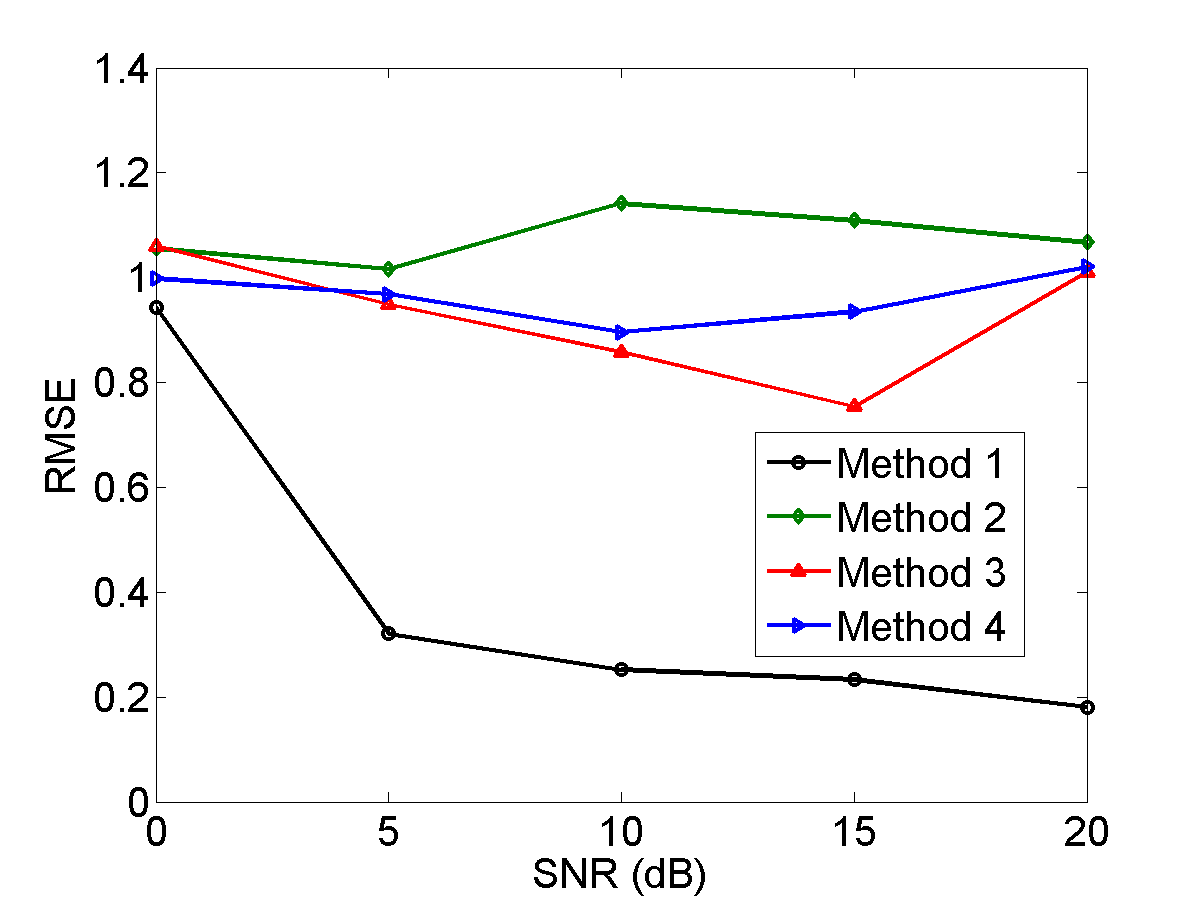}} \\
\end{tabular}
\caption{\small 
Reconstruction results of the signal in \eqref{three_component_signal}. Top-left, top-right and bottom-left: the reconstructed  $z_1(t)$, $z_2(t)$ and $z_3(t)$ by the 2nd-order adaptive FSST.
Bottom-right: RMSEs under different SNRs with various methods (Method 1: 2nd-order adaptive FSST;
Method 2:  regular-PT adaptive FSST;  Method 3: 2nd-order regular-PT adaptive FSST;  Method 4:  conventional 2nd-order FSST).
}
\label{fig:Reconstruction_three_component}
\end{figure}

Due to the page limitation of the paper, we just provide the pictures of the reconstructed components
of the three-component signal $z(t)$ in \eqref{three_component_signal} by the 2nd-order adaptive FSST with $\gs_{est2}(t)$ under the noiseless  environment, 
while we provide RMSEs of four different methods, all in Fig.\ref {fig:Reconstruction_three_component}. In the bottom-right panel of Fig.\ref {fig:Reconstruction_three_component} 
for RMSEs, Method 1, 2, 3 and 4 denote the 2nd-order adaptive FSST with $\sigma_{est2}(t)$,
the regular-PT adaptive FSST with $\sigma_{Re}(t)$, the 2nd-order regular-PT adaptive FSST with $\sigma_{Re2}(t)$ 
and the conventional 2nd-order FSST with constant  $\sigma=0.01$, respectively.
This panel gives the RMSEs of these 4 methods when SNR varies from 0dB to 20dB. Under each SNR, we do Monte-Carlo experiment for 50 runs. 
Obviously, the reconstruction error with the 2nd-order adaptive FSST is less than those with other methods.
Observe that when the noise level is high, for example SNR=0dB,  RMSEs for all methods are large. This is mainly due to the fact that in a high level noise environment,  
the IFs of the modes are hardly estimated by the ridge detection process with the local maxima in the TF plane.

\section{Conclusion}
In this paper, we introduce the adaptive short-time Fourier transform (STFT)  with a time-varying parameter and the adaptive STFT-based synchrosqueezing transform (called the adaptive FSST). 
We also introduce the 2nd-order adaptive FSST. We analyze the support zones of the STFTs of
linear frequency modulation (LFM) signals with the Gaussian window function. We develop the well-separated condition for non-stationary signals by using LFM signals to
approximate non-stationary signals during at local time. We propose a method to select the time-varying parameter automatically.
The experimental results on both synthetic and real data demonstrate
that the adaptive FSST is efficient for the instantaneous frequency estimation, sharp representation in the TF and the separation of multicomponent non-stationary signals with fast-varying frequencies.
We will study the theoretical analysis of the adaptive FSST in our future work. In addition,
 our further study will consider other types of time-varying window functions besides the Gaussian window function. In this paper we consider signals of components without crossover IF curves. In the future, we will consider how to recover components with crossover IF curves. 

\bigskip

\n {\bf Acknowledgments:} 
The authors wish to thank Curtis Condon, Ken White, and Al Feng of the Beckman Institute of the University of Illinois for the bat data in  Fig.\ref{fig:adaptiveSST_bat} and for permission to use it in this paper.

\section*{Appendix}

{\bf Proof of Theorem \ref{theo:recover_STFT_para}.} \quad From \eqref{STFT_para_freqdomain}, we have
\begin{eqnarray*}
&&\int_{-\infty}^\infty \wt V_{x}(t, \eta) d\eta=  \int_{-\infty}^\infty\int_{-\infty}^\infty \wh x(\zeta)\wh g_{\gs(t)}(\eta-\zeta)e^{i2\pi t \zeta}d\zeta d\eta
=\int_{-\infty}^\infty \wh x(\zeta) e^{i2\pi t \zeta}   \int_{-\infty}^\infty\wh g_{\gs(t)}(\eta-\zeta)d\eta d\zeta\\
&&=\int_{-\infty}^\infty \wh x(\zeta) e^{i2\pi t \zeta}   \int_{-\infty}^\infty\wh g_{\gs(t)}(\eta)d\eta  d\zeta
=\int_{-\infty}^\infty\wh g_{\gs(t)}(\eta) e^{i2\pi \cdot 0 \cdot \eta}  d\eta \int_{-\infty}^\infty \wh x(\zeta) e^{i2\pi t \zeta}  d\zeta\\
&&=g_{\gs(t)}(0) x(t)=\frac {g(0)}{\gs(t)}x(t),
\end{eqnarray*}
where exchanging the order of $d\eta$ and $d\zeta$ follows from the Fubini's theorem. This shows \eqref{STFT_recover_para}.

To prove \eqref{STFT_recover_real_para}, note that
for real-valued $x(t)$, since $g_{\gs(t)}(\tau)$ is real-valued, we have
$\wt V_{x}(t, -\eta) =\overline{\wt V_{x}(t, \eta)}.$ 
Hence, from \eqref{STFT_recover_para}, we have
\begin{eqnarray*}
\frac {g(0)}{\gs(t)} x(t)\hskip -0.6cm &&=
 \int_0^\infty \wt V_{x}(t, \eta)  d\eta +\int_{-\infty}^0 \wt V_{x}(t, \eta)  d\eta
= \int_0^\infty \wt V_{x}(t, \eta)  d\eta + \int_0^\infty  \wt V_{x}(t, -\eta) d\eta \\
 && = \int_0^\infty \wt V_{x}(t, \eta)  d\eta + \overline{ \int_0^\infty \wt V_{x}(t, \eta)  d\eta}
=2 {\rm Re} \Big( \int_0^\infty \wt V_{x}(t, \eta)  d\eta\Big).
\end{eqnarray*}
Thus \eqref{STFT_recover_real_para} holds.  \hfill$\blacksquare$

\bigskip

{\bf Proof of Theorem  \ref{theo:2nd_phase_para}.}  Here we will show that
$\go^{adp, 2nd}_s(t, \eta)=c+rt$ for $s(t)$ given by
\begin{equation}
\label{def_chirp_At}
s(t)=A(t) e^{i2\pi \phi(t)}=Ae^{pt+\frac q2 t^2} e^{i2\pi (ct +\frac 12 r t^2)}
\end{equation}
where $p, q$ are two real constants.

From $s'(t)=\big(p+qt+i2\pi (c+rt )\big) s(t)$ and
$$
\wt V_s(t, \eta)=\int_{-\infty}^\infty s(t+\tau) \frac 1{\gs(t)}g(\frac \tau{\gs(t)})e^{-i2\pi \eta \tau} d\tau,
$$
we have
\begin{eqnarray*}
&&\frac {\partial} {\partial t} \wt V_s(t, \eta)=\int_{-\infty}^\infty s'(t+\tau)\;\frac 1{\gs(t)}g(\frac \tau{\gs(t)})e^{-i2\pi \eta \tau} d\tau + \int_{-\infty}^\infty s(t+\tau) (-\frac {\gs'(t)}{\gs(t)^2})g(\frac \tau{\gs(t)})e^{-i2\pi \eta \tau} d\tau\\
&&\qquad + \int_{-\infty}^\infty s(t+\tau) (-\frac {\gs'(t) \tau }{\gs(t)^3}) g'(\frac \tau{\gs(t)}) e^{-i2\pi \eta \tau} d\tau\\
&&=(p+qt+i2\pi (c+rt)\big) \wt V_s(t, \eta)+(q+i 2\pi r)   \int_{-\infty}^\infty \tau s(t+\tau)\;\frac 1{\gs(t)}g(\frac \tau{\gs(t)})e^{-i2\pi \eta \tau} d\tau\\
&&\qquad - \frac {\gs'(t)}{\gs(t)}\wt V_s(t, \eta) -\frac {\gs'(t)}{\gs(t)}\wt V^{\tau g'(\tau)}_s(t, \eta)\\
&&=\big(p+qt+i2\pi (c+rt) -\frac {\gs'(t)}{\gs(t)}\big) \wt V_s(t, \eta)+(q+i 2\pi r) \gs(t) \wt V^{\tau g(\tau)}_s(t, \eta)
 -\frac {\gs'(t)}{\gs(t)}\wt V^{\tau g'(\tau)}_s(t, \eta)
\end{eqnarray*}
Thus, if $\wt V_s(t, \eta)\not=0$, we have
\begin{equation}
\label{2nd_para_derivation}
\frac {\frac{\partial}{\partial t} \wt V_s(t, \eta)}{\wt V_s(t, \eta)}=
p+qt- \frac {\gs'(t)}{\gs(t)}+i2\pi (c+rt) +(q+i 2\pi r)  \gs(t)\; \frac {\wt V^{\tau g(\tau)}_s(t, \eta)}{\wt V_s(t, \eta)}- \frac {\gs'(t)}{\gs(t)} \frac {\wt V^{\tau g'(\tau)}_s(t, \eta)}{\wt V_s(t, \eta)}.
\end{equation}
Taking partial derivative $\frac{\partial}{\partial \eta}$ to both sides of \eqref{2nd_para_derivation},
$$
\frac{\partial}{\partial \eta}\Big(\frac {\frac{\partial}{\partial t} \wt V_s(t, \eta)}{\wt V_s(t, \eta)}\Big)=
(q+i 2\pi r) \gs(t)\; \frac{\partial}{\partial \eta}\Big(\frac {\wt V^{\tau g(\tau)}_s(t, \eta)}{\wt V_s(t, \eta)}\Big)- \frac {\gs'(t)}{\gs(t)}
\frac{\partial}{\partial \eta}\Big(\frac {\wt V^{\tau g'(\tau)}_s(t, \eta)}{\wt V_s(t, \eta)}\Big).
$$
Therefore, if in addition, $\frac{\partial}{\partial \eta}\Big(\frac {\wt V^{\tau g(\tau)}_s(t, \eta)}{\wt V_s(t, \eta)}\Big)\not=0$,   then $(q+i 2\pi r)\gs(t)=P_0(t, \eta)$, where $P_0(t, \eta)$ 
is defined by \eqref{def_R0}.

Back to \eqref{2nd_para_derivation} , we have
$$
\frac {\frac{\partial}{\partial t} \wt V_s(t, \eta)}{\wt V_s(t, \eta)}=
p+qt- \frac {\gs'(t)}{\gs(t)}+i2\pi (c+rt) + P_0(t, \eta) \frac { \wt V^{\tau g(\tau)}_s(t, \eta)}{\wt V_s(t, \eta)}- \frac {\gs'(t)}{\gs(t)} \frac {\wt V^{\tau g'(\tau)}_s(t, \eta)}{\wt V_s(t, \eta)}.
$$
Hence,
$$
\phi'(t)=c+rt ={\rm Re}\Big\{\frac {\frac{\partial}{\partial t} \wt V_s(t, \eta)}{i2\pi \wt V_s(t, \eta)}\Big\}
- {\rm Re}\Big\{ \frac{\wt V^{\tau g(\tau)}_s(t, \eta)}{i2\pi \wt V_s(t, \eta)} P_0(t, \eta)\Big\}
+ \frac {\gs'(t)}{\gs(t)} {\rm Re}\Big\{ \frac {\wt V^{\tau g'(\tau)}_s(t, \eta)}{i2\pi \wt V_s(t, \eta)} \Big\}.
$$
Thus for a signal  $x(t)$
given by \eqref{def_chirp_At}, at $(t, \eta)$ where $\frac{\partial}{\partial \eta}\Big( \frac {\wt V^{\tau g(\tau)}_x(t, \eta)}{\wt V_x(t, \eta)}\Big)\not=0$ and $\wt V_x(t, \eta)\not=0$,
$\go^{adp, 2nd}_x(t, \eta)$ defined by \eqref{2nd_phase_para} is
$\phi'(t)=c+rt$, the IF of $x(t)$. This shows Theorem  \ref{theo:2nd_phase_para}.
\hfill $\blacksquare$


\begin{thebibliography}{10}

\bibitem{Leon_Cohen}  L. Cohen, {\it Time-frequency Analysis}, Prentice Hall, New Jersey, 1995.

\bibitem{Flandrin99} P. Flandrin, {\it Time-frequency/Time-scale Analysis}, Wavelet Analysis and its Applications, vol. 10, Academic Press Inc., San Diego, CA, 1999.


\bibitem{Stankovic13} L. Stankovi$\acute {\rm c}$, M. Dakovi$\acute {\rm c}$,  and T. Thayaparan, {\it Time-Frequency Signal Analysis with Applications}, Artech House, Boston,  2013.


\bibitem{Hlawatsch92} F. Hlawatsch and G.F. Boudreaux-Bartels, ``Linear and quadratic TF signal representations," IEEE Signal Proc. Magazine, vol. 9, no. 2, pp. 21--67, 1992.

\bibitem{Mallat99} S. Mallat, {\it A Wavelet Tour of Signal Processing}, Academic press, 1999.

\bibitem{MODFM16} S. Meignen, T. Oberlin, P. Depalle,  P. Flandrin, and S. McLaughlin,
``Adaptive multimode signal reconstruction from time–frequency representations,"  {\it Phil. Trans. Royal Soc. A,} vol. 374, no. 2065, Apr. 2016.

\bibitem{Choi89} H. Choi and W. Williams, ``Improved TF representation of multicomponent signals using exponential kernels,"  {\it IEEE Trans. Acoustics and Speech}, vol. ASSP- 37, no. 6, pp. 862--871, Jun. 1989.
 
\bibitem{Stank94} L. Stankovi$\acute {\rm c}$, ``A method for TF signal analysis," {\it IEEE Trans. Signal Proc.}, vol. 42, no.1, pp. 225--229, Jan. 1994. 
 
\bibitem{Stank09} S. Stankovi$\acute {\rm c}$, I. Orovic, and C. Ioana, ``Effects of Cauchy integral formula discretization on the precision of IF estimation: unified approach to complex-lag distribution and its L-Form," {\it IEEE Signal Proc. Letters}, vol. 16, no. 4, pp. 307--310, Apr. 2009.

\bibitem{HMB04} H. Hassanpour, M. Mesbah and B. Boashash, ``SVD-based TF feature extraction for newborn EEG seizure," {\it EURASIP Journal on Advances in Signal Proc.}, vol. 16, pp. 2544--2554, 2004.


\bibitem{Stank06} L. Stankovi$\acute {\rm c}$, T. Thayaparan, and M. Dakovi$\acute {\rm c}$, ``Signal decomposition by using the S-method with application to the analysis of HF radar signals in sea-clutter," {\it IEEE Trans. Signal Proc.}, vol. 54, no. 11,
 pp. 4332--4342, Nov. 2006.


\bibitem{Stank18} L. Stankovi$\acute {\rm c}$,  D. Mandi$\acute {\rm c}$, M. Dakovi$\acute {\rm c}$, 
and M. Brajovi$\acute {\rm c}$, ``Time-frequency decomposition of multivariate multicomponent signals,\rq\rq{} {\it Signal Proc.}, vol. 142, pp. 468--479, Jan. 2018.


\bibitem{Huang98}  N.E. Huang, Z. Shen, S.R. Long, M.L. Wu, H.H. Shih, Q. Zheng, N.C. Yen, C.C. Tung,  and H.H. Liu, ``The empirical mode decomposition and Hilbert spectrum for nonlinear and nonstationary
time series analysis,''  {\it Proc. Roy. Soc. London A}, vol. 454, no. 1971, pp. 903--995, Mar. 1998.


\bibitem{A_Flandrin_reassignment95}  F. Auger and P. Flandrin, ``Improving the readability of TF and  TF representations by the reassignment method,''  {\it IEEE Trans. Signal Proc.}, vol. 43, no. 5, pp. 1068--1089, 1995.

\bibitem{Daub_Maes96} I. Daubechies and S. Maes,  ``A nonlinear squeezing of the continuous wavelet transform based on auditory nerve models,''  in A. Aldroubi, M. Unser  Eds. {\it Wavelets in Medicine and Biology}, CRC Press, 1996, pp. 527--546.

\bibitem{Flandrin04} P. Flandrin, G. Rilling, and
P. Goncalves, ``Empirical mode decomposition as a filter bank," {\it IEEE Signal Proc. Letters}, vol. 11,  pp. 112--114, Feb. 2004.


\bibitem{Huang_Wu06_review} N.E. Huang and Z. Wu, ``A review on Hilbert--Huang transform: Method and its applications to geophysical studies,'' {\it Rev. Geophys.}, vol. 46, no. 2,  June 2008.

\bibitem{Rilling08} G. Rilling and P. Flandrin, ``One or two frequencies? The empirical mode decomposition answers," {\it IEEE Trans. Signal Proc.}, vol. 56, pp. 85--95, Jan. 2008.


\bibitem{Wu_Huang09} Z. Wu and N.E. Huang, ``Ensemble empirical mode decomposition: A
noise-assisted data analysis method,''  {\it Adv. Adapt. Data Anal.}, vol. 1,
no. 1, pp. 1--41,  Jan. 2009.

\bibitem{Li_Ji09} L. Li and H. Ji, ``Signal feature extraction based on improved EMD method," {\it Measurement}, vol. 42, pp. 796--803, June 2009.

\bibitem{HM_Zhou09} L. Lin, Y. Wang, and H.M. Zhou, ``Iterative filtering as an alternative algorithm for empirical mode decomposition,''  {\it Adv. Adapt. Data Anal.}, vol. 1, no. 4, pp. 543--560, Oct. 2009.

\bibitem{ZPTL18} J.D. Zheng, H.Y. Pan, T. Liu, Q.Y. Liu, ``Extreme-point weighted mode decomposition," {\it Signal Proc.}
vol. 42,  pp. 366--374, Jan. 2018.

\bibitem{ShPa18} R.R. Sharma and R.B. Pachori, ``Improved eigenvalue decomposition-based approach for reducing cross-terms in Wigner–Ville distribution," {\it Circuits, Systems, and Signal Proc.}, vol. 37,  no. 8, pp. 3330--3350, Aug. 2018.

\bibitem{Oberlin12a} T. Oberlin, S. Meignen, and V.  Perrier,
``An alternative formulation for the empirical mode decomposition,''  {\it IEEE Trans. Signal Proc.},
vol. 60, no. 5,  pp. 2236--2246, May 2012.

\bibitem{Daub_Lu_Wu11} I. Daubechies, J. Lu, and H.-T. Wu, ``Synchrosqueezed wavelet transforms:
An empirical mode decomposition-like tool,''  {\it Appl. Comput. Harmon. Anal.}, vol. 30, no. 2, pp. 243--261, Mar. 2011.

\bibitem{Thakur_Wu11} G. Thakur and H.-T. Wu, ``Synchrosqueezing based recovery of instantaneous frequency from nonuniform samples," {\it SIAM J. Math. Anal.}, vol. 43, no. 5, pp. 2078--2095, 2011.

\bibitem{Wu_thesis} H.-T. Wu, {\it Adaptive Analysis of Complex Data Sets},  Ph.D. dissertation,
Princeton Univ., Princeton, NJ,  2012.


\bibitem{MOM14} T. Oberlin, S. Meignen, and V. Perrier, ``The Fourier-based synchrosqueezing
transform,''  in {\it Proc. 39th Int. Conf. Acoust., Speech,
Signal Proc. (ICASSP)}, 2014, pp. 315--319.

\bibitem{Thakur_etal_Wu13} G. Thakur, E. Brevdo, N. Fu$\check{\rm c}$kar, and H.-T. Wu, ``The synchrosqueezing algorithm for time-varying spectral analysis: Robustness properties and new paleoclimate applications,''  {\it Signal Proc.}, vol. 93, no. 5, pp. 1079--1094, 2013.

 \bibitem{Iatsenko15} D. Iatsenko, P.-V. E. McClintock, and A. Stefanovska, ``Linear and synchrosqueezed TF representations revisited: Overview, standards of use, resolution, reconstruction, concentration, and algorithms," {\it  Digital Signal Proc.},  vol. 42,  pp. 1--26, July 2015.

\bibitem{Meignen17} S. Meignen, D.-H. Pham, and S. McLaughlin, ``On demodulation, ridge detection and synchrosqueezing for multicomponent signals," {\it IEEE Trans. Signal Proc.}, vol. 65, no. 8, pp. 2093--2103, Apr. 2017.

 \bibitem{MOM15} T. Oberlin, S. Meignen, and V. Perrier,``Second-order synchrosqueezing transform or invertible reassignment? Towards ideal TF representations,''  {\it  IEEE Trans. Signal Proc.},
vol. 63, no. 5, pp.1335--1344, Mar. 2015.

\bibitem{OM17} T. Oberlin and S. Meignen, ``The 2nd-order wavelet synchrosqueezing transform,''  in {\it  2017 IEEE International Conference on Acoustics, Speech and Signal Processing (ICASSP)}, March 2017, New Orleans, LA, USA.

\bibitem{FACMF17} D. Fourer, F. Auger, K. Czarnecki, S. Meignen, and P. Flandrin, ``Chirp rate and instantaneous frequency
estimation: application to recursive vertical synchrosqueezing," {\it IEEE Signal Processing Letters}, vol. 24,
no. 11, pp. 1724--1728, 2017.

\bibitem{BMO18} R. Behera, S. Meignen, and T. Oberlin, ``Theoretical analysis of the 2nd-order synchrosqueezing transform,"
{\it Appl. Comput. Harmon. Anal.},   vol. 45, no. 2, pp. 379--404, Sep. 2018.


\bibitem{Pham17} 	D.-H. Pham and S. Meignen, ``High-order synchrosqueezing transform for multicomponent signals analysis - With an application to gravitational-wave signal," {\it IEEE Trans. Signal Proc.}, vol. 65, no. 12, pp. 3168--3178, June 2017.

\bibitem{Li_Liang12} C. Li and M. Liang, ``A generalized synchrosqueezing transform for
enhancing signal TF representation,''  {\it Signal Proc.}, vol. 92, no. 9, pp. 2264--2274, 2012.


\bibitem{Chui_Walt15} C.K. Chui and M.D. van der Walt, ``Signal analysis via instantaneous frequency estimation of signal components,''  {\it Int'l  J.  Geomath.}, vol. 6, no. 1, pp. 1--42, Apr. 2015.

\bibitem{Yang15} H.Z. Yang, ``Synchrosqueezed wave packet transforms and diffeomorphism based spectral analysis for 1D general mode decompositions," {\it Appl. Comput. Harmon. Anal.}, vol. 39, no.1,  pp.33--66, 2015.

\bibitem{S_transform_SST15} Z.-L. Huang, J. Z.  Zhang, T. H.  Zhao, and Y. B. Sun, ``Synchrosqueezing S-transform and its application in seismic spectral decomposition," {\it IEEE Trans. Geosci. Remote Sensing}, vol. 54, no. 2, pp. 817--825, Feb. 2016.

\bibitem{Chui_Lin_Wu15} C.K. Chui, Y.-T. Lin, and H.-T. Wu,  ``Real-time dynamics acquisition from irregular samples - with application to anesthesia evaluation,'' {\it Anal. Appl.}, vol. 14, no. 4,  pp.537--590,   July 2016.

\bibitem{Daub_Wang_Wu15} I. Daubechies, Y. Wang, and H.-T. Wu, ``ConceFT: Concentration of frequency and time via a multitapered synchrosqueezed transform,'' 
{\it Phil. Trans. Royal Soc. A,} vol. 374, no. 2065, Apr. 2016.

\bibitem{Wang_etal14} S. Wang, X. Chen, G. Cai, B. Chen, X. Li, and Z. He, ``Matching demodulation
transform and synchrosqueezing in TF analysis,''
{\it IEEE Trans. Signal Proc.}, vol. 62, no. 1, pp. 69--84, Jan. 2014.

\bibitem{Jiang_Suter17} Q.T. Jiang and B.W.  Suter,  ``Instantaneous frequency estimation based on synchrosqueezing wavelet transform,'' {\it  Signal Proc.}, vol. 138,   pp.167--181, 2017.

\bibitem{YangY14} H.Z. Yang and L.X. Ying, ``Synchrosqueezed curvelet transform for two-dimensional mode decomposition," {\it SIAM J. Math Anal.}, vol 46, no. 3,  pp.2052--2083, 2014.


\bibitem{Chui_Mhaskar15}  C.K. Chui and H.N. Mhaskar,  ``Signal decomposition and analysis via extraction of frequencies,'' {\it Appl. Comput. Harmon. Anal.}, vol. 40, no. 1, pp. 97--136, 2016.

\bibitem{LCJJ17}  L. Li, H.Y. Cai, Q.T.  Jiang and H.B.  Ji,  ``An empirical signal separation algorithm based on linear TF analysis," {\it Mechanical Systems and Signal Proc.}, vol. 121, pp. 791--809, Apr. 2019.

\bibitem{Yang18} H.Z. Yang, ``Statistical analysis of synchrosqueezed transforms," {\it Appl. Comput. Harmon. Anal.}, vol. 45, no. 3, pp. 526--550, Nov. 2018.

\bibitem{ZYLD18} Z.C. Zhang, T. Yu, M.K. Luo, and K. Deng, ``Estimating instantaneous frequency based on phase derivative and linear canonical transform with optimised computational speed," {\it IET Signal Proc.}, vol.12, no.5,  pp. 574--580, Jul. 2018.

 \bibitem{Li_Liang_fault12} C.  Li and M.  Liang, ``Time frequency signal analysis for gearbox fault diagnosis using a generalized synchrosqueezing transform," {\it Mechanical Systems and Signal Proc.}, vol. 26, pp. 205--217, 2012.

\bibitem{WCSGTZ18} S.B. Wang, X.F. Chen, I.W. Selesnick, Y.J. Guo,  C.W. Tong and X.W. Zhang,
``Matching synchrosqueezing transform: A useful tool for characterizing signals with fast varying instantaneous frequency and application to machine fault diagnosis," {\it Mechanical Systems and Signal Proc.}, 
vol. 100, pp. 242--288, Feb. 2018.


\bibitem{Yang_Crystal_15} H.Z. Yang, J.F. Lu, and L.X. Ying,
``Crystal image analysis using 2D synchrosqueezed transforms," {\it Multiscale Modeling $\&$ Simulation}, vol. 13,  no. 4, pp. 1542--1572, 2015.

\bibitem{Yang_Crystal_18} J.F. Lu and H.Z. Yang,  ``Phase-space sketching for crystal image analysis based on synchrosqueezed transforms," {\it SIAM J. Imaging Sci.}, vol. 11, no. 3, pp.1954--1978, 2018.


\bibitem{HLY18} K. He, Q. Li, and Q. Yang,  ``Characteristic analysis of welding crack acoustic emission signals using synchrosqueezed wavelet transform," {\it J. Testing and Evaluation}, vol. 46, no. 6, pp. 2679--2691, 2018.

\bibitem{Wu_breathing14} H.-T. Wu, Y.-H. Chan, Y.-T.  Lin, and Y.-H. Yeh, ``Using synchrosqueezing transform to discover breathing dynamics from ECG signals,'' {\it Appl. Comput. Harmon. Anal.},  vol. 36, no. 2, pp. 354--459,  Mar. 2014.

\bibitem{Wu_sleep15}  H.-T. Wu, R. Talmon, and Y.L.  Lo, ``Assess sleep stage by modern signal processing techniques," {\it IEEE Trans. Biomedical Engineering}, vol. 62, no. 4, 1159--1168, 2015.

\bibitem{Wu_heartbeat17} C.L. Herry, M. Frasch,  A.J. Seely, and H.-T. Wu, ``Heart beat classification from single-lead ECG using the synchrosqueezing transform," {\it Physiological Measurement}, vol. 38, no. 2, Jan. 2017.

\bibitem{Jones94} D.L. Jones and R.G. Baraniuk, ``A simple scheme for adapting TF representations,"  {\it IEEE Trans. Signal Proc.}, vol. 42, no. 12, pp.  3530--3535, Dec. 1994.

\bibitem{Czerwinski97} N. Czerwinski and D.L. Jones, ``Adaptive short-time Fourier analysis,"  {\it IEEE Signal Proc. Letters}, vol. 4, no. 2,  pp. 42--45 , Feb. 1997.

\bibitem{Katkovnik98} V. Katkovnik and L. Stankovi$\acute {\rm c}$, ``Instantaneous frequency estimation using the Wigner distribution with varying and data-driven window length," {\it IEEE Trans. Signal Proc.}, vol. 46, no. 9, pp. 2315--2325, Sep. 1998.

\bibitem{Zhong10} J.G. Zhong and Y. Huang, ``Time-frequency representation based on an
 adaptive short-time Fourier transform," {\it IEEE Trans. Signal Proc.}, vol. 58, no. 10, pp. 5118--5128, Oct. 2010.

\bibitem{Stankovic01}L. Stankovi$\acute {\rm c}$, ``A measure of some TF distributions concentration," {\it Signal Proc.}, vol. 81, no. 3, pp. 621-631, 2001.

\bibitem{Wu17} Y.-L. Sheu, L.-Y. Hsu, P.-T. Chou, and H.-T. Wu, ``Entropy-based time-varying window width selection for nonlinear-type TF analysis," {\it Int'l J. Data Sci. Anal.}, vol. 3,  pp. 231--245, 2017.

\bibitem{Saito17} A. Berrian and N. Saito, ``Adaptive synchrosqueezing based on a quilted short-time Fourier transform,"
arXiv:1707.03138v5, Sep. 2017.

 \bibitem{Baraniuk01} R. Baraniuk, P. Flandrin, A. Janssen, and O. Michel,  ``Measuring TF information content using the R${\rm \acute e}$nyi entropies," {\it  IEEE Trans. Inform. Theory}, vol. 47, no. 4, pp. 1391--1409, 2001.

\bibitem{SP18} V. Sharma and A. Parey, ``Performance evaluation of decomposition methods to diagnose leakage in a reciprocating compressor under limited speed variation," {\it Mechanical Systems and Signal Proc.}, in press, 2018, https://doi.org/10.1016/j.ymssp.2018.07.029.

\end{thebibliography}
\end{document}